\documentclass{aa}

\usepackage{graphicx}
\usepackage{xcolor}
\usepackage[position=top]{subfig}
\usepackage[fleqn]{amsmath}
\usepackage[utf8]{inputenc}
\usepackage{epsfig,amsmath,amssymb}
\usepackage[varg]{txfonts}
\usepackage{multirow}

\usepackage{tcolorbox}

\definecolor{myblue}{rgb}{0.00, 0.0, 0.9}
\definecolor{myred}{rgb}{0.90, 0.0, 0.0}
\definecolor{mygreen}{rgb}{0.0, 0.7, 0.0}
\usepackage[breaklinks,  citecolor=myblue, linkcolor=myred, urlcolor=purple, colorlinks=true, debug, baseurl=' ']{hyperref}

\titlerunning{LOS tomography of the dust polarization sky}

\authorrunning{Pelgrims et al.}

\begin{document}

%

%
\title{Starlight-polarization-based tomography of the magnetized interstellar medium: \\ \textsc{Pasiphae}'s line-of-sight inversion method}

\author{V. Pelgrims
      \inst{1,2}\fnmsep\thanks{pelgrims@physics.uoc.gr}\fnmsep,
      G.~V. Panopoulou
      \inst{3},
      K. Tassis\inst{1,2}
      V.~Pavlidou\inst{1,2},
A. Basyrov\inst{4},
D.~Blinov\inst{1,2},
E.~Gjerl{\o}w\inst{4},
S.~Kiehlmann\inst{1,2},
N.~Mandarakas\inst{1,2},
A.~Papadaki\inst{1,2,5},
R.~Skalidis\inst{1,2},
A.~Tsouros\inst{1,2},
R.~M.~Anche\inst{6,7},\\
H.~K.~Eriksen\inst{4},
T.~Ghosh\inst{8},
J.~A.~Kypriotakis\inst{1,2},
S.~Maharana\inst{1,2,7},
E.~Ntormousi\inst{1,2,9}, 
T.~J.~Pearson\inst{10},\\
S.~B.~Potter\inst{11,12},
A.~N.~Ramaprakash\inst{1,7,10},
A.~C.~S.~Readhead\inst{10},
I.~K.~Wehus\inst{4}}
          
\institute{
Institute of Astrophysics, Foundation for Research and Technology-Hellas, N. Plastira 100, Vassilika Vouton, GR-71110 Heraklion, Greece
\and
Department of Physics, and Institute for Theoretical and Computational Physics, University of Crete, Voutes University campus, GR-70013 Heraklion, Greece
\and
Hubble Fellow, California Institute of Technology, MC350-17, 1200 East California Boulevard, Pasadena, CA 91125, USA
\and
Institute of Theoretical Astrophysics, University of Oslo, P.O. Box 1029 Blindern, NO-0315 Oslo, Norway
\and
Institute of Computer Science, Foundation for Research and Technology-Hellas, GR-71110 Heraklion, Greece
\and
Department of Astronomy/Steward Observatory,
Tucson, AZ, 85721-0065, USA
\and
Inter-University Centre for Astronomy and Astrophysics, Post bag 4, Ganeshkhind, Pune, 411007, India
\and
School of Physical Sciences, National Institute of Science Education and Research, HBNI, Jatni 752050, Odisha, India
\and
Scuola Normale Superiore di Pisa, piazza dei Cavalieri 7, 56126 Pisa
\and
Cahill Center for Astronomy and Astrophysics, California Institute of Technology, 1216 E California Blvd, Pasadena, CA, 91125, USA
\and
Department of Physics, University of Johannesburg, PO Box 524, Auckland Park 2006, South Africa
\and
South African Astronomical Observatory, PO Box 9, Observatory, 7935, Cape Town, South Africa
}

\date{Received 29 July 2022 / Accepted 28 December 2022}

\abstract{
We present the first Bayesian method for tomographic decomposition of the plane-of-sky orientation of the magnetic field with the use of stellar polarimetry and distance. This standalone tomographic inversion method presents an important step forward in reconstructing the magnetized interstellar medium (ISM) in three dimensions within dusty regions.
We develop a model in which the polarization signal from the magnetized and dusty ISM is described by thin layers at various distances, a working assumption which should be satisfied in small-angular circular apertures. Our modeling makes it possible to infer the mean polarization (amplitude and orientation) induced by individual dusty clouds and to account for the turbulence-induced scatter in a generic way.
We present a likelihood function that explicitly accounts for uncertainties in polarization and parallax. We develop a framework for reconstructing the magnetized ISM through the maximization of the log-likelihood using a nested sampling method.
We test our Bayesian inversion method on mock data, representative of the high Galactic latitude sky, taking into account realistic uncertainties from \textit{Gaia} and as expected for the optical polarization survey \textsc{Pasiphae} according to the currently planned observing strategy.
We demonstrate that our method is effective at recovering the cloud properties as soon as the polarization induced by a cloud to its background stars is higher than $\sim 0.1\%$ for the adopted survey exposure time and level of systematic uncertainty.
The larger the induced polarization is, the better the method's performance, and the lower the number of required stars.
Our method makes it possible to recover not only the mean polarization properties but also to characterize the intrinsic scatter, thus creating new ways to characterize ISM turbulence and the magnetic field strength.
Finally, we apply our method to an existing data set of starlight polarization with known line-of-sight decomposition, demonstrating agreement with previous results and an improved quantification of uncertainties in cloud properties.
}

\keywords{ISM: dust, magnetic fields, structure --
   polarization -- 
   Methods: statistical
}

\maketitle

\section{Introduction}

Studies of the interstellar medium (ISM) have relied on two-dimensional (2D) projections on the sky until recently. With the advent of sophisticated techniques
and state-of-the-art facilities, astronomy has entered a new realm in which the third dimension can finally be accessed with accuracy, enabling the mapping of the ISM in three dimensions (3D). Astronomers -- and the public -- will soon be able to experience the Universe in 3D flying through  real-world data sets loaded in virtual environments.

\textit{Gaia} data on stellar distances in particular (e.g., \citealt{Gaia2016}; \citealt{GaiaEDR32021}; \citealt{Bailer-Jones2021}) have allowed for the precise localization in 3D space of more than one billion stars in our Galaxy through the accurate determination of stellar parallaxes. Coupling measurements of stellar parallaxes to reddening, Bayesian inversion methods have already been successful at reconstructing 3D maps of the dust density distribution (e.g., \citealt{Green2019}; \citealt{Lallement2019}; \citealt{Leike2019}; \citealt{Leike2020}), leading to stunning 3D images mapping the dust content of the ISM, in the Solar neighborhood from within the first tens of parsec 
and up to much larger distances within the Galactic disk, already covering a substantial fraction of the Galaxy ($6000 \times 6000 \times 800$ pc$^3$ for \citealt{Lallement2019,Lallement2022}).
Such 3D mappings of the ISM content are of great interest for several areas in astrophysics. They shed new light on the dynamics shaping the Galaxy, breaking degeneracies caused by 2D mapping on the 3D shapes of ISM clouds and cloud complexes, their formation mechanisms, and their history \citep[e.g.,][]{Ivanova2021,Bialy2021,Rezaei2022,Zucker2022,Lallement2022,Konstantinou2022}. Ultimately, 3D images of the dust content of the Galaxy could also help in the characterization of  Galactic foregrounds for observational cosmology and extra-galactic astrophysics (e.g., \citealt{Martinez2018}).

Impressive as they may be, 3D reconstructions of the ISM dust distribution are leaving out an important component of the Galaxy: magnetic fields, which are ubiquitous in the ISM. Magnetic fields are relevant in a variety of processes, from regulating star formation (e.g., \citealt{Mouschovias2006}; \citealt{Hennebelle2019}; \citealt{Li2021}) to shaping large-scale structures in the disk and the halo of the Galaxy (e.g., \citealt{Beck2015}).
Magnetic fields in the Galaxy also affect our ability to study the Universe's structure and history, from its first moments to its later ages. Aspherical dust grains line up their shortest axis with the ambient magnetic field (e.g., \citealt{Andersson2015}). As a result, the thermal radiation emitted by those grains is polarized. This emission constitutes the major limitation in cosmologists' search for primordial B modes, the clear proof of primordial gravitational waves from inflation, and cosmic birefringence in the polarization of the cosmic microwave background (CMB) (e.g., \citealt{BICEPKECKPlanck2015}; \citealt{PlanckXXX2016}; \citealt{PlanckXI2020}; \citealt{DiegoPalazuelos2022}).
This emission also represents a foreground in polarization studies of individual extra-galactic objects (e.g., \citealt{Pelgrims2019}).

Significant effort has been invested in the last two decades to characterize the dust-polarized emission in order to disentangle it from the cosmological signal. However, this task has been proven to be very convoluted. Variations in dust spectral emission distribution, either in the plane of the sky (POS) or along the line of sight (LOS) (\citealt{Tassis2015}; \citealt{PlanckL2017}; \citealt{Pelgrims2021}; \citealt{Ritacco2022}), and unexpected signatures of the dust signal in polarization power spectra (e.g., \citealt{PlanckInt2016XXXVIII}) -- all rooted in the tight connection between dust clouds and the magnetic field -- add many layers of complexity. Various sophisticated techniques are being developed to address these problems. The most direct way of attacking them and of providing confident and accurate solutions requires 3D mapping of the Galactic magnetic field in dusty regions (e.g., \citealt{HerviasCaimapo2022}; \citealt{Pelgrims2022};  \citealt{Konstantinou2022}; \citealt{Huang2022}).

Accessing the LOS structure of the magnetic field from dust emission alone is not feasible. Three-dimensional maps of the dust distribution can help identify LOSs with several clouds and place constraints on their respective significance; however, those maps alone provide no information about the magnetic fields permeating those clouds. While they can be combined with maps of dust-polarized emission in a coherent analysis to model the Galactic magnetic field (GMF) on large scales (\citealt{Pelgrims2020}), they cannot provide significant information on cloud scales, with perhaps some exceptions (e.g., \citealt{Rezaei2020}).

\smallskip

Fortunately, there are other probes that make it possible to infer the structure of the magnetized ISM in 3D. Among those, the linear polarization of stars, measured from the near-infrared (NIR) to the near-ultraviolet, is of particular interest, and can be used to study and model the dusty and magnetized ISM, from  the smallest to the largest scales. 
While starlight usually starts out unpolarized, the same aspherical dust grains that are responsible for the polarized thermal emission induce a polarization to it when partially absorbing it, due to dichroic extinction (e.g., \citealt{Andersson2015}). Starlight polarization has been related to the magnetic field and the ISM in the Galaxy since its early observation (e.g., \citealt{Hiltner1949,Hiltner1951}; \citealt{Davis1951}; \citealt{Heiles2000}).
In comparison to other probes of the magnetized ISM, starlight polarization has the significant advantage that it can provide direct 3D information as soon as stellar distances are known.
The potential of such 3D magnetic tomography to recover information on the LOS structure of the magnetic field has been demonstrated recently by \cite{Pan2019a} using data collected from the RoboPol polarimeter (\citealt{Ramaprakash2019}), while correlation analysis of dust-polarized emission at sub-millimeter wavelengths and starlight polarization data has proven useful to locate the distance to the dominant polarized dust emission component seen at high Galactic latitude (\citealt{Skalidis2019}).

In recent years, several regions of the sky have been mapped with a high density of stellar polarization measurements ($>100$ stars per sq.degree), including a significant portion of the inner Galaxy (in the NIR \citealt{GPIPS2020}), as well as more diffuse regions of the ISM (e.g., \citealt{Pan2019a}; \citealt{Skalidis2022}).
These data sets have paved the way to 3D mapping magnetic fields in the general ISM of the Galaxy \citep[e.g.,][]{Pavel2012}, far from the dense regions of star formation that had been traditionally studied with large stellar samples (e.g., \citealt{Pereyra2004}; \citealt{Sugitani2011}; \citealt{Marchwinski2012}; \citealt{Santos2014}; \citealt{Franco2015}; \citealt{Kwon2015}; \citealt{Eswaraiah2017}). In the near future, the \textsc{Pasiphae} survey (\citealt{Tassis2018}) and the SOUTH POL survey (\citealt{Magalhaes2005}) will enable a leap forward by generating stellar polarization data for millions of stars, covering a large fraction of the sky. 
In conjunction with measurements of stellar distances obtained by \textit{Gaia}, those data sets will pave the way for the characterization of the dusty magnetized ISM in 3D.
Since stellar polarization traces the very same medium (magnetized dust) that produces the dominant CMB polarization foreground, starlight polarization data may offer a unique independent means to model out the polarization signal of the Galaxy, allowing the study of the very first moments of the Universe.

The observed polarization of each single star is the integrated effect of dichroic absorption from all interstellar clouds lying between us and the star. For this reason, in order to derive the complex 3D structure of the magnetized ISM from starlight polarization data and stellar distances, we need to develop methods that invert this LOS integration. So far, no standard method has been established in the literature to accomplish such a task in an automated, Bayesian way. On the one hand, different ad hoc methods have been considered (e.g., \citealt{Andersson2006}; \citealt{Pan2019a} ; \citealt{Doi2021}), but they are not easily scalable to large data sets since they are not well adapted for automation and they do not allow for a straightforward, robust estimation of the credible interval of the reconstruction. 
On the other hand, methods developed for extinction data cannot be used unaltered on starlight polarization data. The main reason is that polarization is a pseudo-vector quantity. This implies that it cannot be described by a single scalar quantity --  two are needed: either the degree of polarization and polarization position angle, or its linear Stokes parameters. Additionally, even if contributions from individual clouds are additive (as is the case for linear Stokes parameters in the case of low amounts of  extinction), polarization increments can be either positive, negative, or null.
Because of these fundamental differences, dedicated, specialized methods need to be developed for the problem of starlight polarization tomography.

\smallskip

In this paper, we present such a specialized Bayesian method, developed for the \textsc{Pasiphae} survey, implemented in Python, and now made publicly available for use by the community\footnote{\url{https://github.com/vpelgrims/Bisp_1/}}.
The inversion method developed here works on a per line-of-sight basis. We defer to future work for information on the extension of the method, which must  take the correlation of the solutions in the plane of the sky into account.

In Sect.~\ref{sec:model} we present our model for the distance dependence of starlight polarization along sightlines of the diffuse ISM, and we explain how we built our data equation and derived the corresponding likelihood.
In Sect.~\ref{sec:method} we provide details on the implementation of our Bayesian method and  validate its performance by applying it to two simple examples of mock data.
In Sect.~\ref{sec:performance} we present extensive testing of the performance of the method in the low signal-to-noise (S/N) regime and identify the method's limitations.
We apply the method to currently available data in Sect.~\ref{sec:app2realdata} and compare the results from our method to the literature.
We finally summarize and discuss our work in Sect.~\ref{sec:conclusion}.
This paper contains two appendices. In Appendix~\ref{sec:mock_data_appendix}, we explain the creation of the mock observations used for performance testing, which are based on actual star samples, realistic estimates of the uncertainties on stellar parallax and polarization, and that rely on a complete toy model for the 3D structure of the ISM along sightlines. In Appendix~\ref{sec:IntrSc}, we explore our toy model of the magnetic field geometry to gain intuition on the effects of turbulence-induced fluctuations in the ISM on the polarization observables.

\section{Model, data equation, and likelihood}
\label{sec:model}
In this section, we lay the foundation for a model that describes the distance-dependence of stellar polarization toward a sightline of the diffuse ISM. We construct a generic data equation and build the corresponding likelihood that relates model parameters and star data.
We first discuss the case of a single cloud along the LOS and then proceed to the generalization to cases with multiple-clouds.

\subsection{Model: Thin-layer magnetized clouds}
We model the LOS polarization induced by an individual cloud to background stars as being dominated by a single thin polarizing dust layer at the cloud distance ($d_{\rm{C}}$).
As already described by many authors (e.g., \citealt{Andersson2015}; \citealt{Hensley2021}; \citealt{Draine2021b}), the polarization induced by a dust cloud to the light of background stars depends on the dust opacity at the frequency of observation, the polarizing efficiency (which relates dust reddening $E(B-V)$ to a maximum polarization fraction) and on the apparent 3D orientation of the magnetic field ($\mathbf{B}$) that permeates the cloud. The latter is described through the inclination angle of the magnetic field lines with respect to the POS ($\gamma_\mathbf{B}$) as well as the position angle of the POS component of the magnetic field ($\psi_\mathbf{B}$).
For a single star behind a cloud, with starlight intensity $I_{\rm{V}}$, the vector of its relative Stokes parameters in the visible $(q_{\rm{V}},\,u_{\rm{V}}) = (Q_{\rm{V}}/I_{\rm{V}},\,U_{\rm{V}}/I_{\rm{V}})$ equals the cloud polarization vector $(q_{\rm{C}},\,u_{\rm{C}})$ given by
\begin{equation}
\begin{pmatrix}
q_{\rm{C}} \\
u_{\rm{C}}
\end{pmatrix}
= P_{\rm{max}}\,\cos^2\gamma_\mathbf{B} \, 
\begin{pmatrix}
\cos[2 \, \psi_\mathbf{B}] \\
\sin[2 \, \psi_\mathbf{B}]
\end{pmatrix} \, .
\label{eq:stokes_reg}
\end{equation}
In this equation, in which we have neglected any possible source of noise, $P_{\rm{max}} \approx 13\% \, E(B-V)$ (\citealt{Pan2019b}; \citealt{PlanckXII2020}) where the reddening $E(B-V)$ generally depends on the dust grain physical properties and on the column density.
Using single-frequency starlight polarization only, we have access to the position angle of the POS component of the magnetic field (related to the electric vector position angle, EVPA, of the stellar polarization) and to the magnitude of the induced polarization, that is the degree of polarization: $p_{\rm{C}} = (q_{\rm{C}}^2 + u_{\rm{C}}^2)^{1/2}$ (related to the degree of stellar polarization). The latter is affected by the dust extinction, the dust polarizing efficiency, the inclination of the magnetic field with respect to the POS, and by possible LOS depolarization caused by turbulence within the cloud.

If the distribution of dust and the magnetic field properties were spatially homogeneous within a cloud, a single stellar measurement would suffice to describe the polarization properties it induces.
Considering an ensemble of stars to constrain the cloud polarization properties makes it possible to take advantage of the number statistics and to sample the distance axis to provide constraints on the cloud distance. The former is critical for the S/N regime that is expected at high Galactic latitudes (\citealt{Skalidis2018}).
Furthermore, observations of interstellar dust reveal fluctuations in the dust distribution on a range of scales (e.g., \citealt{Miville-Deschenes2016}) and fluctuations in the density and the magnetic field are also expected as a result of magneto-hydrodynamic (MHD) turbulence (e.g., \citealt{Goldreich1995}; \citealt{Cho2003}; \citealt{Heiles2005}). Therefore, in order to obtain a realistic description of the polarization properties of a cloud within a region, we consider an ensemble of stellar measurements from LOSs within a finite circular aperture (called "beam" in the remainder of the paper) toward the cloud. We describe the Stokes parameters induced by the cloud to the ensemble of stars with a well-defined mean and a measure of dispersion about that mean. We refer to this dispersion as \textit{intrinsic scatter}, to distinguish it from other sources of dispersion in the measurements, such as noise.

The intrinsic scatter has effects on polarization observables. We explore these effects in Appendix~\ref{sec:IntrSc} where we characterize the variations produced by the intrinsic scatter on $p_{\rm{C}}$ and on $\psi_{\mathbf{B}}$; or more generally in the  $(q_{\rm{V}},\,u_{\rm{V}})$ plane.
In particular, we show that 3D variations of the magnetic field can generate biases and a nonzero cross term in the polarization plane. The biases are irrelevant in our case since we are interested in recovering the mean values. The cross term, however, needs to be accounted for given that it might reach a non-negligible fraction of the variance of the Stokes parameters $(q_{\rm{V}},\,u_{\rm{V}})$ (\citealt{Montier2015}).
Other sources of variance in the polarization properties, such as fluctuations of the dust extinction across the sky, may reduce the importance of the off-diagonal element compared to the diagonal elements. Nevertheless, we retain the off-diagonal terms in our analysis for completeness.

As a first approximation, we assume that the turbulence-induced variations generate a bivariate normal distribution about the mean in the $(q_{\rm{V}},\,u_{\rm{V}})$ plane.
As a result, and in absence of observational noise, our stochastic model for the vector of Stokes parameters of a star $i$ in the background of a cloud is
\begin{align}
    \mathbf{m}_i &=
    \begin{pmatrix}
        q_{\rm{C}} \\
        u_{\rm{C}}
    \end{pmatrix}
    + {\rm{G}}_2(0,C_{\rm{int}})_i \;,
\end{align}
where $q_{\rm{C}}$ and $u_{\rm{C}}$ now denote the mean polarization values induced by the cloud. We denote the mean Stokes vector of the cloud as $\bar{\mathbf{m}} = (q_{\rm{C}}\, u_{\rm{C}})^\dagger$.
${\rm{G}}_2(0,C_{\rm{int}})_i$ is a random realization of a 2D bivariate normal distribution centered on $(0,0)$ with the 2-by-2 covariance matrix, $C_{\rm{int}}$. The latter encodes the variances and covariances induced by sources of intrinsic fluctuations (e.g., turbulence) on the Stokes parameters.
According to this generic description, six parameters are necessary to describe the polarization data of stars toward such a cloud: the distance of the cloud ($d_{\rm{C}}$), the mean Stokes parameters $(q_{\rm{V}},\,u_{\rm{V}})$ and three numbers to characterize the intrinsic-scatter covariance matrix $C_{\rm{int}}$.

\subsection{Data equation}
To write the data equation, we need to account for the fact that a star at distance $d_i$ may either be in the foreground ($d_i < d_{\rm{C}}$) or in the background ($d_i > d_{\rm{C}}$) of the cloud. In the former case no polarization is induced by the cloud and in the latter case the star polarization will be given by the mean polarization of the cloud plus one random realization of the intrinsic scatter. This piecewise-constant behavior is implemented through the use of the Heaviside function ($H(x) = 1$ if $x>0$, $0$ otherwise).

We further add a noise term ($\mathbf{n}_i$) to our stochastic model for the Stokes parameters. We consider that the observational noise which results from photon noise, instrumental polarization, and on-sky instrumental calibration, is described by a bivariate normal distribution ${\rm{G}}_2(0,C_{\rm{obs}})$ with  covariance matrix, $C_{\rm{obs}}$, where the off-diagonal terms can be nonzero. Unlike the intrinsic scatter, the covariance matrix corresponding to the observational uncertainties is generally source dependent; it might depend on the source's brightness, for example.
The variance and covariance in the $(q_{\rm{V}},\,u_{\rm{V}})$ plane result from both the intrinsic scatter and the observational uncertainties. We consider them as independent sources of Gaussian scatter in the polarization plane. Therefore, for stars in the background of a cloud the covariance matrices ($C_{\rm{int}}$ and $C_{\rm{obs}}$) are summed. For stars that are not background to a cloud, only the observational uncertainties are relevant.
The total covariance matrix thus takes the form:
\begin{align}
    \Sigma_i &= C_{{\rm{obs}},i} + C_{\rm{int}}\, H(d_i - d_{\rm{C}})
    \label{eq:cov_total_1C}\,.
\end{align}
As a result, the data equation for the case of a single cloud along the LOS is:
\begin{align}
    \mathbf{s}_i &= \mathbf{m}_i \, H(d_i - d_{\rm{C}}) + \mathbf{n}_i \nonumber \\
                &=
                \begin{cases}
                    \bar{\mathbf{m}} + {\mathrm{G}}_2(0,C_{\rm{int}} + C_{{\rm{obs}},i})_i
                        & \text{if } d_i > d_{\rm{C}}\\
                    \mathbf{0} + {\mathrm{G}}_2(0,C_{{\rm{obs}},i})_i & \text{otherwise,}
                \end{cases}
    \label{eq:data_equation_1C}
\end{align}
where $\mathbf{s}_i$ is the vector of the measured Stokes parameters,
$d_i$ is the distance of the star, and $\mathbf{n}_i$ is the source-dependent noise term.
Denoting $\bar{\mathbf{m}}_i$ the mean polarization induced by any dust cloud between us and the star and using Eq.~\ref{eq:cov_total_1C}, we thus write the vector of measured Stokes parameters as a random draw of a bivariate normal distribution with mean $\bar{\mathbf{m}}_i$ (with value either $\mathbf{0}$ or $\bar{\mathbf{m}}$) and 2-by-2 covariance matrix $\Sigma_i$:
\begin{equation}
    \mathbf{s}_i \leftarrow \mathrm{G}_2(\bar{\mathbf{m}}_i,\Sigma_i) = \frac{1}{2\pi \, |\Sigma_i|^{1/2}} \; \exp \left(-\frac{1}{2} \, \mathbf{\eta}_i^\dagger \, \Sigma^{-1} \, \mathbf{\eta}_i \right)
    \label{eq:s_i_1C}
\end{equation}
in which $|\Sigma_i| = \det \left( \Sigma_i \right)$ is the determinant of $\Sigma_i$ and where we have introduced $\mathbf{\eta}_i = \mathbf{s}_i - \bar{\mathbf{m}}_i$ with $\bar{\mathbf{m}}_i = \mathbf{0}$ or $\bar{\mathbf{m}}$ depending on whether the star is in the foreground of background to the cloud.
In Eqs.~\ref{eq:cov_total_1C} to~\ref{eq:s_i_1C}, both the modeled mean and the modeled covariance terms for the stellar polarization depend on whether the source is in the foreground or the background of the cloud.

\subsection{Likelihood}
The data equation (Eq.~\ref{eq:data_equation_1C}) concerns only the Stokes parameters of the stars. As explicitly written, our model for the Stokes parameters relates to the distance of the stars which, in turn, is also a measured quantity that comes with an uncertainty. This source of uncertainty adds complexity to the problem that we wish to solve, as distance uncertainties might impact the model prediction for $\mathbf{m}_i$, in particular for those stars that are near a cloud.
Star distances are nowadays known with great accuracy (e.g., \citealt{Bailer-Jones2021}) through their parallaxes ($\varpi$) which, to good approximation, have Gaussian uncertainties ($\sigma_\varpi$). For this reason we choose to work in terms of parallaxes rather than distances, the two being related through the inverse relation $\varpi = 1/d$ where parallaxes are measured in arc-seconds and distances in parsec.
To account for this extra source of complexity, we notice that the star distance entering the data equation above should be the true distance of the star, and therefore we use the true parallax of the star in the following. We modify the argument of the Heaviside function from $(d_i - d_{\rm{C}})$ to $(\varpi_{\rm{C}} - \varpi_i^0)$, in which $\varpi_i^0$ denotes the true parallax of a star. We consider that the measured parallax ($\varpi_i$) is a random realization of a Gaussian distribution centered on the true parallax with uncertainty $\sigma_{\varpi_i}$:
\begin{equation}
    \mathbf{\varpi}_i \leftarrow \mathrm{G}(\varpi_i^0,\sigma_{\varpi_i}) = \frac{1}{\sqrt{2\pi} \,\sigma_{\varpi_i}} \exp \left(- \frac{(\varpi_i - \varpi_i^0)^2}{2\, {\sigma_{\varpi_i}}^2} \right) \; .
\end{equation}

\smallskip

In this work we assume that the measurements of the parallax and of the optical polarization of stars are independent and uncorrelated. Further, we assume that the Stokes parameters for star polarization are functions of the true parallax of the star through the generic data equation built in the previous section (Eq.~\ref{eq:data_equation_1C}).
With these notations, the likelihood of the observational data point for star $i$ with measured parallax $\varpi_i$ and Stokes' vector $\mathbf{s}_i$ takes the form:
\begin{align}
    &P\left(\varpi_i,\mathbf{s}_i \, | \, {\mathbf{m}}_i ,\,C_{\rm{int}} \,,\, \varpi_i^0,\sigma_{\varpi_i},C_{\rm{obs},i}\right)
    \nonumber
    \\
    & =\, P\left(\varpi_i \,|\, \varpi_i^0,\sigma_{\varpi_i} \right) \,
        P\left( \mathbf{s}_i \,|\, {\mathbf{m}}_i ,\,C_{\rm{int}} \,,\, C_{\rm{obs},i} \right)
    \nonumber
    \\
    & =\, \frac{1}{\sqrt{2 \pi} \sigma_{\varpi_i}}\, {\rm{e}}^{-\frac{(\varpi_i - \varpi_i^0)^2}{2 {\sigma_{\varpi_i}}^2}} \,
    \frac{1}{2 \pi \, |\Sigma_i|^{1/2}}\, {\rm{e}}^{-\frac{1}{2}{\mathbf{\eta}_i}^\dagger {\Sigma_i}^{-1} {\mathbf{\eta}_i}} 
\end{align}
where, in the last line, we explicitly write the parallax likelihood as a 1D Gaussian with standard deviation equal to the observational uncertainty and the polarization likelihood as a 2D Gaussian with the total covariance matrix that accounts for both the observational uncertainties and the contribution to intrinsic scatter from the crossed cloud (Eq.~\ref{eq:cov_total_1C}).

We are not interested in modeling the true parallax of the star ($\varpi_i^0$). Instead we wish to marginalize over it to define the likelihood of the cloud parameters given an observation for star $i$. This marginalization allows us to separate the likelihood into two parts, one corresponding to the background of the cloud and the other to its foreground as follows:
\begin{align}
&\mathcal{L}_i\left(\varpi_{\rm{C}}, \bar{\mathbf{m}}, C_{\rm{int}} \right) \nonumber \\
    & = \int_{0}^{\infty} {\rm{d}}\varpi_i^0 \; P\left(\varpi_i,\mathbf{s}_i \, | \,\varpi_i^0,\, \bar{\mathbf{m}}_i , \,C_{\rm{int}} ,\, \sigma_{\varpi_i},\,C_{{\rm{obs}},i}\right) \nonumber \\
    & = \underbrace{\int_{0}^{\varpi_{\rm{C}}} {\rm{d}}\varpi_i^0 \;
    P\left(\varpi_i,\mathbf{s}_i \, | \,\varpi_i^0,\, \bar{\mathbf{m}}, \,C_{\rm{int}} ,\,\sigma_{\varpi_i},\,C_{{\rm{obs}},i}\right)}_{\text{background}} \nonumber \\
    & \phantom{\hspace{1.5cm}} \; +
    \underbrace{\int_{\varpi_{\rm{C}}}^{\infty} {\rm{d}}\varpi_i^0 \;
    P\left(\varpi_i,\mathbf{s}_i \, | \,\varpi_i^0,\, \mathbf{0},\, \sigma_{\varpi_i},\,C_{{\rm{obs}},i}\right)}_{\text{foreground}}
    \nonumber \\
    & = P\left(\mathbf{s}_i \, | \, \bar{\mathbf{m}},\,C_{\rm{int}} ,\,C_{{\rm{obs}},i}\right)\;
    \int_{0}^{\varpi_{\rm{C}}} {\rm{d}}\varpi_i^0 \;
    P\left(\varpi_i \, |\, \varpi_i^0,\,\sigma_{\varpi_i}\right)
    \nonumber \\
    & \phantom{\hspace{1.5cm}} \; +
    P\left(\mathbf{s}_i \, | \, \mathbf{0},\,C_{{\rm{obs}},i}\right)\;
    \int_{\varpi_{\rm{C}}}^{\infty} {\rm{d}}\varpi_i^0 \;
    P\left(\varpi_i \, |\, \varpi_i^0,\,\sigma_{\varpi_i}\right)
    \nonumber \\
    & = P\left(\mathbf{s}_i \, | \, \bar{\mathbf{m}},\,C_{\rm{int}} ,\,C_{{\rm{obs}},i}\right)\;
    \frac{1}{2}\,\left(1 + {\rm{erf}}\left(\frac{\varpi_{\rm{C}} - \varpi_i}{\sqrt{2} \sigma_{\varpi_i}}\right) \right)
    \nonumber \\
    & \phantom{\hspace{1.5cm}} \; +
    P\left(\mathbf{s}_i \, | \, \mathbf{0},\,C_{{\rm{obs}},i}\right)\;
    \frac{1}{2}\,\left(1 - {\rm{erf}}\left(\frac{\varpi_{\rm{C}} - \varpi_i}{\sqrt{2} \sigma_{\varpi_i}}\right) \right)
    \; .%
    \label{eq:star_likelihood_1C}
\end{align}

\smallskip

For a given sample of stars with polarization measurements and with known parallaxes and uncertainties, and under the assumption that the data are independent, the likelihood of the cloud parameters for a given LOS is given by the product of the likelihoods of the data points:
\begin{equation}
    \mathcal{L}\left( \varpi_{\rm{C}}, {\bar{\mathbf{m}}}, C_{\rm{int}} \right)
    = \prod_{i=1}^{N_{\rm{star}}} \mathcal{L}_i\left( \varpi_{\rm{C}}, {\bar{\mathbf{m}}}, C_{\rm{int}} \right) \;.
    \label{eq:total_likelihood}
\end{equation}
This is the total likelihood function that we need to maximize to constrain our model parameters given the data.

\subsection{Multicloud case}
The generalization of the single-cloud model to the case with multiple independent clouds along the LOS is straightforward (we take $N_c$ to be the number of clouds along the LOS).
We consider that the Stokes parameters of a star in the background of multiple clouds result from the addition of the contributions from individual clouds.
This approximation, which is correct in the low polarization regime (e.g., Appendix~B of \citealt{Pat2010}), is well motivated for translucent LOSs through the diffuse ISM given that dust clouds can be considered as weak polarizers. The validity of this approximation may need to be revised for denser regions of the ISM, such as the Galactic plane or other LOSs through very dense molecular clouds. 

In this work we assume the linearity of the polarization signal and defer to future work the addition of more complex cases.
Hence, in the low-polarization regime, the mean polarization vectors ($\bar{\mathbf{m}}_i^{[k]}$) of clouds along the LOS are additive. The same is true for the covariance matrices from the intrinsic scatter.
Here, we have introduced the superscript $[k]$ to label clouds from $[1]$ to $[N_c]$, for the nearest and the farthest cloud, respectively.
As for the case of a single cloud, the total covariance ($\Sigma_i$) on the Stokes parameters for a star $i$ depends on the observational uncertainties and on the sum of all sources of intrinsic scatter intervening along the LOS, from the star to the observer.
Thus, assuming $N_c$ independent clouds along the LOS, the data equation for a star $i$ is:
\begin{align}
    \mathbf{s}_i = & \, \mathbf{m}_i + \mathbf{n}_i \nonumber \\
    = & \, \bar{\mathbf{m}}_i + {\mathrm{G}}_2\left(0,\Sigma_i\right)
    \label{eq:data_equation},
\end{align}
where
\begin{equation}
    \bar{\mathbf{m}}_i = \sum_{j \leq k} \bar{\mathbf{m}}^{[j]}
\hspace{.3cm} \text{and} \hspace{.3cm}
    \Sigma_i = C_{{\rm{obs}},i} + \sum_{j \leq k} C_{\rm{int}}^{[j]},
    \label{eq:cov_total}
\end{equation}
in which we implicitly assume that star $i$ lies behind cloud $[k]$ and in front of cloud $[k+1]$, if $k < N_c$ and behind all clouds if $k=N_c$.

\smallskip

To determine the likelihood for the set of cloud parameters given star data we marginalize over the true parallax of the star. This step allows us to separate the likelihood into $N_c + 1$ terms, building the likelihood of a mixture model where the different terms correspond to a different (increasing) number of foreground clouds:
\begin{align}
&\mathcal{L}_i\left(\lbrace \varpi_{\rm{C}}^{[k]}, \bar{\mathbf{m}}^{[k]}, C_{\rm{int}}^{[k]} \rbrace \right) \nonumber \\
    & = P_{0,i} \, P\left(\mathbf{s}_i = \mathbf{0} \,|\, C_{{\rm{obs}},i}\right) \nonumber \\
    & \phantom{=} \; + P_{1,i} \, P\left(\mathbf{s}_i = \bar{\mathbf{m}}^{[1]} \,|\, C_{{\rm{obs}},i} + C_{\rm{int}}^{[1]}\right) \nonumber \\
    & \phantom{=} \; + ... \nonumber \\
    & \phantom{=} \; + P_{N_c,i} \, P\left(\mathbf{s}_i = \sum_{k=1}^{N_c} \bar{\mathbf{m}}^{[k]} \,|\, C_{{\rm{obs}},i} + \sum_{k=1}^{N_c} C_{\rm{int}}^{[k]}\right).
    \label{eq:star_likelihood}
\end{align}
The coefficients $P_{k,i}$ result from the integration of the probability density function of the star parallax in the inter-cloud ranges of parallax and take the form
\begin{equation}
    P_{k,i} = \frac{1}{2}\,\left({\rm{erf}}\left(\frac{\varpi_{\rm{C_k}} - \varpi_i}{\sqrt{2} \sigma_{\varpi_i}}\right) - {\rm{erf}}\left(\frac{\varpi_{\rm{C_{k+1}}} - \varpi_i}{\sqrt{2} \sigma_{\varpi_i}}\right) \right) \;.
\end{equation}
Finally, from an ensemble of stars, the likelihood of the cloud parameters for a given LOS is given by the product of the likelihoods of the data points, as in Eq.~\ref{eq:total_likelihood}.

\section{The Bayesian inversion method: Implementation and validation}
\label{sec:method}
In this section we first describe the implementation of our maximum-likelihood method to decompose the polarization properties of clouds along the LOS using stellar polarization and distance.
We validate our method using mock stellar data sets based on a toy model of discrete clouds along the LOS and explore the sensitivity of the log-likelihood with respect to the different model parameters.
We then provide a solution on how, facing real observations, we can select the correct model, that is, to choose the correct number of clouds that exist along a sightline.

\subsection{Implementation}
\begin{figure*}
    \centering
    \begin{tabular}{cc}%
    single-cloud LOS & two-cloud LOS \\
    \includegraphics[trim={0cm 1.7cm 0cm 0cm},clip,width=8cm]{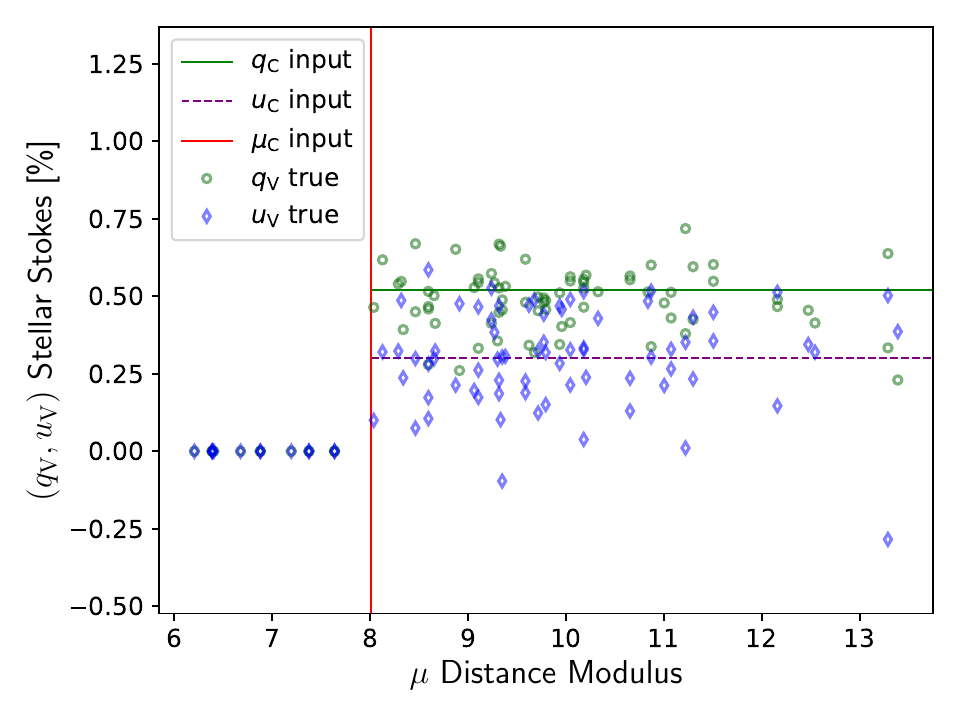}
        & \includegraphics[trim={0.2cm 1.8cm 0.1cm -0.4cm},clip,width=8cm]{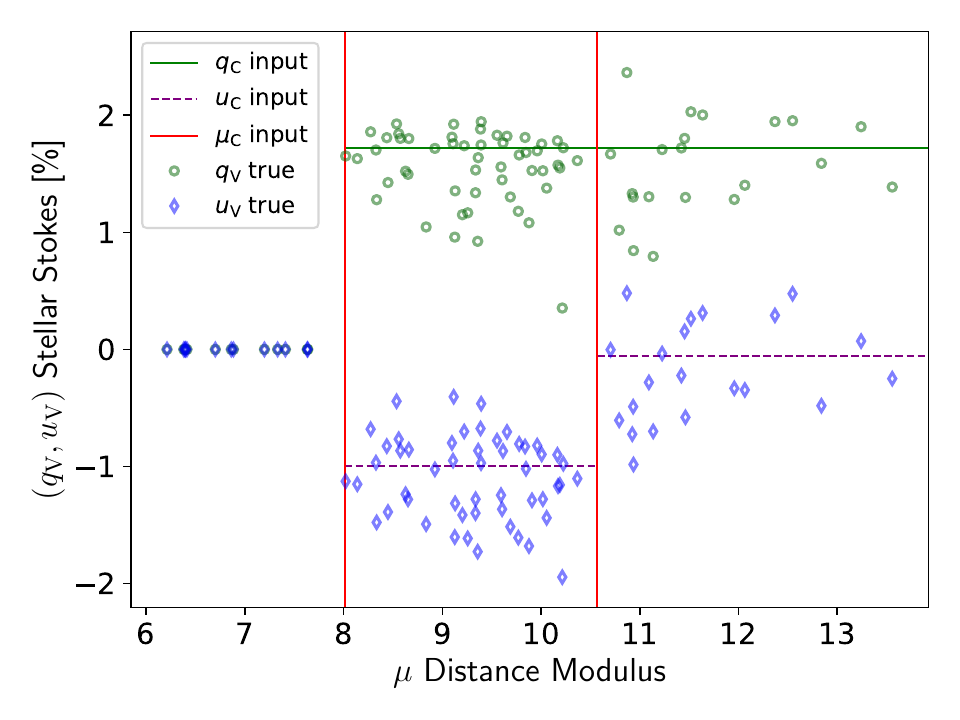} \\
    \includegraphics[trim={0cm 0.4cm 0cm 0.2cm},clip,width=8cm]{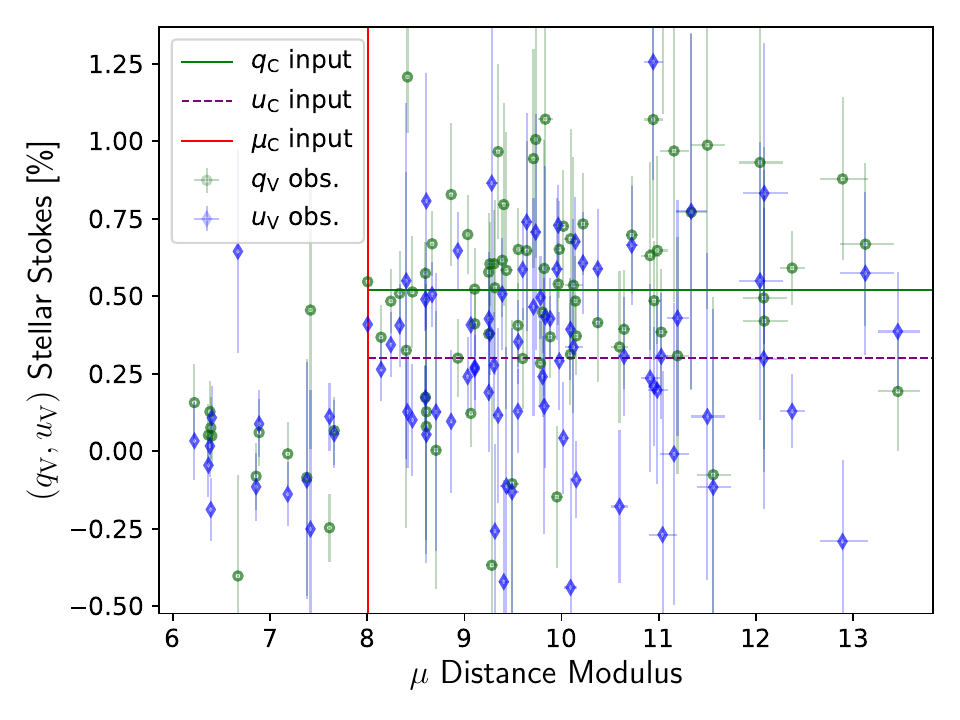}
        &   \includegraphics[trim={0cm 0.4cm 0cm 0.2cm},clip,width=8cm]{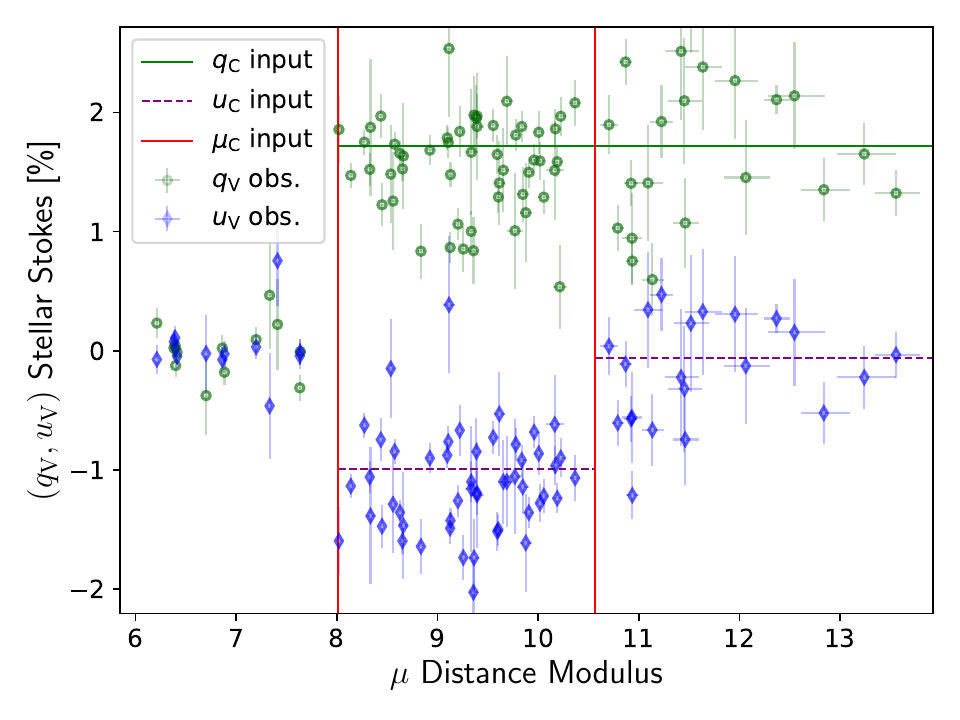}
    \end{tabular}\\[-1.5ex]
    \caption{Example of a single-cloud (left) and two-cloud (right) simulated data set. We show the stellar $q_{\rm{V}}$ (green circles) and $u_{\rm{V}}$ (blue diamonds) Stokes parameters as a function of distance modulus ($\mu_i$). Top and bottom panels show the same data set. Top panels do not include observational noise (they are the "true" data points) while bottom panels do include uncertainties in both parallax and polarization (shown with errorbars). 
    The vertical red lines indicate the input distance modulus of the clouds.
    The horizontal green and dashed-purple lines indicate the input (cumulative) mean Stokes parameters ($q_{\rm{C}}$ and $u_{\rm{C}}$, respectively) before the inclusion of intrinsic scatter and observational noise, i.e., $\bar{\mathbf{m}}_i$ in Eq.~\ref{eq:stokes_reg}.
    }
    \label{fig:1Layer_expl}
\end{figure*}
\begin{table*}
    \begin{center}%
    \caption{\label{tab:ExampleSetup} Setups for the simulated data set used in Sect.~\ref{sec:method} and illustrated in Fig.~\ref{fig:1Layer_expl}.\\[-1.ex]}
    {\small
    \begin{tabular}{lr|ccccc c ccccc}
    \hline \hline \\[-.5ex]
         \multicolumn{2}{l}{Data set} & \multicolumn{5}{c}{Cloud parameters} & & \multicolumn{5}{c}{"true" values} \\ \\[-1.ex]
        & cloud \# & $d_{\rm{C}}$ & $P_{\rm{max}}$ & $\gamma_{\mathbf{B}}$ & $\psi_{\mathbf{B}}$ & $A_{\rm{turb}}$
         & & $q_{\rm{C}}$ & $u_{\rm{C}}$ & $C_{{\rm{int}},qq}$ & $C_{{\rm{int}},uu}$ & $C_{{\rm{int}},qu}$ \\
         & & [pc] & [$\%$] & [$^\circ$] & [$^\circ$] & $-$
         & & $[\%]$ & $[\%]$ & $[\%]^2$ & $[\%]^2$ & $[\%]^2$ \\ \\[-1.ex]
     \hline \\[-.5ex]
     \multicolumn{2}{l}{Single-cloud} & & & & & & & & & & & \\
        & cloud~1 & 400 & 0.8 & 30 & \phantom{-}15 & 0.2 
        & & 0.49 & \phantom{-}0.30 & 0.01 & 0.02 & -0.01 \\ \\[-.5ex]
         \multicolumn{2}{l}{Two-cloud} & & & & & & & & & & & \\
        & cloud~1 & 400 & 2.0 & 5 & -15 & 0.2
        & & 1.56 & -1.01 & 0.10 & 0.12 & \phantom{-}0.10 \\
        & cloud~2 & 1300 & 1.0 & 15 & \phantom{-}45 & 0.2 
        & & 0.00 & \phantom{-}0.91 & 0.17 & 0.18 & \phantom{-}0.16 \\  \\[-1.ex]
     \hline\\[-1.5ex]
    \end{tabular}
    }
    \tablefoot{The labels of the parameters follow the notations given in the text.
    }
    \end{center}
\end{table*}
To maximize the likelihood function and estimate the posterior distributions of our model parameters we rely on a numerical method. We choose to use the code \texttt{dynesty} (\citealt{Speagle2020}) to sample the parameter space using the nested sampling method (\citealt{Skilling2004}). This code has already proven to be powerful and reliable in solving astrophysical problems similar to ours (e.g., \citealt{Zucker2019} ; \citealt{Alves2020}; \citealt{Zucker2022}). The algorithm uses sampling points (named `live points' in \texttt{dynesty}'s definition) to explore the parameter space in a dynamic nested sampling scheme and estimate the posterior distributions on model parameters. It has two main advantages compared to other sampling methods: first, it returns an estimate of the model evidence and second, it includes a well-defined stopping criterion, suitable for automation of the fitting process.

The code \texttt{dynesty} takes as input the function of the log-likelihood that has to be maximized and the definition of functions that implement our prior knowledge on the model parameters. We implemented both uniform and Gaussian priors for the cloud parallaxes and cloud mean polarization.
The Gaussian priors are defined through their means and standard deviations while uniform priors are defined by their lower and upper limits. We strongly recommend using flat priors for the element of the covariance matrix encoding the effects from turbulence. This is because the diagonal elements of the matrix must be positive and can approach zero depending on the (unknown) orientation of the magnetic field in 3D (see Appendix~\ref{sec:IntrSc}). In this case, the diagonal elements ($C_{{\rm{int}},qq}$ and $C_{{\rm{int}},uu}$) are first proposed independently in their respective ranges and then the off-diagonal elements are drawn such that the semi-positive-definiteness of the covariance matrix is guaranteed, that is, $C_{{\rm{int}},qu}$ is drawn from a uniform distribution in the range $(-\sqrt{C_{{\rm{int}},qq}\,C_{{\rm{int}},uu}},\,\sqrt{C_{{\rm{int}},qq}\,C_{{\rm{int}},uu}})$; excluding the limits.
In the case of multiple clouds, the prior function makes internally the distinction between clouds, ranked by their distances. We have hard-coded a lower limit on the number of stars that can exist between two clouds. We fix this limit at five.

Given a set of data points $\lbrace \varpi_i,\, \mathbf{s}_i = (q_{\rm{V}},\,u_{\rm{V}})_i \rbrace$ and corresponding uncertainties $(\sigma_{\varpi_i},\,C_{{\rm{obs}},i})$, the log-likelihood function that \texttt{dynesty} has to maximize is given by 
\begin{align}
    \log \left[ \mathcal{L}\left(\lbrace \varpi_{\rm{C}}^{[k]}, {\bar{\mathbf{m}}}^{[k]}, C_{\rm{int}}^{[k]} \rbrace \right) \right] = & \log \left[ \prod_{i=1}^{N_{\rm{star}}} \mathcal{L}_i\left( \lbrace \varpi_{\rm{C}}^{[k]}, {\bar{\mathbf{m}}}^{[k]}, C_{\rm{int}}^{[k]} \rbrace \right) \right] \nonumber \\
    = & \sum_{i=1}^{N_{\rm{star}}} \log \left[ \mathcal{L}_i\left( \lbrace \varpi_{\rm{C}}^{[k]}, {\bar{\mathbf{m}}}^{[k]}, C_{\rm{int}}^{[k]} \rbrace \right) \right]
    \label{eq:log-likelihood_general}
\end{align}
which requires the number of clouds as an entry and where the $\mathcal{L}_i$'s are given by Eq.~\ref{eq:star_likelihood}.
The number of clouds populating the LOS is a priori unknown. We discuss in Sect.~\ref{sec:modelselection} how we intend to determine it.
We implemented the likelihood functions for up to five clouds along the LOS. Even though the generalization of the implementation to higher number of clouds is trivial we do not deem it necessary given that only few LOSs at intermediate and high Galactic latitudes are expected to show a large number of components with significant contribution to the polarization signal (\citealt{Pan2020}).

\subsection{Mock data for two example LOSs}
To test our method we developed a simple but complete implementation of our layer model to generate mock data sets with realistic number of stars, stellar distance and brightness distributions, and uncertainties both on parallax and polarization. This implementation, which includes a self-consistent prescription for the intrinsic scatter, is presented in detail in Appendix~\ref{sec:toymodel}.
Our toy model for generating stellar polarization has five free parameters per cloud: the cloud parallax ($\varpi_{\rm{C}} = 1/d_{\rm{C}})$, the maximum degree of polarization ($P_{\rm{max}}$), the inclination ($\gamma_{\mathbf{B}_{\rm{reg}}}$) and position ($\psi_{\mathbf{B}_{\rm{reg}}}$) angles of $\mathbf{B}_{\rm{reg}}$ and, finally, the relative amplitude of fluctuations in magnetic field orientation ($A_{\rm{turb}}$).
We draw the reader's attention to the fact that, apart from the cloud parallax, these parameters are not the same as the model parameters entering our data equation. Additionally, our toy-model is stochastic due to the presence of the intrinsic scatter. Therefore, as demonstrated in Appendix~\ref{sec:IntrSc}, the mean values of the Stokes parameters of a cloud and the values characterizing the covariance induced by the intrinsic scatter must be read from the simulated data before observational noise in parallax and polarization are introduced.
To do so, we segment the `clean' data sets at the input cloud distance(s) and we estimate the mean and covariance of the polarization induced by the cloud to the polarization of stars behind the cloud (but in front of the more distant cloud, if any). We refer to these estimates as the "true" values in the remainder of this paper.

We show in Fig.~\ref{fig:1Layer_expl} two examples of simulated data for a single-cloud case (left) and a two-cloud case (right) applied to a sample of stars typical to intermediate to high Galactic latitudes for a circular sky area with a diameter of $0.5^\circ$ (see Appendix~\ref{sec:gums}). We show the relative Stokes parameters as a function of distance modulus ($\mu_i = 5\,\log(d_i) -5$, where $d_i$ is the star distance in parsec). The top row shows the simulated data before noise in parallax and polarization is added and the bottom row shows the same simulated data sets when noise is added.
The parameter values used to generate both mock data from our toy model (see Appendix~\ref{sec:toymodel}) are reported in Table~\ref{tab:ExampleSetup}.
The two-cloud LOS is chosen such that ($i$) the far-away cloud alone induces about half the polarization of the nearby cloud alone, ($ii$) that the presence of the far-away cloud is only clear in one of the two Stokes parameters due to the POS orientation of $\mathbf{B}$ permeating the second cloud, and ($iii$) that there are at least 20 stars in the background of each cloud. The simulated data sets include uncertainties on stellar parallaxes based on \textit{Gaia} performance (see Appendix~\ref{sec:mock_data_appendix}), realistic uncertainties on individual Stokes parameters as expected for WALOP-N (the northern instrument that will be used for the \textsc{Pasiphae} survey) with 5 minutes of exposure time (see Appendix~\ref{sec:mock_realnoise} and Fig.~\ref{fig:Sigma_qu-vs-Rmag}), and include a prescription for the intrinsic scatter.
For completeness, we report in Table~\ref{tab:ExampleSetup} the true values of the mean Stokes parameters and of the elements of the covariance matrix corresponding to the intrinsic scatter corresponding to the two examples shown in Fig.~\ref{fig:1Layer_expl}. These values correspond to the parameters entering our data equation and, ultimately, should be retrieved from the application of an inversion method to the data.

For the remainder of this section, we use these two examples of mock starlight polarization data to explore the sensitivity of our log-likelihood with respect to the different sampled parameters and to validate our implementation.

\begin{figure*}
    \centering
    \includegraphics[trim={0cm .4cm 0cm 1cm},clip,height=3.5cm]{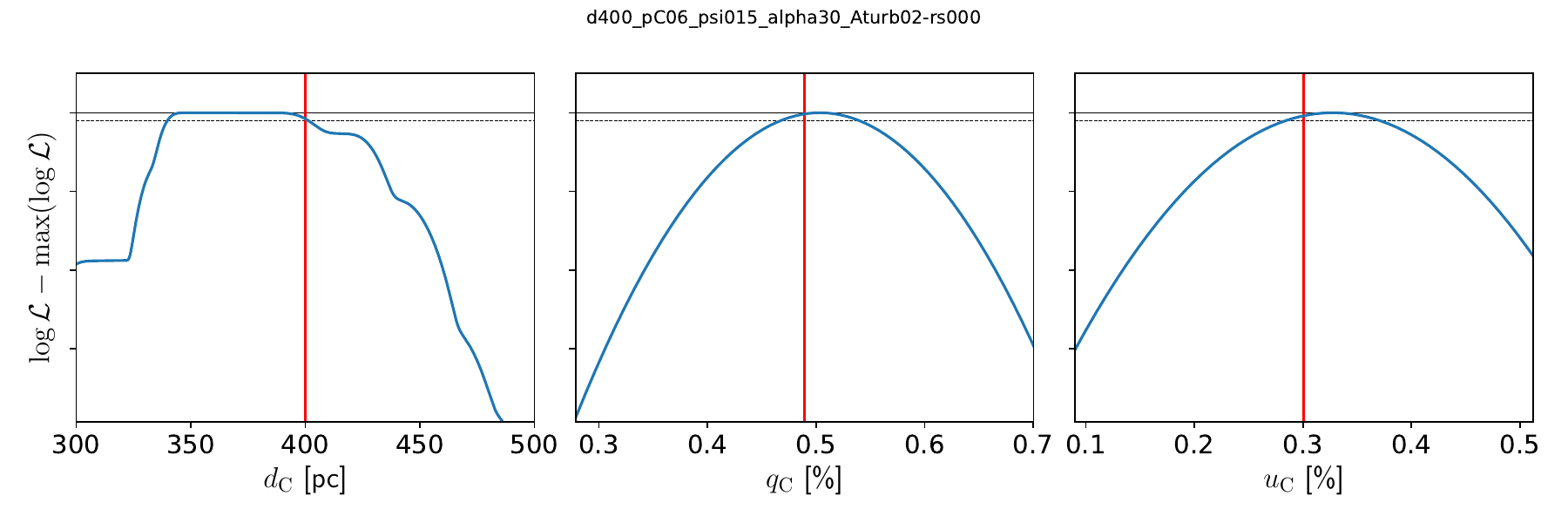} \\
    \includegraphics[trim={0cm .4cm 0cm 1cm},clip,height=3.5cm]{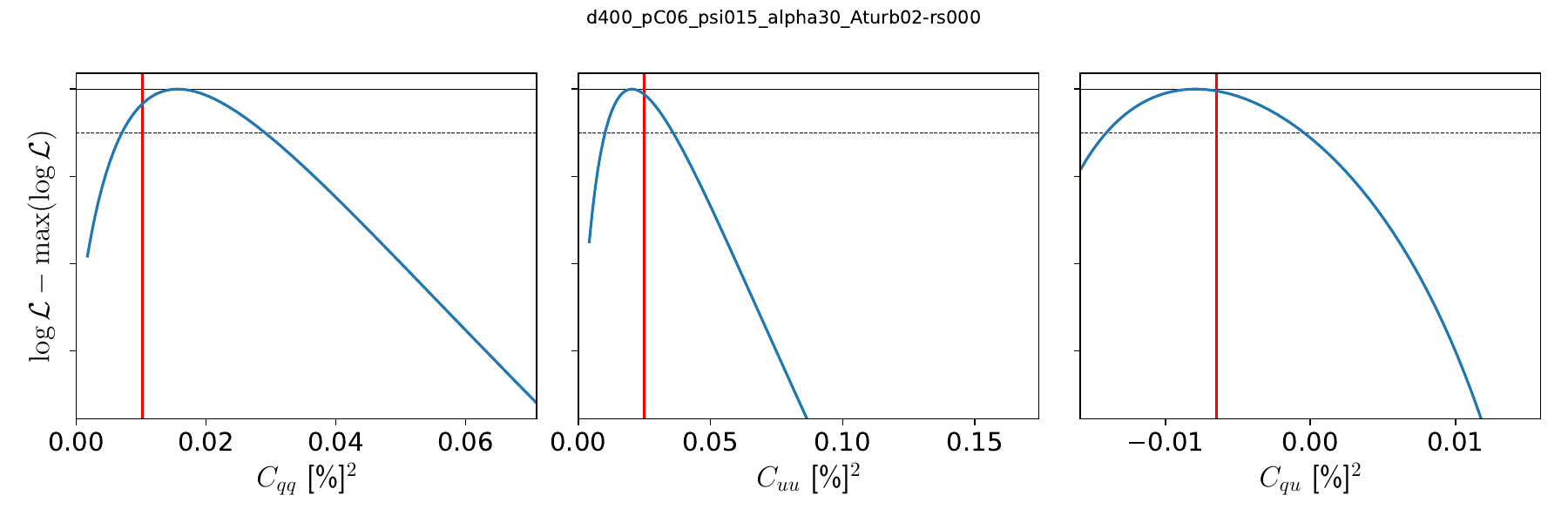}
    \caption{Curves of ($\log(\mathcal{L}) - \text{max}(\log(\mathcal{L}))$) corresponding to 1D likelihood scans through the parameter space for the mock data set with a single cloud along the LOS. For each scan only the explored parameter varies, while all other parameters are kept fixed to their true values. The log-likelihood ($\log(\mathcal{L})$) is estimated at each point. The horizontal solid and dashed line show the values of 0 and -1, respectively, providing an approximate estimate of the location of the 68 per cent credible interval. In the top (bottom) row the vertical axis ranges from -40 to 5 (-7.5 to 0.3). The red vertical line on each panel indicates the so-called true value reported in Table.~\ref{tab:ExampleSetup}.
    }
    \label{fig:1DlikelihoodScan_logL}
\end{figure*}
\begin{figure*}
    \centering
    \includegraphics[trim={0cm .4cm 0cm 1cm},clip,height=3.5cm]{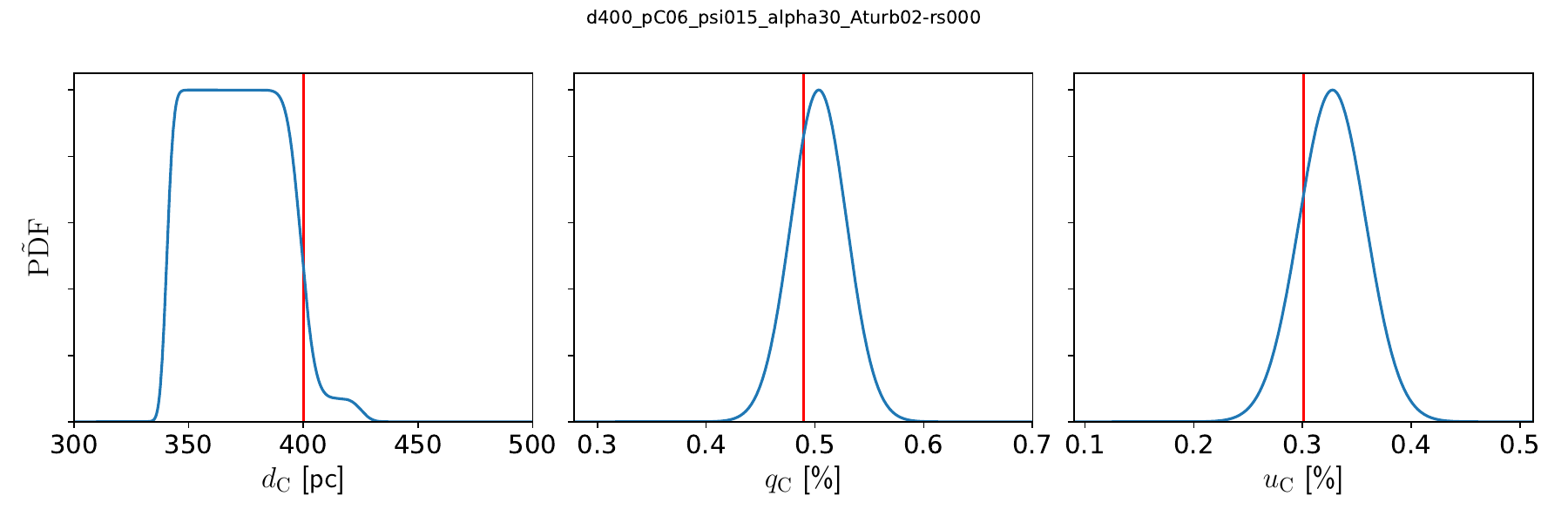} \\
    \includegraphics[trim={0cm .4cm 0cm 1cm},clip,height=3.5cm]{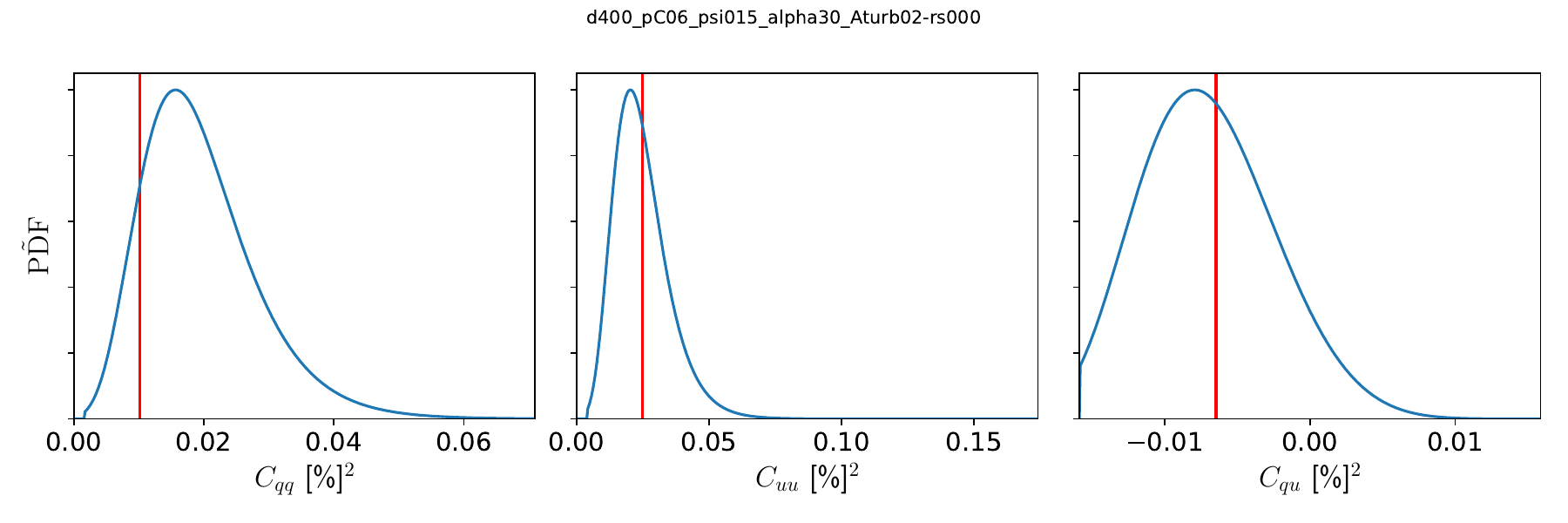}
    \caption{Conditional probability distributions ($\tilde{\text{PDF}}$) corresponding to the ($\log(\mathcal{L}) - \text{max}(\log(\mathcal{L}))$) curves in Fig.~\ref{fig:1DlikelihoodScan_logL}}
    \label{fig:1DlikelihoodScan_PDF}
\end{figure*}

\subsection{1D conditional posterior distributions}
As a first test of implementation, we perform 1D likelihood scans through the parameter space. This exercise also allows us to study the sensitivity of the likelihood, and therefore of the observables, to each model parameter individually. We present in Fig.~\ref{fig:1DlikelihoodScan_logL} the conditional log-likelihood curves corresponding to the scans using the one-cloud LOS mock data shown in Fig.~\ref{fig:1Layer_expl} (bottom left). We show in Fig.~\ref{fig:1DlikelihoodScan_PDF} the conditional probability distribution function (PDF) corresponding to these scans. They are not estimates of the 1D marginalized posterior distributions obtained from a fit since all parameters are not varied during the scans but are fixed to their true values.
An actual fit to this particular simulated data set is performed in Sect.~\ref{sec:sanitychecks}.
The validation of our method and its implementation for several realistic cases will be presented in the Sect.~\ref{sec:performance} where the performance and the limitations of the method will be assessed.

\begin{figure}
    \centering
    \includegraphics[trim={0cm .4cm 0cm 1cm},clip,height=6cm]{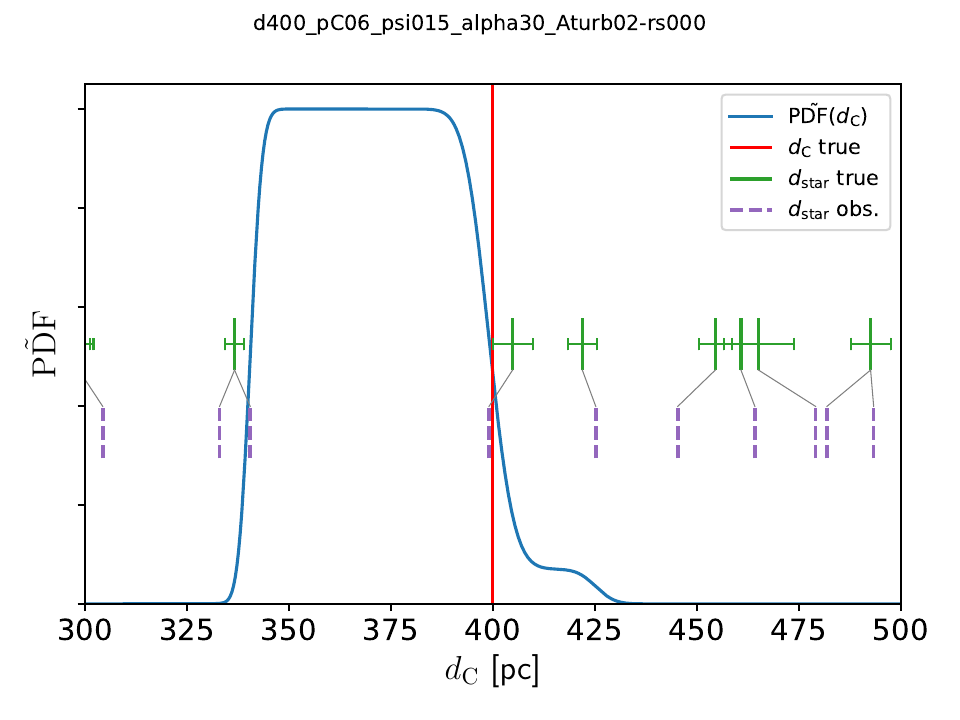}\\[-1.5ex]
    \caption{Conditional probability distribution ($\tilde{\text{PDF}}$) corresponding to the 1D likelihood scan through $\varpi_{\rm{C}}$ for the one-cloud mock data set using the one-layer model. The vertical (continuous) green and (dashed) purple segments indicate the true and observed distances of the stars, respectively. The gray oblique lines link the two. The green horizontal errorbars indicate the 68\% confidence level on star distances obtained from $\sigma_{\varpi_i}$. The vertical red line indicates the input cloud parallax.
Due to randomization on parallaxes, some stars with very similar (true) parallaxes (green) are dispersed.
    }
    \label{fig:1DlikelihoodScan_PDFplxe}
\end{figure}

First, we note that the input parameter values always fall in the interval where $\log(\mathcal{L}) - \max(\log(\mathcal{L})) > -1$ meaning that the input values fall inside the approximated 68\% credible interval.
Second, we see that the shapes of the conditional log-likelihood curve, and of its corresponding conditional PDF, from the exploration of the cloud distance shown in Figs.~\ref{fig:1DlikelihoodScan_logL} and~\ref{fig:1DlikelihoodScan_PDF} are somewhat surprising while the curves obtained for the other parameters look quite conventional. The very reasons for the unconventional shapes of the cloud-distance curves come from the unevenly distributed constraints (the stars) on the parallax (distance) space, their unequal uncertainties along that axis, the smearing in the foreground and the background that the latter can generate, and, last but not least, the unequal constraining power of each star in the fit since polarization uncertainties are unequal. A star with large polarization uncertainties will constrain the fit less, and thus the position of the cloud along the LOS, as compared to a star with small polarization uncertainties. We illustrate part of this complexity in Fig.~\ref{fig:1DlikelihoodScan_PDFplxe} in which we repeat the top left panel of Fig.~\ref{fig:1DlikelihoodScan_PDF} but where the true and observed parallaxes are indicated by vertical segments.
It is clear that the likelihood of having a cloud with any distance between two distant constraints is constant and that the steepness of the variations depends on the parallax uncertainties.
In general, standard statistics are not appropriate for characterizing the posterior distributions on cloud distances and dedicated metrics have to be considered to quantify success and goodness of fit. We address this point in Sect.~\ref{sec:goodnessofrec}.

\subsection{Sanity checks}
\label{sec:sanitychecks}
\begin{figure*}
    \centering
    \includegraphics[trim={0cm 0.cm 0cm 0cm},clip,height=15cm]{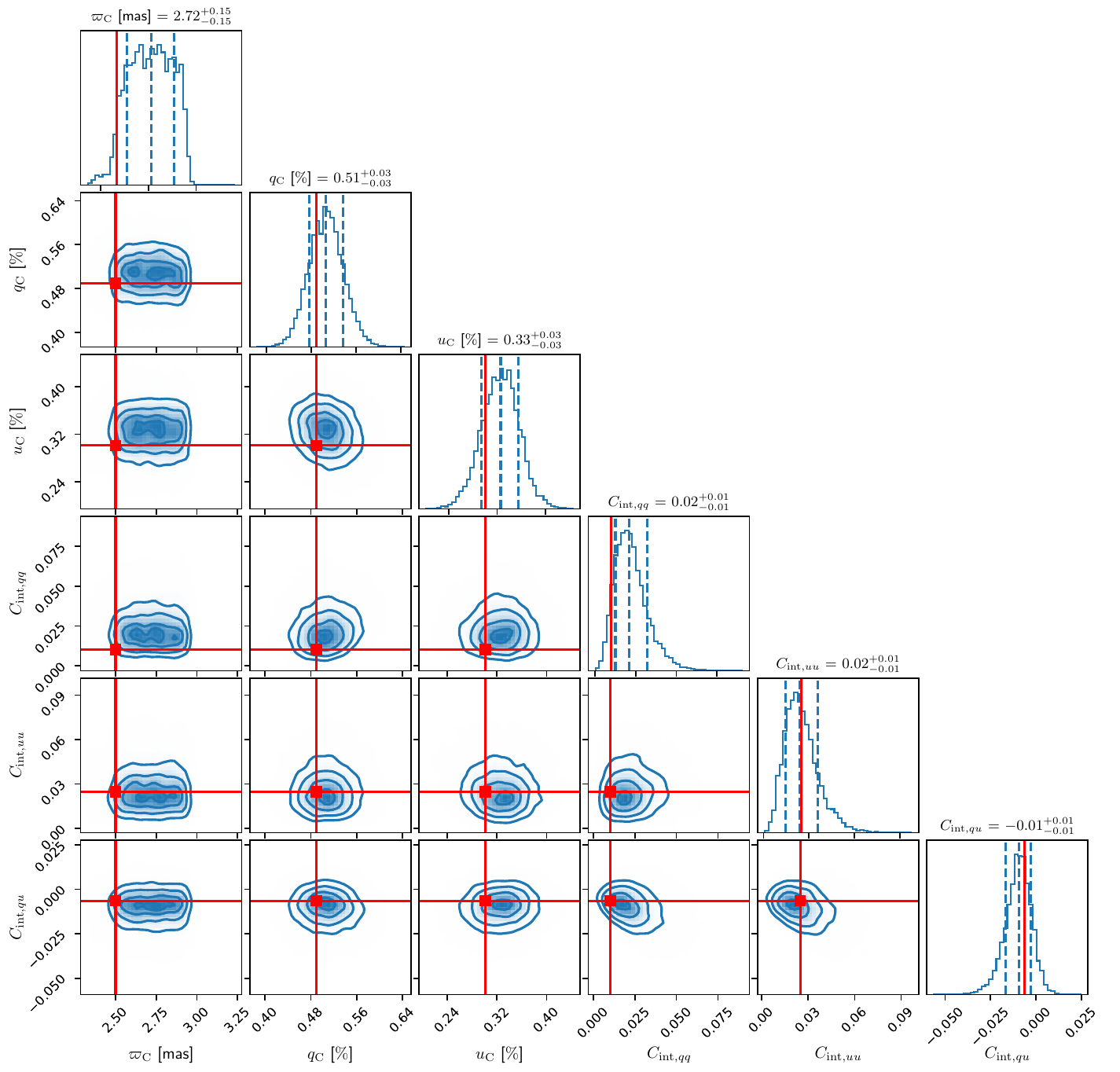}\\[-1.5ex]
    \caption{Performance of the one-layer model in fitting the single-cloud mock dataset. 1D and 2D marginalized posterior distributions for the sampled model parameters obtained by log-likelihood maximization. The red lines indicate the true parameter values. Cloud parallax ($\varpi_{\rm{C}}$) is given in milli-arcseconds (mas), mean Stokes parameters ($q_{\rm{C}},\,u_{\rm{C}}$) in per cent and the elements of the covariance matrix encoding the effect of the turbulence-induced intrinsic scatter are given in per cent to the square (i.e., multiplied by 10,000).
    The dashed vertical lines indicate the 16, 50, and 84 percentiles of the 1D marginalized distributions and the values for the 68\% confidence interval can be read from the title on each of the diagonal panels.
    }
    \label{fig:1Layer_corner}
\end{figure*}

We check that our implementation of the model and of the maximization of the log-likelihood (Eq.~\ref{eq:log-likelihood_general}) through the nested sampling method is effective, by first applying our inversion method to the single-cloud LOS data used above and using a model with a single layer.
We use 1000 live points, start with loose uniform priors on all parameters (the same as used in Sect.~\ref{sec:performance_1L} and reported in Table~\ref{tab:Priors}) and sample the parameter space until an uncertainty of about 0.1 is achieved on the log of the model evidence $\mathcal{Z}$ (see Eq.~\ref{eq:evidence} below). Then the sampling is stopped and the samples are post-processed to generate 1D and 2D  marginalized posterior distributions of the model parameters. The resulting histograms are shown on a corner plot format (\citealt{corner}) in Fig.~\ref{fig:1Layer_corner}.
In this example, the obtained maximum likelihood value is $\log \hat{\mathcal{L}} = 784.55$ and the evidence is $\hat{\mathcal{Z}} = 765.54 \pm 0.18$, all the input polarization-related parameters are found within the 68\% credible interval of the estimated posterior distributions and 97\% of the posterior on cloud parallax is contained between the true parallaxes of the two stars that directly bracket the input cloud parallax value. The method, therefore, demonstrates high accuracy in this example. We emphasize that not only do we recover the cloud distance and the mean values of the Stokes parameters but also an accurate estimate of the polarization covariance from the intrinsic scatter.

\begin{figure*}
    \centering
    \includegraphics[trim={0cm 0.cm 0cm 0cm},clip,height=15cm]{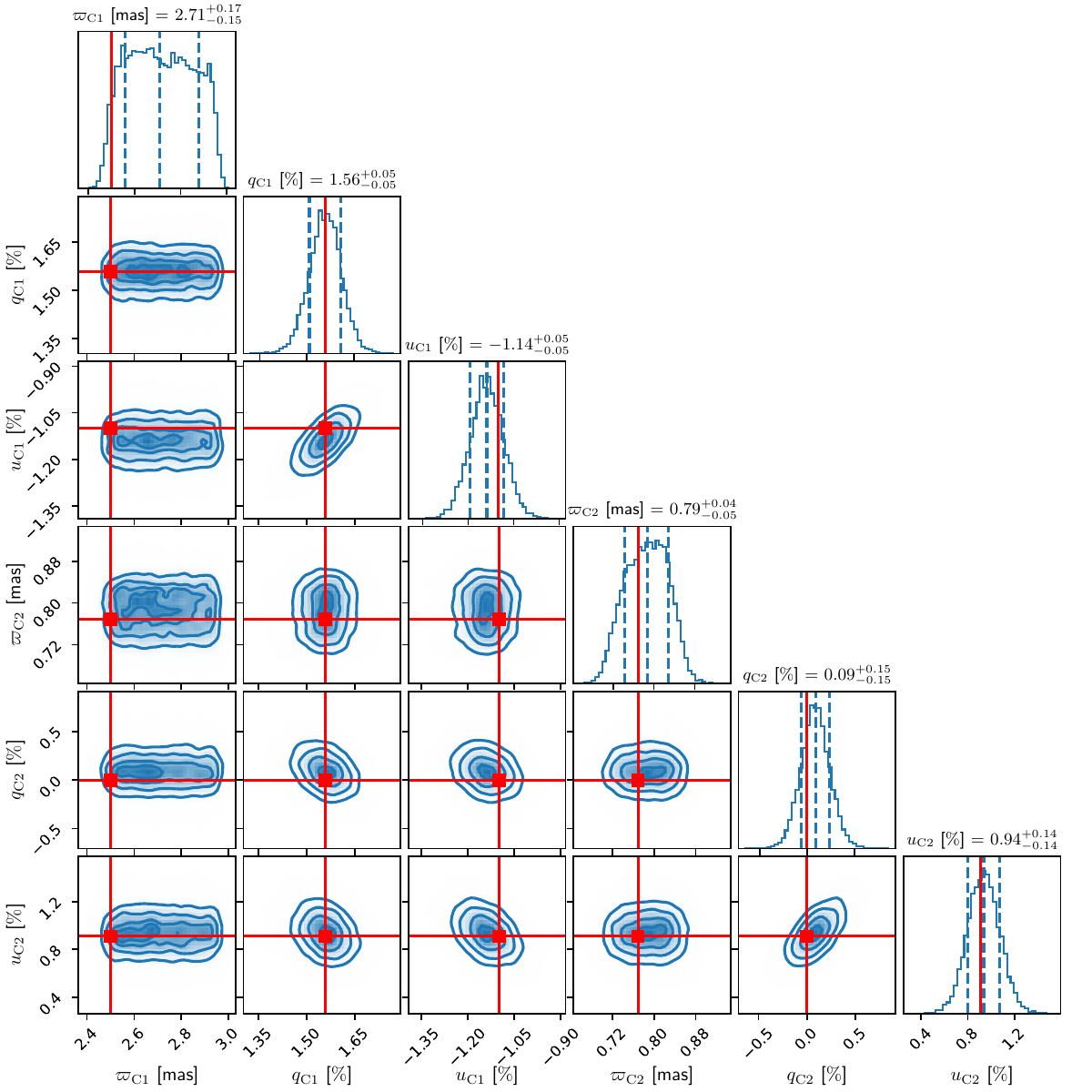}\\[-1.5ex]
    \caption{Same as in Fig.~\ref{fig:1Layer_corner} but showing a restricted sample of the model parameters of a two-layer-model fit to the two-cloud LOS shown in example in Fig.~\ref{fig:1Layer_expl}. We show the estimated 1D and 2D marginalized posterior distributions only for the parallax and Stokes parameters of the 2 modeled clouds.
    }
    \label{fig:2Layer_corner_plxqu}
\end{figure*}

Similar conclusions are reached from the application of our fitting method to the two-cloud LOS example presented above using a model with two layers. In this case the priors set for each cloud are the same as used for the single-cloud LOS above and we use the same setup to analyze the data.
We show in Fig.~\ref{fig:2Layer_corner_plxqu} the posterior distributions reconstructed by our fit for the parallaxes and Stokes parameters of the two clouds. The Stokes parameters are found to be well within the 68\% confidence interval of the estimated posteriors and 98\% and 83\% of the posteriors on clouds' parallaxes are contained between the true parallaxes of the two stars that directly bracket the values of their respective input-cloud parallaxes. We do not show the full corner plot, including the constraints on the intrinsic scatter parts, only because visualization of 12 parameters lead to poor insights; the posterior distributions are similarly very good, all true values falling within the 68\% confidence interval of the estimated posteriors. The very fact that the posterior on the parallax of the second cloud appears less tight than that of the first cloud makes sense given that the second cloud is farther away from the observer where stars have larger uncertainties on their parallax and are generally fainter, thus showing larger uncertainties on Stokes parameters, than closer stars. Additionally, the presence of foreground clouds (stars) add noise to the reconstruction of the background cloud.

\begin{figure}
    \centering
    \includegraphics[trim={0.1cm 5.2cm 0cm 0cm},clip,height=5.5cm]{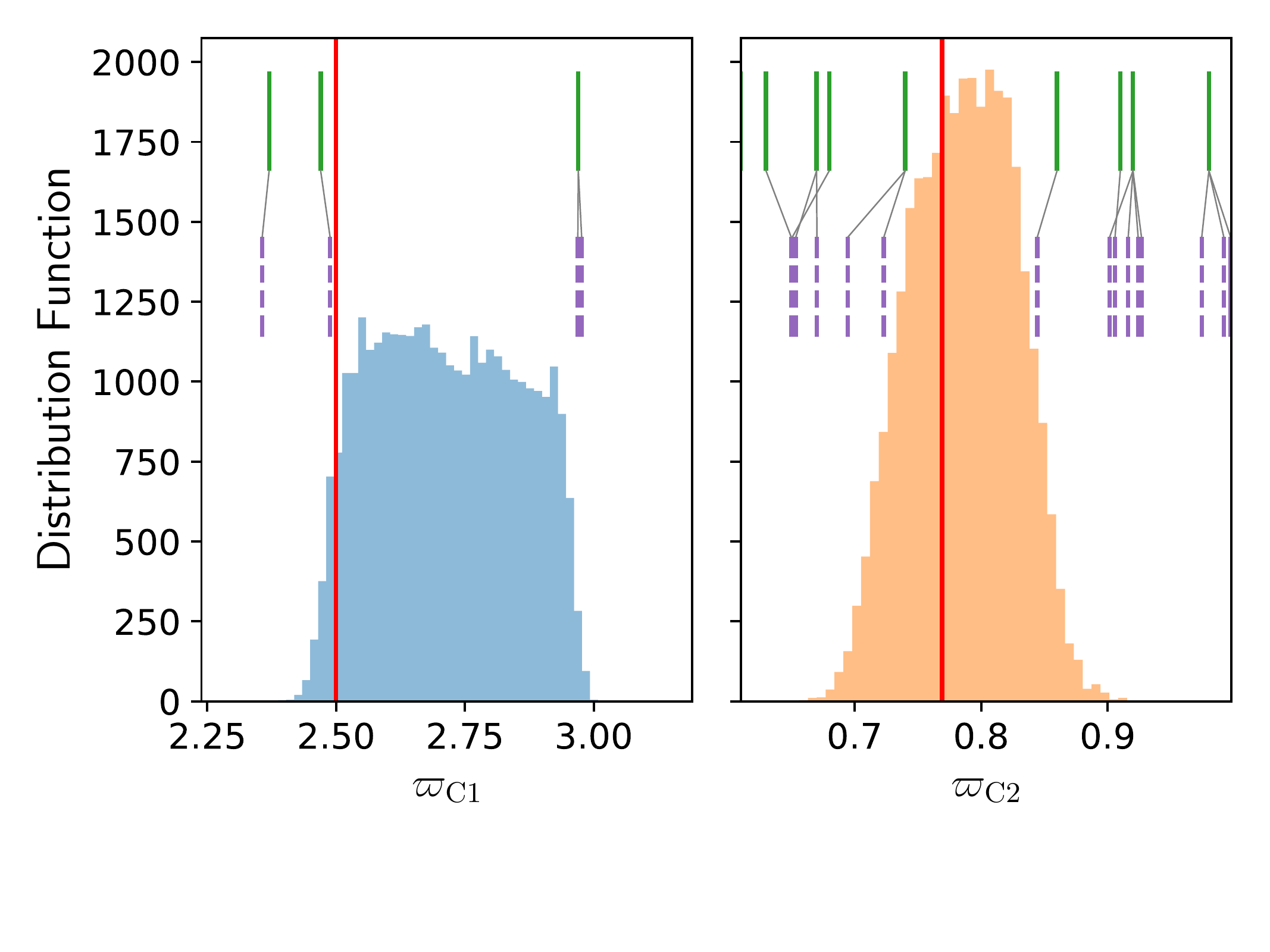}\\[-1.2ex]
    {\small \hspace{1.cm} $\varpi_{\rm{C}_1}$ [mas] \hspace{2.2cm} $\varpi_{\rm{C}_2}$ [mas] \hspace{1.cm}}\\[-1.5ex]
    \caption{Estimated posterior distributions for $\varpi_{\rm{C}1}$ (left) and $\varpi_{\rm{C}2}$ (right) in [mas] with marked observed and input parallaxes of surrounding stars with (dashed) purple and (continuous) green vertical segments, respectively, similar to Fig.~\ref{fig:1DlikelihoodScan_PDFplxe}. Vertical red lines indicate the input parallaxes of the clouds.
    As before, some stars with very similar (true) parallaxes (green) are dispersed after randomization.
    }
    \label{fig:2Layer_postplx}
\end{figure}
Figure~\ref{fig:2Layer_postplx} presents the estimated posterior distributions obtained for cloud parallaxes. The shapes of the posterior distributions are well understood when parallaxes of surrounding stars are considered, in a way similar to Fig.~\ref{fig:1DlikelihoodScan_PDFplxe}, illustrating the inhomogeneous distribution of constraints along that dimension.

\smallskip

\begin{figure*}
    \centering
    \begin{tabular}{cc}%
    {\hspace{0.7cm}} single-cloud LOS & {\hspace{0.7cm}} two-cloud LOS \\
    \includegraphics[trim={.5cm 0.4cm 0cm 0.2cm},clip,width=8cm]{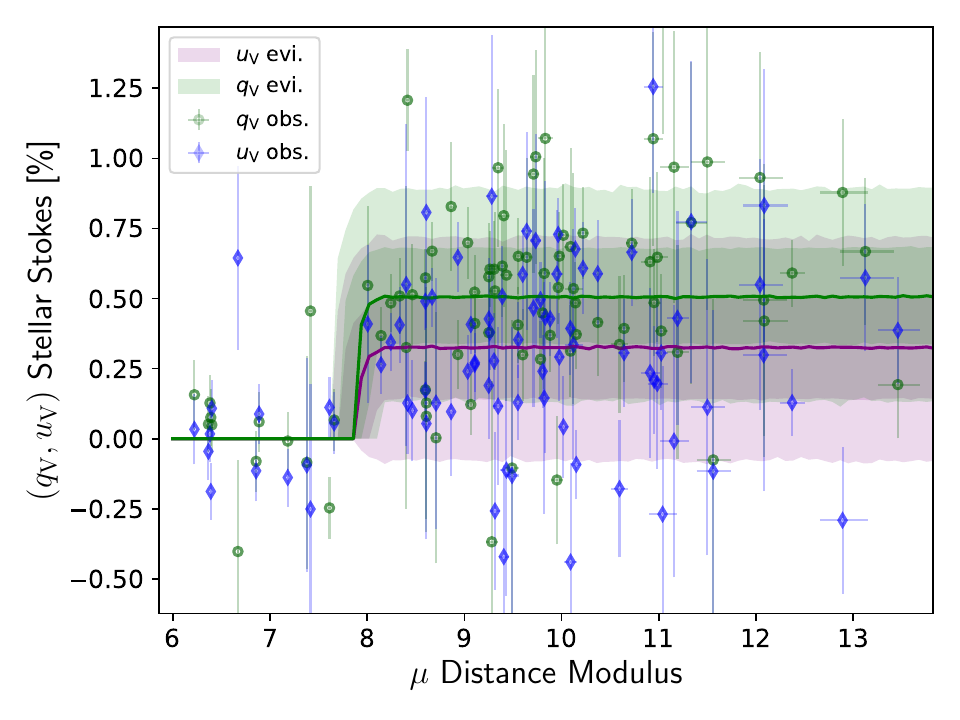}
        & \includegraphics[trim={0.5cm 0.4cm 0cm 0.2cm},clip,width=8cm]{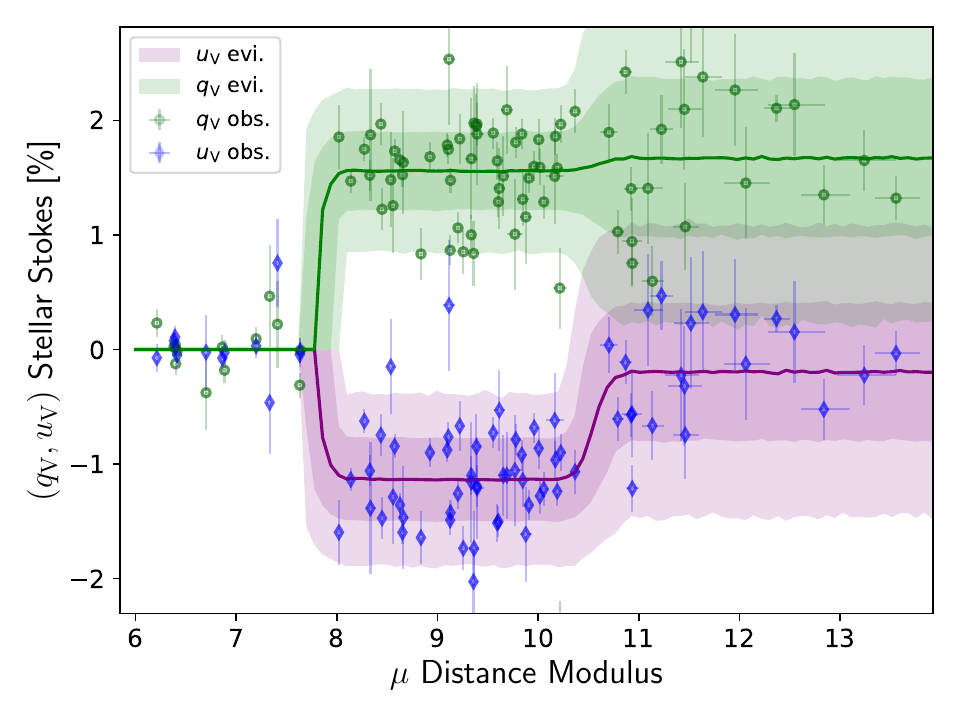}
    \end{tabular}\\[-1.5ex]
    \caption{Representation of the model evidences obtained for the single-cloud LOS (left) and two-cloud LOS (right).
    The data points are the same as in Fig.~\ref{fig:1Layer_expl}. The shaded areas illustrate the distributions of $(q_{\rm{V}},\, u_{\rm{V}})$ obtained at every $\mu$ value through re-sampling of the posterior distributions obtained from the maximum-likelihood analysis of the data, as explained in the text.
    }
    \label{fig:1L-2L_results}
\end{figure*}
In Fig.~\ref{fig:1L-2L_results} we reproduce the bottom panels of Fig.~\ref{fig:1Layer_expl} where we also represent the model evidences obtained from our maximum-likelihood analysis of the data. Namely, for any distance in the range spanned by the data stars, the $(q_{\rm{V}},\, u_{\rm{V}})$ are estimated by re-sampling the posterior distributions of the model parameters. The modeled $(q_{\rm{V}},\, u_{\rm{V}})$ are computed taking into account the intrinsic scatter and the correlation that it induces between the Stokes parameters. Given that our model is stochastic, we generate ten random draws for each of the 1000 sets of the model parameters randomly extracted from the posteriors. This leads to a bivariate distribution of $(q_{\rm{V}},\, u_{\rm{V}})$ for any distance modulus. The shaded areas span the ranges of $[2.5,\, 97.5]$ (light) and $[16,\,84]$ (dark) percentiles of the distributions for each $\mu$. The medians of the distributions are given by the continuous line.
The figure shows that that the data points and the model agree very well.

Another way of examining the agreement between data and model is to look at the residuals of the data points (in the polarization plane) as compared to the model predictions. Given that our model is stochastic and that (\textit{i}) the polarization must be regarded as a bivariate quantity and (\textit{ii}) measurement uncertainties are heteroscedastic; we are interested in the distribution of the Mahalanobis distances of the individual measurements in the beam with respect to their respective modeled means and computed with their respective total covariance matrices. The distribution of the Mahalanobis distances squared is expected to follow a $\chi^2$ distribution with two degrees of freedom. We checked (not shown) that this is the case. Our distributions of Mahalanobis distances have a median of about 1.4 and no outliers are found.

\subsection{Model selection}
\label{sec:modelselection}
When dealing with real data, we generally do not know how many dust clouds exist along the LOS. In our approach, the number of clouds is an implicit parameter of the model.
We first choose the number of clouds with which to model the data and then perform the maximum-likelihood analysis of the data. The latter provides an estimate of the posterior distributions on model parameters, an estimate of the evidence and an estimate of the maximum log-likelihood value. In itself, this approach cannot assess the validity of the chosen model.
However, if we fit different models (i.e., with different input number of clouds) we can compare the results and decide, based on statistical arguments, which model is preferred by the data. This provides us with a standalone method to infer the number of clouds along the LOS and to proceed to its Bayesian decomposition in terms of dust layers, using polarization data of stars with measured distances only.

We consider two criteria to decide on the model that should be preferred given the data.
The first criterion is based on the evidence returned by the nested sampling method (\citealt{Skilling2004}) and the second is the Akaike information criterion (AIC) (\citealt{Akaike1974}).

The evidence, directly estimated through the use of \texttt{dynesty}, results from the integration over the full parameter space ($\Omega_\Theta$) of the likelihood of the model parameters ($\Theta$) given the data (multiplied by the prior on the model parameters $\pi(\Theta)$):
\begin{equation}
    \mathcal{Z} \equiv \int_{\Omega_\Theta} d\Theta \, \mathcal{L}(\Theta) \, \pi(\Theta)\;.
    \label{eq:evidence}
\end{equation}
For an ensemble of models, the `best' model is the one that maximizes the evidence.

The AIC originates from information theory and is a measure of the amount of information that is lost by representing the data by a given model. It is based on the maximum likelihood value and includes a penalty for the  number of model parameters. For a model $j$ with $M$ parameters, if the maximum likelihood is denoted by $\hat{\mathcal{L}}_j$, the AIC is given by:
\begin{equation}
    {\rm{AIC}}_{j} = 2\,M - 2\,\log(\hat{\mathcal{L}}_j) \; .
\end{equation}
For an ensemble of models, the `best' model is the one that minimizes the loss of information.
The use of the AIC in model comparison is also attractive as it makes it possible to quantify the probability that a given model may minimize the information loss in comparison to the model that actually minimizes it in our data analysis experiment. Given a set of models $\left\lbrace m \right\rbrace$, the probability that model $j$ minimizes the information loss is given by (\citealt{Boisbunon2014}):
\begin{equation}
    P_{j|\{m\}} = \exp\left((\min_m{\left\lbrace {\rm{AIC}}_{m} \right\rbrace}  - {\rm{AIC}}_{j} )/2 \right) \; .
\end{equation}
In a conservative approach only models with small probability ($\lesssim 1\%$) should be disregarded. We use the above criteria in Sect.~\ref{sec:app2realdata}.

\section{Performance}
\label{sec:performance}
In the previous sections we have introduced a Bayesian method to decompose the starlight polarization data in terms of independent clouds along the LOS by maximizing a dedicated log-likelihood function. We have demonstrated on two examples that the method is effective in performing the decomposition and in recovering the true values of the modeled data.
In this section we aim to investigate the performance of the method and identify the limits of its applicability for the case of \textsc{Pasiphae}-like observations.

The performance of the method at recovering cloud parameters is expected to depend on: (\textit{i}) the amplitude of the polarization signal that a given cloud induces to the light of background stars, (\textit{ii}) the number of stars effectively sampling a given cloud (i.e., stars in the background of that cloud but in the foreground of any potential farther one), (\textit{iii}) the noise level of the stellar polarization measurements, (\textit{iv}) the precision on the star parallaxes (which is also dependent on the star distances), (\textit{v}) the degree of intrinsic scatter, and (\textit{vi}) on the number of clouds along the LOS.

\smallskip

To determine the performance of the method we need to rely on a metric that quantifies the goodness of the fit in both distance and polarization.
Our metrics are introduced below, after we introduce the simulated starlight polarization data which rely on actual stellar magnitudes and parallaxes from \textit{Gaia}. Then, we study the performance of our method by applying it to several single-cloud-LOS and two-cloud-LOS cases that could be typical of LOSs at intermediate and high Galactic latitudes that will be targeted by \textsc{Pasiphae}. Finally we explore the use of the criteria to select the most likely model to test their efficiency.

\smallskip

We expect, and have checked (not shown), that the method performs best when the polarization induced by the ISM is large compared to the uncertainties in the stellar polarization (but still within the weak polarization limit assumed for the diffuse ISM).
We therefore wish to test our method in conditions of low S/N in polarization to determine its limitations.
At intermediate and high Galactic latitude ($|b| > 30^\circ$) the 80-th percentile of the distribution of stellar polarization in \cite{Berdyugin2014} is at about 0.3\%.
This value is also in broad agreement with the distribution obtained from the extrapolation of the \ion{H}{I} column density of low-velocity clouds in those sky areas (calculated based on the data by \citealt{Pan2020}).
Hence, to generate the mock starlight polarization samples on which to apply our inversion method, we explore the parameter space of our toy model (see below and Appendix~\ref{sec:toymodel}) so that $p_{\rm{C}}$ ranges from 0\% to about 0.3\%.

\subsection{Simulated data}
\label{sec:simulatedData}
To study the performance of our inversion method we rely on simulations of starlight polarization data. We construct realistic stellar samples from the Gaia Universe Model Snapshot\footnote{\url{https://dpac.obspm.fr/gaiasimu/html/}} (GUMS) database.
As fully described in Appendix~\ref{sec:gums}, we obtain stellar distance and photometry data at high Galactic latitudes within circular beams having two different diameters: 0.5$^\circ$ and $1^\circ$. We apply two selection criteria based on stellar brightness and S/N in parallax. We retain stars with SDSS $r$-band magnitude $r < $ 16 mag, which is the expected limiting magnitude of the \textsc{Pasiphae} survey (\citealt{Tassis2018}). To ensure high precision in stellar distance information, we keep stars with S/N in parallax larger than five ($\varpi/\sigma_\varpi \geq 5$). The main characteristics of the stellar samples are shown in Fig.~\ref{fig:R_dist_cuts}, after the application of these selection criteria. We show the distributions of star distance (modulus) and SDSS $r$-band magnitude for a $1^\circ$ beam and a $0.5^\circ$ beam.

To proceed further in the modeling of starlight polarization data for our performance tests, we select one sample per beam size. We choose the samples with the mean number of stars, sampling actual data at high Galactic latitudes.
For each star in the two samples we have estimates of the actual parallax, parallax uncertainty, and magnitude in the SDSS-$r$ band.
Then, applying our toy model (Appendix~\ref{sec:toymodel}) to these star samples we can generate simulated starlight polarization data with realistic uncertainties for any desired setup of the magnetized ISM. The latter is determined by the number of dust clouds along the considered LOS and by five parameters per cloud (see Appendix~\ref{sec:toymodel}).
Two examples of simulated starlight polarization data obtained for a one-cloud and a two-cloud ISM setup applied to the mean star sample within a beam of $0.5^\circ$ in diameter are shown in Fig.~\ref{fig:1Layer_expl} with, however, a larger induced polarization signal than for the performance tests carried out below.

\subsection{Metric to assess the goodness of LOS decomposition}
\label{sec:goodnessofrec}
Our primary focus in the LOS decomposition of the starlight polarization data is to retrieve the distance of the ISM clouds along the LOS and to infer their mean polarization properties. The quality of any reconstruction should therefore focus on these two aspects.

First, to quantify the accuracy with which we recover the cloud parallaxes (distances), we have to deal with the nonregular -often multi modal- shapes of the posterior distributions.
In some cases, the cloud parallax at maximum-likelihood may not correspond to the mode of the posterior distribution.
This becomes pathological once the cloud parallax at maximum log-likelihood belongs to a peak of the posterior distribution that has a small amplitude as compared to the dominant peak or when the bulk of the posterior distribution is squeezed onto one of the limits imposed by the prior. This may happen when the constraining power of the stars is not sufficient for the log-likelihood hyper-surface to show a strong global maximum at maximum-likelihood value; cases generally corresponding to weak ISM polarization signal, or low star density around the dust cloud. In such an occurrence, the fit fails and should be disregarded.

We found that (i) the relative difference between the cloud parallax at maximum log-likelihood with the parallax of the closest star to the input cloud, and (ii) the fact that the cloud parallax at maximum log-likelihood appears in (one of) the main modes of its posterior distribution are two criteria that jointly allow us to assess the quality of our fit on $\varpi_{\rm{C}}$ in addition to the rejection of fits with posterior distribution squeezed on one of the prior limits. If the maximum-likelihood parallax value belongs to (one of) the main modes, then the fit is valid and the relative difference tells us about the accuracy of the recovered cloud parallax.
To decide whether the maximum-likelihood parallax belongs to a significant peak, we analyze the posterior distribution using the peak-finder algorithm \texttt{find\_peaks} of the Python library \texttt{Scipy}
which identifies all local maxima through simple comparison of neighboring values.
Applying it to the marginalized parallax posterior distribution we identify the peaks and their boundaries (i.e., we find local maxima of the marginalized PDF and the range between the two adjacent local minima of each of those). We compute the fraction of the PDF corresponding to each peak. If the fraction associated to the peak to which the maximum-likelihood belongs is higher than a given threshold, then we consider the peak as (one of) the dominant one(s) and the fit as valid. We refer to this, and the detection of squeezed posterior distribution on one of the prior limit, as the criterion on $\hat{\varpi}_{\rm{C}}$. We use a threshold of 0.5 in what follows and make sure that our results do not depend sensitively on this choice.

Second, to quantify the accuracy with which we recover the mean Stokes parameters of the clouds, we rely on the computation of the Mahalanobis distance of the true values with respect to the bivariate posterior distributions of the cloud Stokes parameters. If $\mathbf{c}_\star = (q_{\rm{C}} \,u_{\rm{C}})^\dagger$ is the vector of true mean polarization of the cloud, $\hat{\mathbf{c}} = (\hat{q}_{\rm{C}} \,\hat{u}_{\rm{C}})^\dagger$ is the value at maximum-likelihood, and $\hat{\Sigma}$ is the covariance matrix computed from the re-sampling of the estimated posteriors, the Mahalanobis distance is given by:
\begin{equation}
    d_{\rm{Maha}}(\mathbf{c}_\star|\hat{\mathbf{c}},\hat{\Sigma}) = \sqrt{ (\mathbf{c}_\star - \hat{\mathbf{c}})^\dagger \, {\hat{\Sigma}}^{-1} \, (\mathbf{c}_\star - \hat{\mathbf{c}})} \;.
    \label{eq:dMaha}
\end{equation}

It is also useful to introduce the Mahalanobis distance between the bivariate posterior distribution on the cloud parallax and the point $(0,\,0)$ of the polarization plane. This distance, labeled $d_{\rm{Maha}}(\mathbf{0}|\hat{\mathbf{c}},\hat{\Sigma})$ obtained by substituting the true polarization of a cloud by zero, is a measure of the significance with which a cloud polarization is detected given the data and the sensitivity of the inversion method.

The distance $d_{\rm{Maha}}(\mathbf{c}_\star|\hat{\mathbf{c}},\hat{\Sigma})$ determines whether the `true' Stokes parameters are located, and centered, within the respective bivariate distribution. However, if the reconstruction method fails at recovering the cloud distance for example, the posterior distributions estimated from the sampling method can become arbitrarily large. In such a scenario, it is useful to consider both the absolute Euclidean distance between the pair of Stokes parameters at maximum-likelihood value and the corresponding true values and the effective size of the estimated posteriors.
We use the Euclidean distance defined as
\begin{equation}
    {\rm{L}_2} = \sqrt{(q_{\rm{C}} - \hat{q}_{\rm{C}})^2 + (u_{\rm{C}} - \hat{u}_{\rm{C}})^2} \;,
\end{equation}
and we infer the sizes of the posterior distributions on $q_{\rm{C}}$ and $u_{\rm{C}}$ by considering the ratio ($\xi$) of the posterior sizes that are expected in ideal cases to the measured sizes from the estimated posteriors. If the cloud Stokes parameters were constrained only from $N_{\rm{bg}}$ stars all with uncertainties of 0.1\% (minimal value in our simulated data), the expected size of the posterior is $(0.1\,[\%] / \sqrt{N_{\rm{bg}}})$. We call this estimate ideal in the sense that, in addition to considering the smallest possible polarization uncertainties, it neglects possible correlation between $q_{\rm{C}}$ and $u_{\rm{C}}$, the presence of intrinsic scatter, the presence of foreground stars -which add noise to the reconstruction-, and scrambling along LOS distance from parallax uncertainties.
Within the low S/N regime explored in this section, we found that the parameter
\begin{equation}
    \xi = 2\, \frac{0.1 \, [\%]}{\sqrt{N_{\rm{bg}}}\, \sqrt{\hat{\sigma}_{q_{\rm{C}}} \, \hat{\sigma}_{u_{\rm{C}}}}} \; ,
\end{equation}
where $\hat{\sigma}_{q_{\rm{C}}}$ and $\hat{\sigma}_{u_{\rm{C}}}$ denote the standard deviation (in \%) of the respective estimated posteriors, generally takes values around unity for well-behaved fits and takes small values when the cloud polarization is not well-constrained.

\smallskip

In summary, a reliable reconstruction of the ISM structure along the LOS is expected if, simultaneously, we have a well behaved posterior on cloud parallax with small relative difference (between the cloud parallax at maximum-likelihood and the parallax of the star closest to the input), a low ${\rm{L}}_2$ distance and $\xi$ above a threshold which we choose to be 0.5.

\subsection{One-layer cases}
\label{sec:performance_1L}
In this subsection we explore the behavior and the performance of our inversion method for several cases of a single cloud along the LOS.
We made sure that our inversion method works equally well for any position angle of the magnetic field and will not make any distinction in what follows even though we let it vary to generate our simulated data.
We want to infer the power of our method at recovering the cloud distance and at recovering the Stokes parameters when the amplitude of the polarization signal varies, the distance of the cloud varies, and the amplitude of the intrinsic scatter varies. 
For a given sample of stars, the choice of cloud distance affects the number of stars in the foreground and the background that are available to constrain the model. The amplitude of the (mean) polarization signal depends on both $P_{\rm{max}}$ and $\gamma_{\mathbf{B}_{\rm{reg}}}$ and both should therefore be explored. The amplitude of the intrinsic scatter directly affects the amount of scatter in the $(q_{\rm{V}},\,u_{\rm{V}})$ plane for background stars only.

\medskip

Given a sample of stars, we choose the cloud distances such that 90\%, 70\%, 50\%, 30\%, and 10\% of the stars are in the background of the cloud and hence, useful to constrain the polarization properties of the cloud. The lower the fraction of stars in the background ($f_{\rm{bg}}$), the larger the distance of the cloud, and the larger the parallax uncertainties of stars in the distance neighborhood of the cloud. Thus, we expect larger relative differences in cloud parallax, and consequently larger ${\rm{L}}_2$ and smaller $\xi$, at small $f_{\rm{bg}}$ than at large $f_{\rm{bg}}$. In practice, however, this picture may be changed due to the imposed prior on cloud parallax and, in particular, due to the lower limit on which the cloud parallax posterior may be squeezed. Those cases, that anyway do not pass the criterion on $\hat{\varpi}_{\rm{C}}$, may show small relative differences on cloud parallaxes simply because of the limit of the prior on $\varpi_{\rm{C}}$ is close to the input cloud parallax value. In such cases, though, the posterior on the mean polarization Stokes parameter should be broader than expected (i.e., showing low $\xi$ values).

For a fixed level of intrinsic scatter with $A_{\rm{turb}} = 0.2$ and the mean star sample with a $1^\circ$ diameter beam, and for all values of $f_{\rm{bg}}$, we first study the impact of $p_{\rm{C}}$, the degree of polarization induced by the cloud to background stars, on the quality of the reconstruction of the ISM structure along the LOS. For this sample made of 345 stars the $f_{\rm{bg}}$ cuts correspond to cloud distances of about 270, 565, 790, 1050, and 1712 pc.  
To avoid mixing possible dependence on $P_{\rm{max}}$ and $\gamma_{\mathbf{B}_{\rm{reg}}}$, we impose the regular component of the magnetic field to be in the POS ($\gamma_{\mathbf{B}_{\rm{reg}}} = 0^\circ$) and only vary $P_{\rm{max}}$ from 0.05\% to 0.3\% in steps of 0.05\%. We further consider 10 realizations for each pair of ($f_{\rm{bg}},\,P_{\rm{max}}$) varying the POS position angle of  $\mathbf{B}_{\rm{reg}}$.
This generates a set of 300 simulated samples of starlight polarization data to which we apply our inversion method. Each sample is analyzed through our maximum-likelihood method using the one-layer model. We consider relatively loose flat priors on the six parameters with $\{q_{\rm{C}},\, u_{\rm{C}}\} \in [-2\%,\,2\%]$, $\{C_{{\rm{int}},qq},\, C_{{\rm{int}},uu}\} \in [0,\,10^{-4}]$, $|C_{{\rm{int}},qu}| \leq 10^{-4}$, and $\varpi_{\rm{C}}$ in the range corresponding to distances between 100 and 3500 pc. A value of $\{C_{{\rm{int}},qq} = 10^{-4}$ corresponds to an intrinsic scatter of 1\% on Stokes $q$.
For convenience, the definitions of these priors are repeated in Table~\ref{tab:Priors}.
We consider 1000 live points and sample the parameter space until we reach a tolerance on the estimated log-evidence below 0.1. Generally in our cases, this corresponds to an uncertainty on the log-evidence on the order of one part per 10,000.
\begin{table*}
    \begin{center}%
    \caption{\label{tab:Priors} Model parameters and range limits of uniform priors.\\[-1.ex]}
    {\small
    \begin{tabular}{lr|ccccc}
    \hline \hline \\[-.5ex]
         \multicolumn{2}{l}{Cases} & \multicolumn{5}{c}{Model parameters} \\ \\[-1.ex]
         & cloud \# & $\varpi_{\rm{C}}$ & $q_{\rm{C}}$ & $u_{\rm{C}}$ & $C_{{\rm{int}},qq}$ & $C_{{\rm{int}},uu}$ \\
         &  & [mas] & $[\%]$ & $[\%]$ & $[\%]^2$ & $[\%]^2$ \\ \\[-.5ex]
     \hline \\[-.5ex]
     \multicolumn{2}{l}{Sect.~\ref{sec:performance_1L}} & & & & & \\
        & cloud~1 & [0.286,~10] & [-2,~2] & [-2,~2] & [0,~1] & [0,~1] \\ \\[-.5ex]
         \multicolumn{3}{l}{Sect.~\ref{sec:performance_2L}, \ref{sec:modselect_1L}, and \ref{sec:modselect_2L}} & & & & \\
        & cloud~1 & [1.667,~10] & [-2,~2] & [-2,~2] & [0,~1] & [0,~1] \\
        & cloud~2 & [0.286,~3.334] & [-2,~2] & [-2,~2] & [0,~1] & [0,~1] \\  \\[-.5ex]
        \hline\\[-1.5ex]
    \end{tabular}
    }
    \tablefoot{Labels of the model parameters follow the notations given in the text: $\varpi_{\rm{C}}$ is the cloud parallax, $q_{\rm{C}}$ is the cloud's mean Stokes parameter $q$, $u_{\rm{C}}$ is the cloud's mean Stokes parameter $u$, $C_{{\rm{int}},qq}$ and $C_{{\rm{int}},uu}$ are the diagonal elements of the intrinsic-scatter covariance matrix. In all cases, the flat prior on $C_{{\rm{int}},qu}$, the off-diagonal element of the same matrix, is defined by ($- \sqrt{C_{{\rm{int}},qq}\, C_{{\rm{int}},uu}}$,~$\sqrt{C_{{\rm{int}},qq}\, C_{{\rm{int}},uu}}$) such that the semi-positive-definiteness of the covariance matrix is guaranteed (thus, excluding the limits).
    The parallax values of 0.286, 1.667, 3.334, and 10 correspond to distance values of 3500, 600, 300, and 100 pc, respectively.
    }
    \end{center}
\end{table*}

We subsequently analyze the posterior distributions, flag reconstructions with odd posterior on $\varpi_{\rm{C}}$ (i.e., those not passing the criterion on $\hat{\varpi}_{\rm{C}}$), and evaluate the three quantities with which we intend to qualify the goodness of the reconstruction: the relative difference between the $\hat{\varpi}_{\rm{C}}$ and the parallax of the star that is nearest to the input cloud, the ${\rm{L}}_2$ distance between $(\hat{q}_{\rm{C}},\,\hat{u}_{\rm{C}})$ and the true $(q_{\rm{C}},\,u_{\rm{C}})$, and the parameter $\xi$ which compares the sizes of the posteriors on $(q_{\rm{C}},\,u_{\rm{C}})$ with the ideal ones (see above).
For the reconstructions that pass the criterion on $\hat{\varpi}_{\rm{C}}$, we checked that there is no bias in the recovered mean polarization properties of the cloud; $(\hat{q}_{\rm{C}},\,\hat{u}_{\rm{C}})$ is always found close to the input $(q_{\rm{C}},\,u_{\rm{C}})$ within uncertainties.

\begin{figure}
    \centering
    \includegraphics[trim={0.2cm 0cm 0.2cm 0cm},clip,width=.98\linewidth]{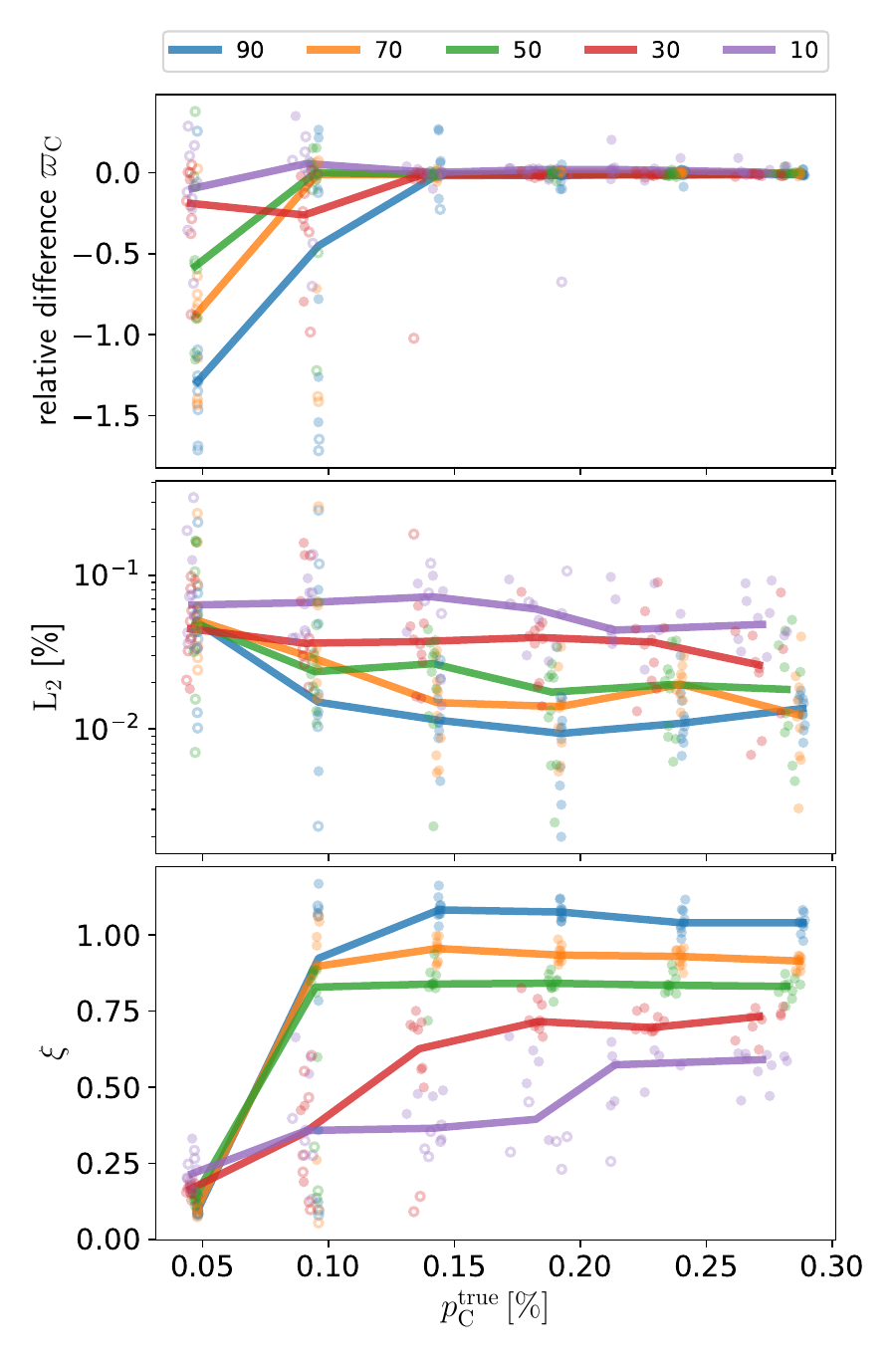}\\[-2.5ex]
    \caption{Performance of the inversion method as a function of the true polarization signal for different cloud distances, probed by the values of $f_{\rm{bg}}$, shown in the legend and given in per cent. The method is applied to simulated data from the $1^\circ$ aperture star sample (345 stars). Top: relative difference between the cloud parallax at maximum log-likelihood with the parallax of the closest star to the input cloud ($2\,(\hat{\varpi}_{\rm{C}} - \varpi_{\rm{closest}})/(\hat{\varpi}_{\rm{C}} + \varpi_{\rm{closest}})$).
    Middle: ${\rm{L}}_2$, distance between the true and estimated mean polarization vector. Bottom: $\xi$, the ratio between ideal and estimated posterior size on clouds' Stokes parameters. 
    $P_{\rm{max}}$ varies in multiples of 0.05\%, $\gamma_{\mathbf{B}_{\rm{reg}}} = 0^\circ$ and $A_{\rm{turb}} = 0.2$ and 10 simulated samples are obtained by varying $\psi_{\mathbf{B}_{\rm{reg}}}$. For the same $f_{\rm{bg}}$, the solid lines connect the median of the 10 reconstructions on simulated samples with the same $P_{\rm{max}}$.
    Filled (empty) symbols correspond to fits that pass (do not pass) the $\hat{\varpi}_{\rm{C}}$ criterion explained in the text.
    }
    \label{fig:Performance_Beam60-G00At02}
\end{figure}
We present in Fig.~\ref{fig:Performance_Beam60-G00At02} our results on parallax relative differences (top), ${\rm{L}}_2$ (middle) and $\xi$ (bottom) as a function of the true $p_{\rm{C}}$ for the five cloud distances corresponding to the five $f_{\rm{bg}}$ thresholds. The different colors indicate different $f_{\rm{bg}}$ values, the solid lines connect the medians obtained for each $f_{\rm{bg}}$ at the different $P_{\rm{max}}$ values (all fits). The scatters and offsets in $p_{\rm{C}}^{\rm{true}}$ about multiples of 0.05\% observed in the figure originate from the presence of the intrinsic scatter. Filled (empty) symbols indicate fits that pass (do not pass) the criterion on $\hat{\varpi}_{\rm{C}}$. 

As a general trend, we see that the performance on both cloud distance and cloud polarization properties is generally good for the cases with  $p_{\rm{C}}^{\rm{true}} \gtrsim 0.1$\%. 
Below this threshold a significant fraction of reconstructions lead to pathological posterior distributions on cloud parallax. In practice those reconstructions should be disregarded. Pathological reconstructions also extend to larger $p_{\rm{C}}^{\rm{true}}$ for cases corresponding to $f_{\rm{bg}} = 10\%$. This is due to both a lower number of stars to constrain the model parameters and larger parallax uncertainties. Running the same performance test using starlight polarization samples in the $0.5^\circ$ diameter beam, we find that it is the number of stars in the background of the cloud that is the dominant factor, as pathological cases are found at larger $f_{\rm{bg}}$ on the order of 30\% in tested cases (not shown). It is difficult to constrain the distance of a (relatively) far away cloud beyond which only about 30 stars provide constraints.

When $p_{\rm{C}}^{\rm{true}} \gtrsim 0.1\%$, the relative differences in cloud parallax obtained for most of the reconstructions with well-behaved $\varpi_{\rm{C}}$ posteriors are found below the level of 5\% regardless of the cloud distance. The Euclidean distances ${\rm{L}}_2$ indicate a good accuracy at recovering the cloud mean polarization, with about 90\% of the valid reconstructions leading to an ${\rm{L}}_2$ below 0.05\%. The larger the $f_{\rm{bg}}$, the smaller ${\rm{L}}_2$, thus the more accurate the reconstruction. This is an expected feature since for lower $f_{\rm{bg}}$, the reconstructions can be noisier due to both the presence of (an increasing fraction of) foreground stars and a smaller amount of stars constraining the cloud polarization properties. This is also seen in the $\xi$ panel. For the valid reconstructions, the posterior distributions get larger ($\xi$ decreases) when $f_{\rm{bg}}$ decreases, thus lowering the significance with which a cloud can be detected given the data.
The sizes of the posteriors on the Stokes parameters seem to be well determined irrespective of $p_{\rm{C}}^{\rm{true}}$ as soon as the recovery of the cloud distance is successful.

\begin{figure}
    \centering
    \includegraphics[trim={0.4cm 0.4cm 0.4cm 0cm},clip,width=.98\linewidth]{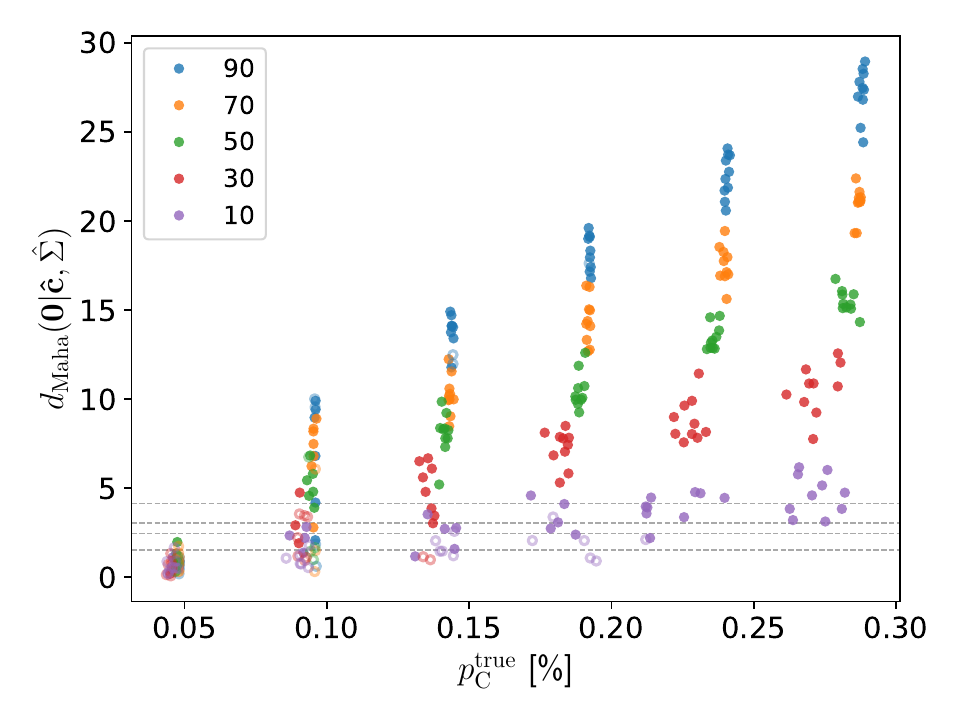}\\[-1.5ex]
    \caption{Significance of the cloud detection ($d_{\rm{Maha}}(\mathbf{0}|\hat{\mathbf{c}},\hat{\Sigma})$) as a function of the input cloud polarization and for the several $f_{\rm{bg}}$ values corresponding to the same reconstructions characterized through Fig.~\ref{fig:Performance_Beam60-G00At02}.
    The horizontal lines indicate threshold values corresponding to detection levels with 68\%, 95\%, 99\%, and 99.98\% probabilities of finding a distance lower than threshold. Symbol and color conventions are the same as in Fig.~\ref{fig:Performance_Beam60-G00At02}.
    }
    \label{fig:dMahaNull-G00At02}
\end{figure}

We show in Fig.~\ref{fig:dMahaNull-G00At02} the significance of the detection of the cloud defined as the Mahalanobis distance of the zero polarization with respect to the estimated posterior distributions on the cloud polarization as a function of $p_{\rm{C}}^{\rm{true}}$. As expected, the significance with which a cloud can be detected depends both on the amplitude of the polarization signal that it induces to the background stars and on the fraction of stars in the background. The larger the $p_{\rm{C}}^{\rm{true}}$ and the larger the $f_{\rm{bg}}$, the larger the detection significance.

We repeat the performance analysis presented above with a level of intrinsic scatter given by $A_{\rm{turb}} = 0.1$. Owing to the fact that the intrinsic scatter is explicitly accounted for in the maximized log-likelihood, the performance on cloud parallax and mean polarization is only slightly better in this case than in the case of $A_{\rm{turb}} = 0.2$.

We finally repeat our performance analysis using starlight polarization data simulated from the mean star sample in a $0.5^\circ$ diameter beam. The sample is made of 85 stars. The general picture depicted above remains similar. However, the decreased number of stars (a factor of about four) naturally implies a less accurate and precise reconstruction of the ISM along the LOS.
The cloud parallaxes can be recovered with a relative difference lower than 15\% for a majority of good reconstructions but we found more reconstructions that do not pass the criterion on $\hat{\varpi}_{\rm{C}}$. The drop in precision is due to the loss in density of stars that effectively sample the distance axis along the LOS. In what concerns the polarization properties, only 55\% (80\%) of the reconstructions that pass the criterion on $\varpi_{\rm{C}}$ and with $p_{\rm{C}}^{\rm{true}} > 0.1\%$ show an ${\rm{L}}_2$ distance lower than 0.05\% (0.1\%). No bias is observed and the sizes of the posteriors are well defined. Due to the reduced number of stars (and the low polarization considered here) several reconstructions lead to cloud detections that do not exceed the level of 2$\sigma$ ($d_{\rm{Maha}}(\mathbf{0}|\hat{\mathbf{c}},\hat{\Sigma}) \gtrsim 2.45$). However, for $f_{\rm{bg}} \geq 0.5$ (corresponding to cloud distances lower than $\approx 750$ pc) and $p_{\rm{C}}^{\rm{true}} \gtrsim 0.1\%$, most of the reconstructions are successful and the cloud is detected at high significance ($d_{\rm{Maha}}(\mathbf{0}|\hat{\mathbf{c}},\hat{\Sigma}) > 4.12$).

\smallskip

\begin{figure}
    \centering
    \includegraphics[trim={0cm 0cm 0cm 0cm},clip,width=.98\linewidth]{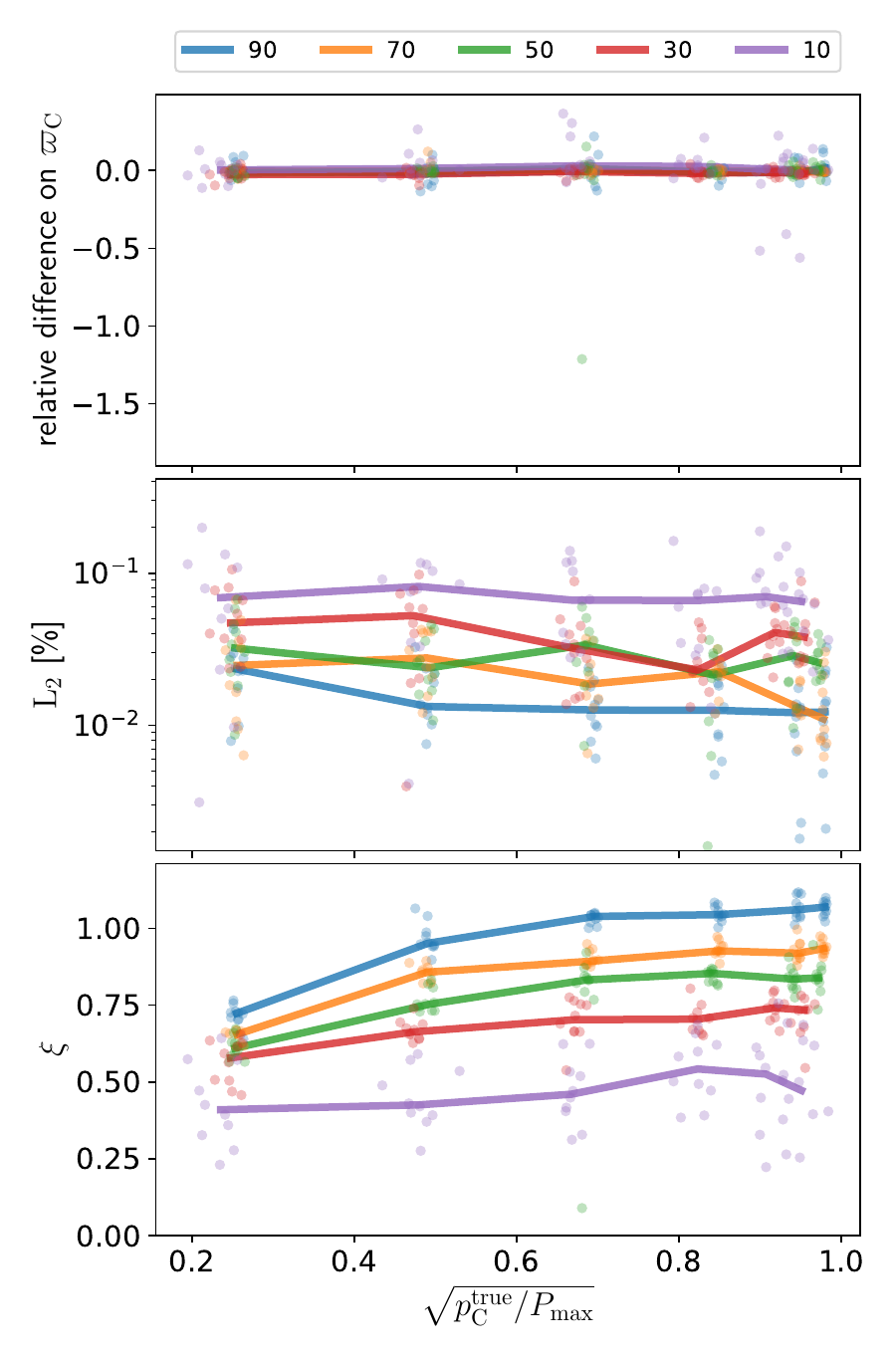}\\[-2.5ex]
    \caption{Performance of the inversion method as a function of $\sim\gamma_{\mathbf{B}}$, the inclination angle of the magnetic field with respect to the POS. Conventions are the same as in Fig.~\ref{fig:Performance_Beam60-G00At02}.
    In absence of intrinsic scatter the abscissa would correspond to $\cos(\gamma_{\mathbf{B}_{\rm{reg}}})$.
    \label{fig:Performance-vs-G_Beam60-pC020At02}
    }
\end{figure}

\begin{figure}
    \centering
    \includegraphics[trim={0.4cm 0.4cm 0.4cm 0cm},clip,width=.98\linewidth]{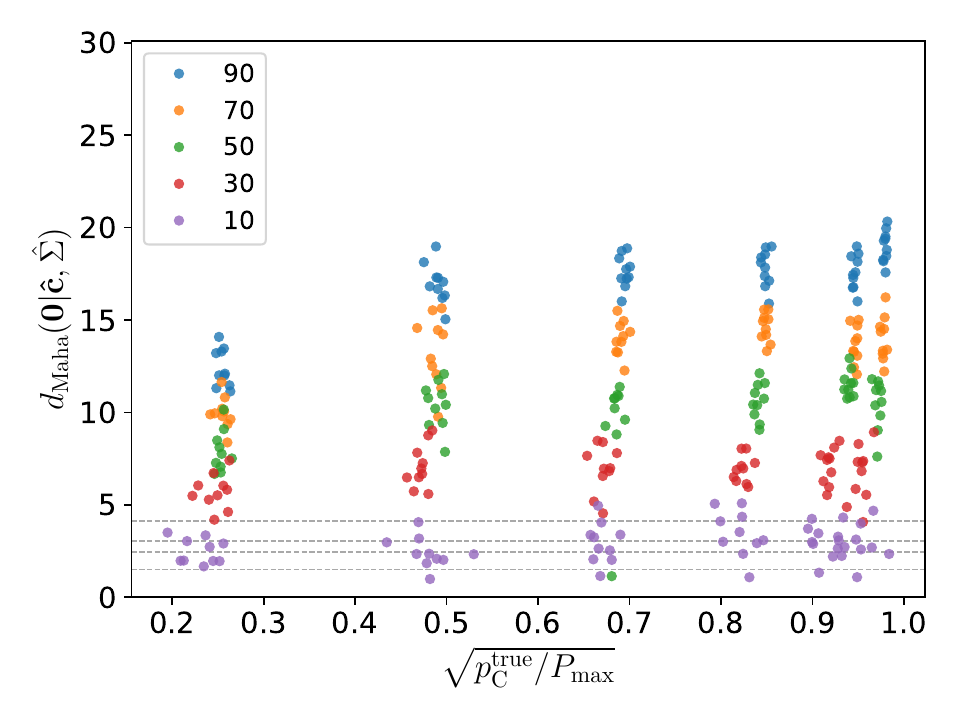}\\[-1.5ex]
    \caption{Significance of the cloud detection in polarization as a function of $\sim\cos(\gamma_{\mathbf{B}})$. Conventions are the same as in Fig.~\ref{fig:dMahaNull-G00At02}.
    }
    \label{fig:dMahaNull-pC020At02}
\end{figure}

We now investigate the possible effects of the inclination angle of the magnetic field with the POS on the performance of our inversion method.
For this purpose we create a new set of simulated starlight polarization samples for the $1^\circ$ diameter beam. We fix $A_{\rm{turb}} = 0.2$, explore the five thresholds in $f_{\rm{bg}}$, sample the inclination angle $\gamma_{\mathbf{B}_{\rm{reg}}}$ from $0^\circ$ to $75^\circ$ in steps of $15^\circ$ and adapt the $P_{\rm{max}}$ values so that the observed $p_{\rm{C}}$ values are similar and approximately about 0.2\% using $P_{\rm{max}} = 0.2 / \cos^2(\gamma_{\mathbf{B}_{\rm{reg}}}) \, [\%]$. We generate 10 simulated samples for each pair of ($\gamma_{\mathbf{B}_{\rm{reg}}}$, $P_{\rm{max}}$) by varying $\psi_{\mathbf{B}_{\rm{reg}}}$. As before, we then apply our inversion method and study its performance.
The results are shown in Fig.~\ref{fig:Performance-vs-G_Beam60-pC020At02}.
For fixed $f_{\rm{bg}}$, the relative differences in parallax and the ${\rm{L}}_2$ distances show very little dependence on the inclination angle. A mild dependence is observed for $\xi$: the larger the inclination angle the larger the size of the posteriors.
However, the panels in ${\rm{L}}_2$ and $\xi$ clearly demonstrate that it is the number of stars in the background of the cloud that is decisive for the performance of the inversion method.
For a similar degree of polarization, a low number of stars in the background of a far away cloud prevents us from recovering the cloud parallax with great accuracy and leads to loose constraints on its polarization properties. The latter is also seen in Fig.~\ref{fig:dMahaNull-pC020At02} where we show the significance with which the polarization of the cloud is detected compared to zero.
A mild dependence of the significance as a function of the inclination angle is observed although we made sure that the mean degree of polarization is similar for all $\gamma_{\mathbf{B}_{\rm{reg}}}$. The more the magnetic field is inclined on the LOS the less the significance of the detection in the polarization plane. This is due to the fact that the scatter in starlight polarization is larger for higher inclination angles.

\medskip

To summarize, for the case of a single cloud along the LOS we find that our inversion method is effective in recovering the cloud properties (parallax and mean polarization) when the polarization signal induced by the cloud to background stars is at least at the level of $\approx$0.1\%.
This threshold corresponds to the minimum uncertainty in Stokes parameters of individual stellar measurements that we introduce in our mock observations to account for systematic uncertainties.
Below this threshold the cloud parallax is not well recovered and we only recover loose constraints on the cloud polarization properties.
Above this threshold the cloud parallax is recovered with high accuracy which increases with both the amplitude of the polarization signal and the fraction of stars in the background of the cloud. The latter also implies that the distance of far away clouds is generally recovered with less accuracy than for nearby clouds. Meanwhile, useful distance (parallax) constraints can be placed for all clouds nearer than about 750 pc even when considering star samples in a $0.5^\circ$ diameter beam. We find that as soon as the cloud parallax is well recovered, the cloud polarization is recovered without bias and detected with a significance that depends on both the amplitude of the input polarization signal and the number of stars in the background of the cloud. The precision on cloud mean polarization can be as low as $0.05\%$ for realistic cases, owing to the number of constraining stars. We find that, for fixed amplitude of induced polarization signal, the inclination of the magnetic field with respect to the POS (within the investigated range) does not affect the precision on the recovered cloud mean polarization degree but slightly changes the significance with which the polarization is detected. For realistic settings we additionally do not find a strong dependence of the performance of our method on the level of intrinsic scatter.

In conclusion, as soon as the polarization signal induced by a cloud to background stars is larger than the minimum uncertainties on individual polarization measurements and that at least about 30 stars are in the background of the cloud, our inversion method is effective in constraining the cloud parallax and its mean polarization properties.
When the polarization signal is weaker or the number of background stars is smaller than that, the data are too limited to allow for the presence of the cloud to be identified with confidence. In these cases, the parameters are poorly constrained leading to unreliable reconstructions. These correspond to the pathological cases which we have shown how to identify.
For most of these cases, the AIC values corresponding to a zero-layer model (i.e., no cloud along the LOS) are found to be lower than the AIC obtained with the one-layer model, suggesting no evidence for any cloud along the LOS given the data.

\subsection{Two-layer cases}
\label{sec:performance_2L}
In this subsection, we test the performance of the method at recovering the structure of the magnetized ISM along the LOS when there are two clouds. Specifically, we are interested in exploring the limiting conditions that could lead to an imprecise reconstruction. In this respect, for a fixed level of stellar polarization uncertainties, we may generally expect that the overall performance of our inversion method will depend on the respective distance of the two clouds (affecting the number of stars in the foreground and the background of each cloud but also in between the clouds), the relative amplitude and weakness of the polarization signals as compared to noise and intrinsic scatter, and possibly the difference in magnetic field orientations.

Given the results presented in the previous subsection we might expect to be unable to recover the cloud properties (distance and polarization) when the far-away cloud has low polarization and or large distance. The detection of this second cloud may also be more difficult due to the additional scatter induced by the presence of a foreground cloud that not only modifies the zero-mean but also adds dispersion due to the intrinsic scatter. The contributions from the intrinsic scatter in the two clouds will also add for the stars in the background of the second cloud.
Additionally, in cases where there are too few stars in between the two clouds or too few stars in the background of the far-away cloud, we expect that cloud properties of the nearby cloud might not be well retrieved because most of the polarization signal will then be attributed to only one of the two clouds.
Generally, if the properties of one of the two clouds are not well recovered, this will automatically affect the reconstruction of the second cloud and therefore might compromise the overall reconstruction.
We attempt to assess and quantify any such dependencies in this subsection by applying our inversion method to simulated starlight polarization data obtained from our toy model.

We narrow the range of possible ISM configurations by focusing on most realistic sky conditions.
We fix the distance of the nearby cloud at 350 pc from the Sun. At high Galactic latitudes, this corresponds roughly to the maximum distance of the dusty and magnetized shell of the Local Bubble that surrounds the Sun (e.g., \citealt{Skalidis2018}; \citealt{Pelgrims2020}) and therefore the maximum distance at which we expect a LOS to have intersected the nearest cloud. In our star sample with $1^\circ$ diameter beam, this leads to about 55 and 290 stars in the foreground and background of the nearby cloud, respectively.
Additionally, we fix the inclination angle of the magnetic field permeating the nearby cloud to $30^\circ$, fix its $P_{\rm{max}}$ so that $p_{\rm{C}} = 0.2\%$ and adopt a position angle of $22.5^\circ$ to distribute equally the polarization signal over both Stokes parameters. According to the results obtained in the previous subsection, the properties of such a cloud are expected to be well recovered by our inversion method if this cloud is alone along the LOS.

We add a second cloud along the LOS and vary its distance, degree of polarization, and magnetic field position angle. We choose to fix the distance of the second cloud so that 90\%, 70\%, 50\%, 30\%, and 10\% of the stars in the background of the nearby cloud are also background to the far-away cloud. The fraction of stars in the background of the nearby cloud that are also background to the more distant cloud is denoted $f_{\rm{bg2}}$. Namely, the second cloud is placed at distance of about 520, 685, 900, 1150, and 1800 pc. The number of stars in between the two clouds are respectively $\sim$30, 90, 145, 200, and 260.
For each cloud distance, we then vary the degree of polarization that the second cloud induces to its background stars and the relative position angle of the magnetic field in the far-away cloud with respect to the one in the nearby cloud ($\Delta\psi_{1,2}$) as follows: $p_{\rm{C2}}$ ranges from 0.1\% to 0.25\% in steps of 0.05\%, and $\Delta\psi_{1,2}$ takes values from $0^\circ$ to $150^\circ$ in steps of $30^\circ$. We fix the inclination angle of the magnetic field in the far-away cloud to be zero.
The level of intrinsic scatter in both clouds is set to $A_{\rm{turb}} = 0.2$. Lowering this value in one or both of the clouds might result in slightly better performance.

We generate four random realizations of the simulation settings just described. In total, we consider 480 configurations for the magnetized ISM structure along the LOS, to each of which corresponds a mock starlight polarization sample of 345 stars.
We apply our inversion method to each of these samples using a two-layer model. For all cloud parameters, except for the parallax, we consider flat priors as used before. For the parallaxes of the clouds, we define a flat prior so that the distance of the nearby cloud can range from 100 to 600 pc and the distance of the far-away cloud can range from 300 to 3500 pc.
As before, we use 1000 live points and we sample the parameter space until we reach a tolerance on the estimated log-evidence below 0.1. Then, we analyze the resulting posterior distributions and characterize the results as before to infer the performance of the method.

\begin{figure}
    \centering
    \includegraphics[trim={0.2cm 0cm 0.2cm 0cm},clip,width=.98\linewidth]{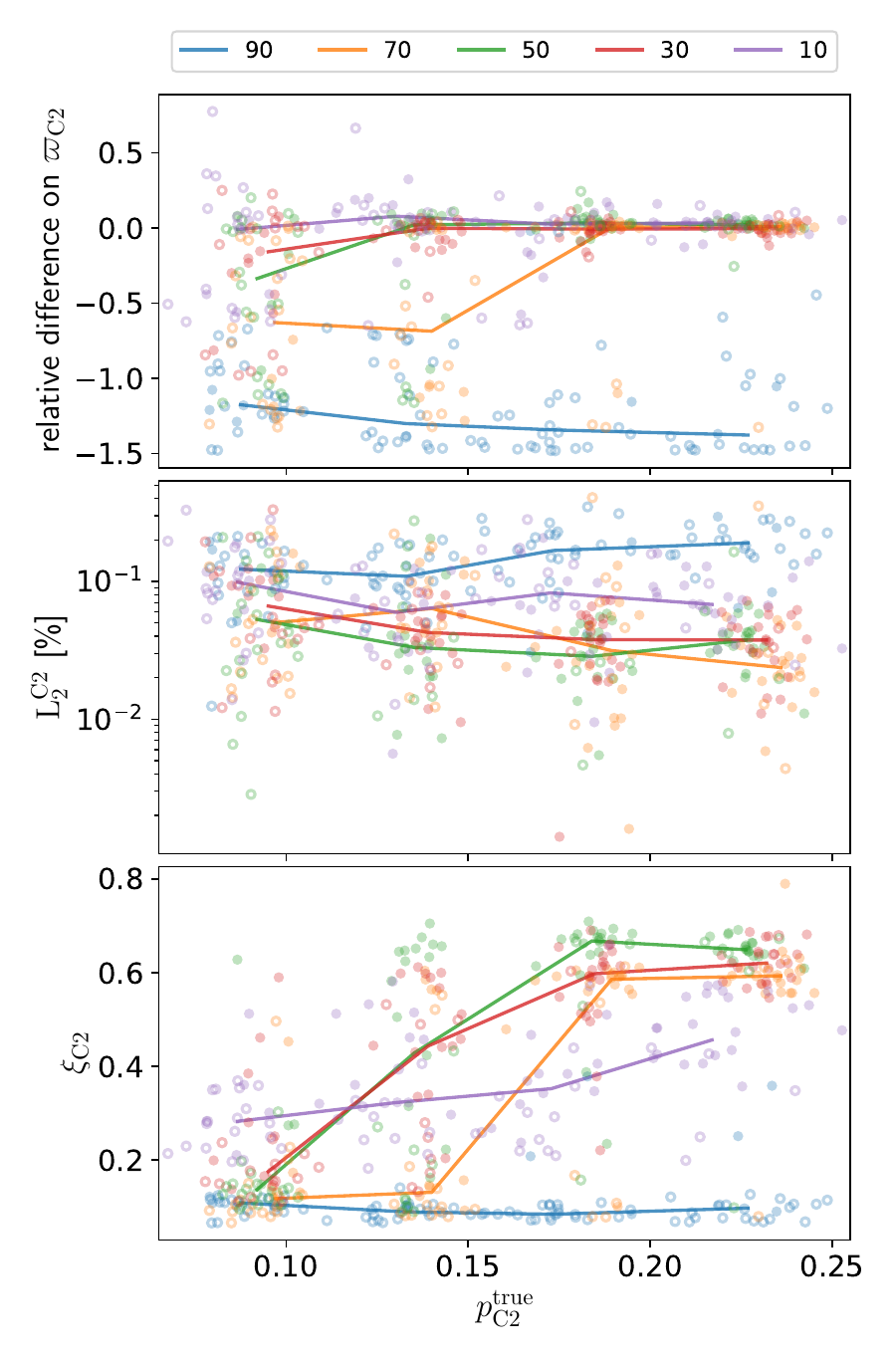}\\[-2.5ex]
    \caption{Performance of the inversion method to retrieve properties of the distant cloud for two-cloud cases, as a function of $p_{\rm{C}2}$ and for all $f_{\rm{bg}2}$ values. Conventions are the same as in Fig.~\ref{fig:Performance_Beam60-G00At02} except that filled symbols correspond to reconstruction in which both $\varpi_{\rm{C}1}$ and $\varpi_{\rm{C}2}$ pass the selection criterion.
    }
    \label{fig:Performance2C_C2-vs-pC2_Beam60-At02}
\end{figure}

In Fig.~\ref{fig:Performance2C_C2-vs-pC2_Beam60-At02} we show the relative differences between the parallax of the second cloud obtained at maximum-likelihood value as compared to the parallax of the closest star to the input cloud parallax, the ${\rm{L}}_2$ distances on second cloud mean polarization (${\rm{L}}_2^{\rm{C}2}$) and the ratio $\xi$ of ideal-to-actual size of the posterior distributions on mean polarization for the second cloud ($\xi_{\rm{C}2}$).
Similarly to Fig.~\ref{fig:Performance_Beam60-G00At02}, we show the results for all values of $f_{\rm{bg}2}$, the fraction of stars in the background of the nearby cloud that are also in the background of the far-away cloud, and present the data as a function of the true degree of polarization input to the far-away cloud.
In this case, the filled (empty) symbols correspond to fits in which the posterior distributions of both cloud parallaxes pass (do not pass) the $\hat{\varpi}_{\rm{C}}$ criterion.

\smallskip

We first notice that a large fraction of fits do not pass the joint selection criterion on cloud parallaxes. In fact, while most of the posteriors on $\varpi_{\rm{C}1}$ pass the criterion (more below), about 45\% of the reconstructions lead to posterior distributions on $\varpi_{\rm{C}2}$ that do not pass it. Invalid posterior distributions generally appear for large $f_{\rm{bg}2}$ values and for low $p_{\rm{C}2}^{\rm{true}}$ values, irrespective of $f_{\rm{bg}2}$.
As soon as $p_{\rm{C}2} \gtrsim 0.1\%$ and $f_{\rm{bg}2} \lesssim 70\%$, the posteriors on cloud parallaxes are valid and lead to reasonable constraints for more than 75\% of the cases (this is for the cases that we have investigated, probing the limits of the method capabilities).

\smallskip

Generally, for both valid and invalid posterior distributions on $\varpi_{\rm{C}2}$, the parallax of the nearby cloud is well constrained close to the input values. This is true except for a few cases which correspond to large values of $f_{\rm{bg}2}$, $p_{\rm{C}2} \gtrsim p_{\rm{C}1}$, and particular $\Delta\psi_{1,2}$ which make it impossible to differentiate the bivariate distribution of $(q_{\rm{V}},\,u_{\rm{V}})$ in the inter-cloud region from the bivariate distribution in the background of the more distant cloud. In such cases, because of the low number of stars sampling only the nearby cloud ($\approx30$), because of the high level of uncertainties in individual polarization measurements, and because of the presence of the intrinsic scatter, the fit attributes all the polarization to only one of the two clouds which turns out to be the nearby one in order to account for the dispersion seen in the $(q_{\rm{V}},\,u_{\rm{V}})$ plane. When $p_{\rm{C}2} \approx p_{\rm{C}1}$, the cloud is placed much closer than the input value when $\Delta\psi = 90^\circ$ or much farther than the input value when $\Delta\psi = 0^\circ$.
The reason is that when $\Delta\psi \approx 90^\circ$, the mean polarization of all the stars in the background of the second cloud approximate zero while in the case $\Delta\psi \approx 0^\circ$ the majority of the those stars have a degree of polarization around 0.4\%. The cloud distance is adapted to ensure the best compromise to minimize the residuals of individual star polarization and thus, to maximize the likelihood.
When the polarization of the second cloud is lower, this dependence on $\Delta\psi$ tends to vanish simply because the induced polarization is somewhat negligible as compared to the polarization (and scatter) induced by the foreground cloud which are loosely constrained by only 30 stars. The tendency to drive the cloud parallax far away from the input value is thus not significant and the parallax of the nearby cloud is found close to its input value.
In any case, when the second cloud is too close to the nearby cloud, the second-cloud parallax is not well recovered and it exhibits a pathological posterior. Such a fit should therefore be disregarded.

\smallskip

For reconstructions that satisfy the $\hat{\varpi}_{\rm{C}}$ criterion for both clouds, we find that the relative differences on $\varpi_{\rm{C}1}$ are below 15\% for 98\% of the studied cases (or below 5\% in 58\% of the cases) and that of $\varpi_{\rm{C}2}$ are below 15\% for 80\% of the studied cases (or below 5\% in 60\% of the cases).
For the nearby cloud, the Euclidean distances (${\rm{L}}_2^{\rm C1}$) between the true Stokes parameters and those at maximum-likelihood differ by less than 0.05\% in degree of polarization in 90\% of the cases (or below 0.1\% in 99\% of the cases) and the $\xi$ values corresponding to these reconstructions are generally at about one or greater. Thus we find very good accuracy and precision on the retrieved mean polarization of the foreground cloud.
For the far-away cloud, the ${\rm{L}}_2^{\rm{C}2}$ are below 0.05\% in 59\% of the cases (or below 0.1\% in 84\% of the cases) and the $\xi$ values are lower than one but above 0.5 for 64\% of the valid reconstructions (regarding the $\hat{\varpi}_{\rm{C}}$ criterion). The fact that $\xi_{\rm{C}2}$ is generally lower than $\xi_{\rm{C}1}$ is related to the presence of the intrinsic scatter already induced by the nearby cloud in addition to the one in the second cloud. We note that none of these is accounted for in the ideal estimate of the posterior size while computing the $\xi$ ratio. Therefore we find that the mean polarization properties of the far-away cloud are also generally well retrieved as soon as $p_{\rm{C}2} \gtrsim 0.1 \%$ and if $f_{\rm{bg}2} \in [70\%,\,30\%]$, that is, if the number of stars in the inter-cloud region is large enough and if there are not too few stars in the background of the second cloud to constrain its properties.

\smallskip

In summary, we find that the inversion method works efficiently for a large range of possible configurations of the magnetized ISM with two clouds along the LOS. It fails at recovering the ISM structure if the mean polarization of the second cloud is comparable to the systematic uncertainty in the starlight polarization (i.e., if $p_{\rm{C}}\approx$0.1\%), if the number of stars in the inter-cloud region is smaller than $\approx 30$, or if the number of stars in the background of the far-away cloud is lower than $\approx 30$. In those cases, the method generally leads to satisfying constraints on the nearby cloud unless the magnetic fields in both clouds are mostly parallel or perpendicular to each other and there are too few stars ($\lesssim 30$) constraining the nearby cloud polarization properties.

\subsection{Test on model selection}
\label{sec:testmodelselection}
Finally, for our tests of method performance, we want to explore the use and relevance of our model selection criteria, using statistical evidence, the AIC, or both, to choose among the different trial models (i.e., to choose how many clouds are on the LOS), as described in Sect.~\ref{sec:modelselection}.
We limit our investigations to single- and two-layer cases in the low polarization regime and defer the exploration of more complex cases that might occur toward denser ISM regions than those anticipated for the \textsc{Pasiphae} survey to further dedicated work.

We generate mock starlight polarization catalogs from the  star sample within the $1^\circ$ beam, placing either one or two clouds along the LOS as described below. We then apply our inversion method testing the one-layer and the two-layer models one after the other. We check the results of each reconstruction and flag those with pathological posterior distributions on cloud parallax, we record the evidence of the model given the data and estimate the value of the AIC from the obtained maximum log-likelihood, and finally proceed to model comparison.

\subsubsection{One-cloud mock samples}
\label{sec:modselect_1L}
We first consider the case when one dust cloud is introduced along the LOS. We use our toy model to produce the set of mock starlight polarization samples.
We fix the distance of the cloud at 350 pc, impose a $p_{\rm{C}}^{\rm{true}} \simeq 0.2\%$ with $\gamma_{\mathbf{B}_{\rm{reg}}} = 30^\circ$, and a level of intrinsic scatter given by $A_{\rm{turb}} = 0.2$. We then vary the position angle of the magnetic field from $0^\circ$ to $162^\circ$ in steps of $18^\circ$ and finally generate 10 random realizations of each setup. In total, this makes 100 mock samples.

When using the one-layer model, we find that all reconstructions are valid in terms of the shape of the posterior distribution.
Only 12 reconstructions from the two-layer model are valid in this respect, i.e., that posteriors on both $\varpi_{\rm{C}1}$ and $\varpi_{\rm{C}2}$ pass the selection criterion, implying that the one-layer model should already be favored against the two-layer model in 88\% of cases.

We find that the evidence is higher for the one-layer model than for the two-layer model in all cases. 20\% show an AIC lower for the two-layer model than for the one-layer model but the probability that, among the two tested models, the one-layer model actually minimizes the loss of information never goes below 36\%.
However, when limiting ourselves to the 12 samples for which we obtain valid reconstructions with the two-layer model, the AIC values are always lower for the one-layer model than for the two-layer model. The probability that, instead, the two-layer model actually minimizes the loss of information ($P_{2L | \{1L,2L\} }$) reaches a value as high as 20\% for one case but more than half (a quarter) of the samples show this probability below the 5\% (1\%) threshold.
We check that for the cases with $P_{2L | \{1L,2L\} } \gtrsim 1\%$ the significance of the far away cloud does not reach the 2$\sigma$-detection threshold (i.e., $d_{\rm{Maha}}(\mathbf{0}|\hat{\mathbf{c}},\hat{\Sigma}) < 2.45$ in all cases), meaning that the contribution to polarization from the model-retrieved far-way cloud is compatible with zero.

\subsubsection{Two-cloud mock samples}
\label{sec:modselect_2L}
We consider mock samples with two clouds along the LOS.
The first cloud is fixed at 350 pc distance from the Sun and has $p_{\rm{C}1} \simeq 0.2\%$ with an inclination angle of $30^\circ$ and $A_{\rm{turb}} = 0.2$, as for the 1-cloud mock samples used above. The POS position angle of the magnetic field in the nearby cloud ($\psi_{\mathbf{B}_{\rm{reg}}}$) is fixed at $22.5^\circ$.
The second cloud is placed at a distance such that $f_{\rm{bg}2} = 70\%$. This corresponds to a distance of about 685 pc which implies that only about 90 stars are located in the inter-cloud region, thus directly constraining the nearby cloud's polarization properties. We fix the inclination angle to $0^\circ$, vary $\Delta\psi_{1,2}$ from $0^\circ$ to $150^\circ$ in steps of $30^\circ$ as before, and consider two values for the far-away cloud polarization: $p_{\rm{C}2} \simeq 0.15\%$ and $0.2\%$. We consider 10 random realizations of each setup making a total of 120 mock samples; 60 for each $p_{\rm{C}2}$ value.
According to previous tests, we know that a significant fraction of runs with $p_{\rm{C}2} \simeq 0.15\%$ will end with pathological posteriors cloud parallaxes but that this fraction should decrease, but not vanish, for $p_{\rm{C}2} \simeq 0.2\%$ as inferred from Fig.~\ref{fig:Performance2C_C2-vs-pC2_Beam60-At02}.

In the cases with $p_{\rm{C}2} \simeq 0.15\%$, we find that only 24 out of 60 trials lead to nonpathological posterior distributions of the cloud parallaxes.
This implies that due to the low polarization induced by the second cloud (with respect to the level of noise and intrinsic scatter), we could detect the presence of the cloud in only about 40\% of the cases. However, out of the 24, 23 have evidence higher for the one-layer model than for the two-layer model. According to the evidence, the two-layer model is to be favored in only one case out of 60. On the other hand, the AIC values are lower for the two-layer model than for the one-layer model (thus favoring the two-layer model) for 20 cases out of the 24 and the probability that instead the two-layer model should actually minimize the loss of information is higher than 1\%, and as high as 10\% for the four remaining cases. The comparison of the AIC values thus suggests that the two-layer model is to be favored in more than 80\% of the cases that pass the selection criterion on $\varpi_{\rm{C}}$. The hypothesis of a two-layer model is never rejected in those cases.

In the case with $p_{\rm{C}2} \simeq 0.2\%$, we find that 44 out of 60 trials lead to nonpathological $\varpi_{\rm{C}}$ posterior distributions.
Out of 44, 14 exhibit an evidence larger for the one-layer model than for the two layer-model, thus suggesting that the one-layer model is to be favored in about 30\% of the valid reconstructions. However, the AIC values indicate that the one-layer model has to be preferred in only one case out of 44 and yet, for this case, $P_{2L | \{1L,2L\} } > 5\%$ which implies that the possibility for a two-layer model cannot be disregarded.
Therefore, while the evidence favors the two-layer model in only about 70\% of the cases that pass the criterion on $\varpi_{\rm{C}}$, the AIC criterion does it in more than 97\% of the cases and never rejects the hypothesis of a two-layer model.

\subsubsection{Conclusion}
When testing the one-layer and two-layer models on mock samples generated with a 1-cloud model, we find that both the evidence and the AIC criterion favor the one-layer model. Hints for the presence of a spurious second cloud (corresponding to case where $P_{2L | \{1L,2L\} } \gtrsim 1\%$) may be encountered in less than 10\% of the cases but none exhibit a significant detection in the polarization plane.

When testing the one-layer and two-layer models on mock samples generated with 2-clouds in both regimes where pathological reconstructions are dominant and subdominant, we find that comparison of AIC values is more sensitive to the presence of a second cloud than is the comparison of the evidences as it never leads to the rejection of the correct model but instead favors the correct model for a very large fraction of the cases that pass the $\hat{\varpi}_{\rm{C}}$ selection criterion.

As a result, to select between models (number of clouds along the LOS), and after we disregard reconstructions with pathological posterior distributions, we prefer the use of the AIC values rather than the evidences. The comparison of the AIC values never suggests rejecting the correct model. It might, however, suggest spurious detection in some cases which, however, do not lead to an erroneously significant detection of a cloud.

\smallskip

We performed those tests in the low-polarization regime with a somewhat high level of scatter.
We expect our conclusions to hold and the performance to improve when the polarization induced by intervening clouds increases with respect to the level of scatter in the polarization data (for example, a deeper survey than assumed here, or for regions of the sky with higher polarization).

\section{Application to observational data}
\label{sec:app2realdata}
In this section we apply our Bayesian inversion method to existing observations of stellar polarization toward the two LOSs with known number of clouds presented by \cite{Pan2019a}. We first describe the data, then apply our method, and finally discuss our results in comparison with the previously obtained decomposition.

\subsection{Archival polarization data in two LOSs of the diffuse ISM}
\cite{Pan2019a} demonstrated that it is possible to perform a tomographic decomposition of the POS magnetic field by combining a large number of stellar polarization data obtained with the RoboPol polarimeter (\citealt{Ramaprakash2019}) and \textit{Gaia} distances. They selected two neighboring LOSs in the diffuse ISM which have different number of clouds along the LOS, as determined by their \ion{H}{I} spectra. One sightline exhibits two distinct peaks (velocity components) corresponding to two clouds overlapping on the POS (2-cloud region). The other sightline exhibits a single prominent peak (1-cloud region). They obtained stellar polarimetry for $\sim 100$ stars in each of these  0.32$^\circ$--wide circular regions, and were able to recover the polarization properties of each cloud separately.

We choose to work with this data set as it is the only one, to the best of our knowledge, that approaches the expected spatial density of stellar measurements of the \textsc{Pasiphae} survey in the optical, while targeting the diffuse ISM. We caution that despite this similarity, the aforementioned data set is not directly representative of the kind of data expected from \textsc{Pasiphae}, for the following reasons. First, the selected regions in \cite{Pan2019a} are at intermediate latitude, with mean stellar polarization of $\sim 2\%$, thus a factor of several higher than the average polarization expected at high Galactic latitudes. Second, their observing strategy did not mimic a survey with fixed exposure time, so that the uncertainties do not reflect exactly those expected by \textsc{Pasiphae}.
Nevertheless, the aforementioned data set remains the most appropriate one to test our Bayesian method and compare directly to the results of the existing tomographic decomposition presented in \cite{Pan2019a}.

\begin{figure}
    \centering
    \begin{tabular}{c}
        \includegraphics[trim={0.5cm 1.6cm 0.5cm 0.3cm},clip,width=.95\linewidth]{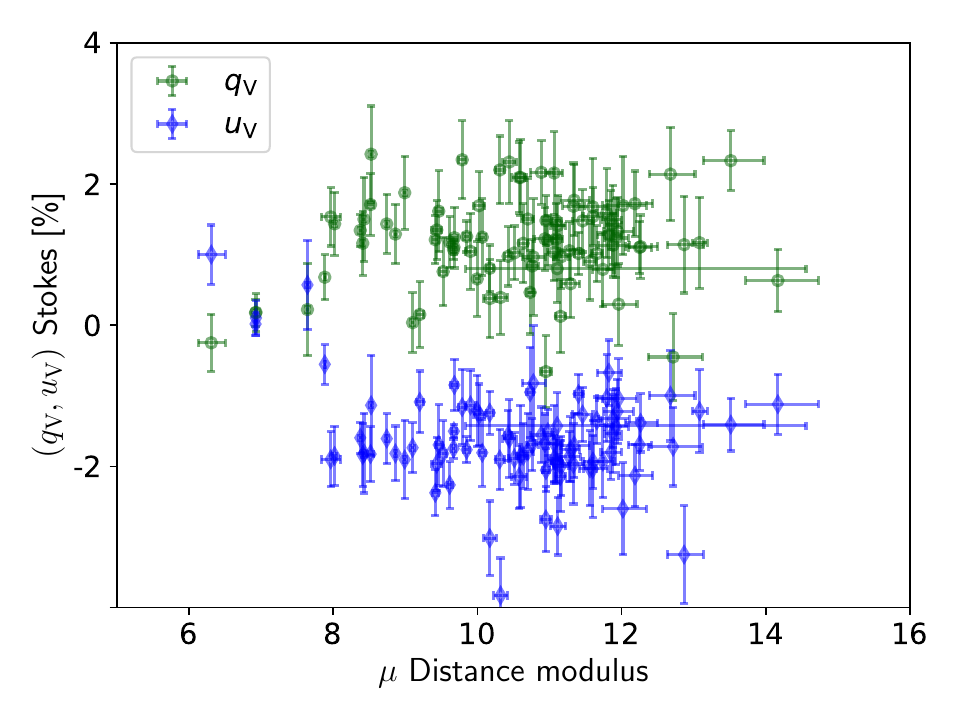} \\[-1.4ex]
        \includegraphics[trim={0.5cm 0.4cm 0.5cm 0.3cm},clip,width=.95\linewidth]{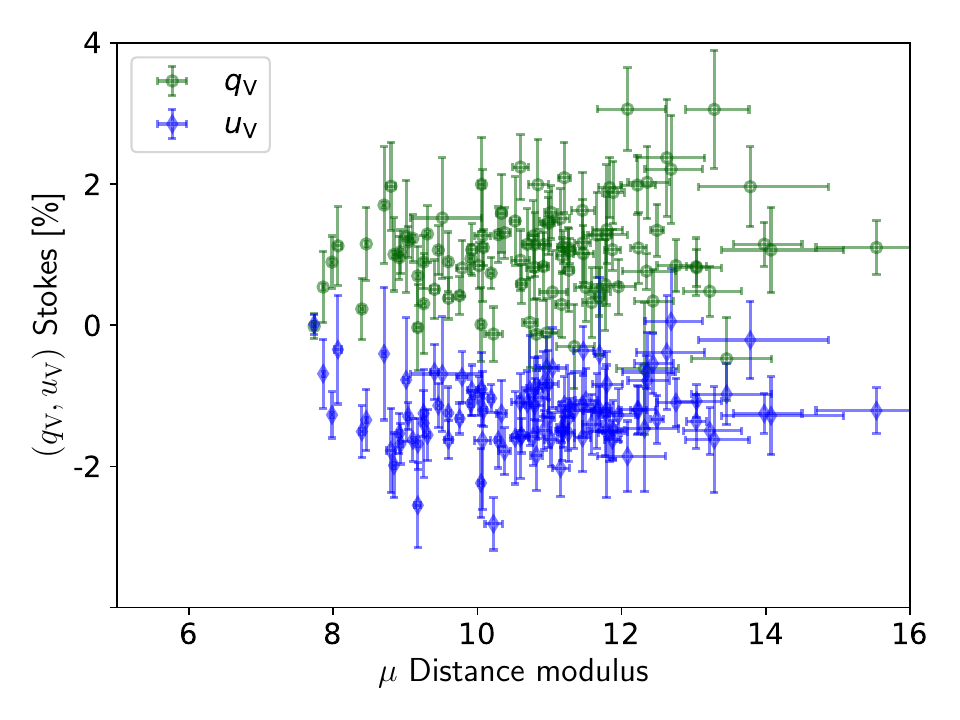}
    \end{tabular}\\[-1.5ex]
    \caption{$(q_{\rm{V}},\,u_{\rm{V}})\,-\,\mu$ plane for the polarization data in (\citealt{Pan2019a}). Top and bottom panels correspond to their 1-cloud and 2-cloud regions, respectively. The relative Stokes parameters $q_{\rm{V}}$'s are shown by dark green circles and the $u_{\rm{V}}$'s are shown by blue diamonds. Vertical and horizontal errorbars show the 68\% confidence level on measurements on polarization and distance modulus, respectively. For visualization purposes we omit to show two stars at distance modulus 1.6 and 18.6 in the 1-cloud region.}
    \label{fig:PanopData}
\end{figure}

In Fig.~\ref{fig:PanopData}, we show their data in the $(q_{\rm{V}},\,u_{\rm{V}})-\mu$ plane, also showing uncertainties in both polarization and parallaxes (propagated to distance moduli). As noted in this previous work, the stellar data show systematic differences in the stellar polarization properties between the two regions, as a result of differences in the ISM across the sky (e.g., dust column density).

\subsection{Results and comparison to previous work}
We apply our inversion method to those two regions separately, trying models with one to three layers.
We use flat priors on our six parameters. In the case of multiple clouds, our implementation choice guarantees that these are at least five stars in between two successive clouds. In the initialization of the flat priors, we further impose that all cloud parallaxes must lie in the parallax ranges defined so that the cloud's minimum distance is 100 pc and a cloud's maximum distance is such that there are at least ten stars in the background. The limits on cloud parallax thus depend on the studied sample.
The limits defining the parameter space of the polarization are rather arbitrary. We set a large range for the mean Stokes parameters of the clouds, from -5\% to 5\%, and consider that the elements of the intrinsic-scatter covariance matrix can go as high as $0.8\times 10^{-4}$ for the diagonal elements (this corresponds to a standard deviation on individual Stokes at the value of almost 1\%). Again, internally, the prior function ensures that $C_{\rm{int},qu}^2 < C_{\rm{int},qq} \, C_{\rm{int},uu}$ for the covariance matrix to be invertible.
The resulting statistics are given in Table~\ref{tab:ResultsPanopData} and we discuss the results for each region next.

\begin{table*}[]
    \centering
    \caption{\label{tab:ResultsPanopData}
    Statistics for model comparison from inference of data in (\citealt{Pan2019a}).}
    \begin{tabular}{c | cccc | cccc}
    \hline \hline \\[-1.5ex]
         & \multicolumn{4}{c}{1-Cloud LOS} & \multicolumn{4}{c}{2-Cloud LOS} \\
        Model &
            $\hat{\mathcal{Z}}$ & $\log\hat{\mathcal{L}}$ & AIC & $P_{j|\{m\}}$ [\%] &
                $\hat{\mathcal{Z}}$ & $\log\hat{\mathcal{L}}$ & AIC & $P_{j|\{m\}}$ [\%] \\ \\[-1.5ex]
    \hline \\[-.5ex]
    1-layer  & 671.1$\pm$0.2    & 688.7 &       -1365.4 & 100 & 757.6$\pm$0.2 & 775.0 & -1538.0 & 100 \\
    2-layer  & 661.1$\pm$0.3    & 690.6 &       -1357.1 & 1.5 & 750.8$\pm$0.2 & 780.1 & -1536.3 & 42.5 \\
    3-layer  & 655.6$\pm$0.3    & 686.9 &       -1337.8 & 1.0 \,10$^{-4}$ & 743.9$\pm$0.6 & 777.4 & -1518.9 & 6.9\,10$^{-3}$ \\ [+1.5ex]
    \hline
    \end{tabular}
\end{table*}

\subsubsection{1-cloud region}
The values of the AICs obtained for the set of tested models inform us that the one-layer model is preferred by the data in the sense that it minimizes the loss of information. The probability $P_{j|\{m\}}$ that the two-layer model is actually the model that should minimize the loss of information takes the value of 1.5\% and that probability for the three-layer model is negligible. That is, there might be an indication for a second cloud in the data of the 1-cloud region but it is marginal.
Assuming there is a second cloud along that LOS, the data is not sufficient to make the significance of the second cloud to show up. However, we have seen in Sect.~\ref{sec:testmodelselection} that such a low value might point to spurious detection. It is therefore safe to conclude that there is most likely only one cloud along this LOS.

The analysis of the posterior distributions of the best model leads to the following results for the cloud. The distance of the cloud is $382.7^{+10.9}_{-23.8}$ pc and the Stokes parameters are  $(q_{\rm{C}},\,u_{\rm{C}}) = ( (1.21\pm0.05)\%, \, (-1.70\pm0.05)\%)$. Here, to summarize the posterior distributions, we report the medians of the distributions and their 16 and 84 percentiles.
By re-sampling the posterior distributions of the Stokes parameters we obtain the distributions of degree of polarization and position angle characterized by: $p_{\rm{C}} = (2.09 \pm 0.05)\%$ and $\psi_{\mathbf{B}} = (-27.2 \pm 0.7)^\circ$.
These values are fully consistent with those obtained by \cite{Pan2019a} for their 1-cloud region ($d_{\rm{C}} \in [346,\, 393]$ pc, $p_{\rm{C}} = (2.04 \pm 0.04)\%$ and $\psi_{\mathbf{B}} = (-27.5 \pm 0.6)^\circ$.

For this fit, the parameter values corresponding to the maximum-likelihood are well centered in the distributions, with $d_{\rm{C}}=380.9$ pc and $(q_{\rm{C}},\, u_{\rm{C}}) = (1.21\%, \, -1.71\%) $. As we have seen in the previous section, situations with significant offsets between parameter values at maximum-likelihood and posterior distributions may happen when the data are not sufficient enough to provide strong constraints on the cloud distance. This is not the case here.

\subsubsection{2-cloud region}
According to our Bayesian inversion method, the one-layer model is the model preferred by the data available for the 2-cloud region. Both the evidence and the AIC criteria agree, as reported in Table~\ref{tab:ResultsPanopData}. However, the probability that the two-layer model actually minimizes the loss of information is very large ($\gtrsim 40\%$) and therefore we cannot reject the hypothesis that a second cloud exists along this LOS. The probability that the three-layer model minimizes the loss of information is negligible. The fact that both the evidence is smaller and the AIC is larger for the two-layer model than for the one-layer model, likely results from the fact that (i) fewer stars sample the second cloud and (ii) the signal of the second cloud is weak as compared to the signal of the dominant cloud and close to the noise level.
\begin{table}
    \begin{center}
    \caption{\label{tab:Panop2CBeam_bestfit} Summary of posteriors on cloud parameters obtained for the 2-cloud region.}
    \begin{tabular}{l|c | c c}
    \hline \hline \\ [-1.5ex]
    \multirow{2}{*}{Parameter}   & 1-layer   & \multicolumn{2}{c}{2-layer} \\
                & cloud 1   & cloud 1 & cloud 2 \\  [+1.ex]
    \hline \\
    $d_{\rm{C}}$ [pc]   & $370.1{\tiny{_{-12.1}^{+ 15.7}}}$ & $371.2{\tiny{_{- 12.7}^{+ 15.5}}}$ & $2329.0{\tiny{_{-426.2}^{+358.1}}}$ \\[+.5ex]
    $q_{\rm{C}}$ [\%]   & $1.00 \pm 0.05$ & $0.96 \pm 0.06$ & $0.21 \pm 0.18$ \\[+.5ex]
    $u_{\rm{C}}$ [\%]   & $-1.28 \pm 0.04$& $-1.33 \pm 0.05$ & $0.19 \pm 0.11$ \\[+.5ex]
    $p_{\rm{C}}$ [\%]   & $1.62 \pm 0.04$ & $1.64 \pm 0.05$ & $0.32{\tiny{_{-0.12}^{+0.15}}}$ \\[+.5ex]
    $\psi_{\rm{C}}$
        [$^\circ$]      & $-26.0 \pm 0.9$ & $-27.1 \pm 1$ & $20.6{\tiny{_{-12.6}^{+18.3}}}$ \\[+1.5ex]
    \hline
    \end{tabular}
    \tablefoot{Values indicates the median of the posterior distributions and the uncertainties are computed from the posteriors and indicate the offset from the median to the 16 and 84 percentiles.
    $p_{\rm{C}}$ and $\psi_{\rm{C}}$ are built from posteriors on $q_{\rm{C}}$ and $u_{\rm{C}}$. We show results obtained with the 1-layer and 2-layer models only.}
    \end{center}
\end{table}

Summary statistics on the cloud model parameters are given in Table~\ref{tab:Panop2CBeam_bestfit} for the case of the one-layer and two-layer model.
For the fits obtained with the one- and two-layer models, the parameter values corresponding to the maximum-likelihood are well centered in their respective posterior distributions. We have
$d_{\rm{C1}}=369.2$ pc and $(q_{\rm{C1}},\, u_{\rm{C1}}) = (1.00\%, \, -1.28\%) $ for the one-layer model and
$d_{\rm{C1}}=360.2$ pc and $(q_{\rm{C1}},\, u_{\rm{C1}}) = (0.93\%, \, -1.33\%) $
and
$d_{\rm{C2}}=2380.0$ pc and $(q_{\rm{C2}},\, u_{\rm{C2}}) = (0.30\%, \, 0.14\%) $
for the two-layer model.
The fits and the results are robust and the data are sufficient to generate relatively stable and significant extrema in the log-likelihood hyper-surface.
Therefore, we conclude that it is somewhat likely that there are 
two clouds in the 2-cloud region.
The fact that the best-fit parameters of the nearby cloud from the two-layer model and those from the one-layer model are very similar demonstrates that the polarization signal is dominated by the nearby cloud. This explains why there is no absolute strong evidence for a second cloud and that the one-layer model is ranked first according to the AIC and evidence criteria.

Our method returns results that are consistent with those obtained by \cite{Pan2019a}.
To study the 2-cloud region, they fixed the nearby cloud distance at 360 pc, found that the far-away cloud distance that maximizes the detection is $d_{\rm{C}2} \approx 1700$ pc to which they assigned an uncertainty of $\pm 440$ pc. Using $d_{\rm{C}2} \approx 1700$ pc, they found that the nearby and faraway cloud polarization properties were $(p_{\rm{C}1},\, \psi_{\mathbf{B}_{\rm{C}1}}) = ((1.65 \pm 0.04)\%,\, (-27.3 \pm 0.8)^\circ)$ and $(p_{\rm{C}2},\, \psi_{\mathbf{B}_{\rm{C}2}}) = ((0.28 \pm 0.07)\%,\, (36 \pm 8)^\circ)$, respectively.

\medskip

\subsubsection{Results comparison}
\begin{figure}
    \centering
    \includegraphics[trim={-.4cm 0.4cm 0.4cm 0.2cm},clip,width=.86\linewidth]{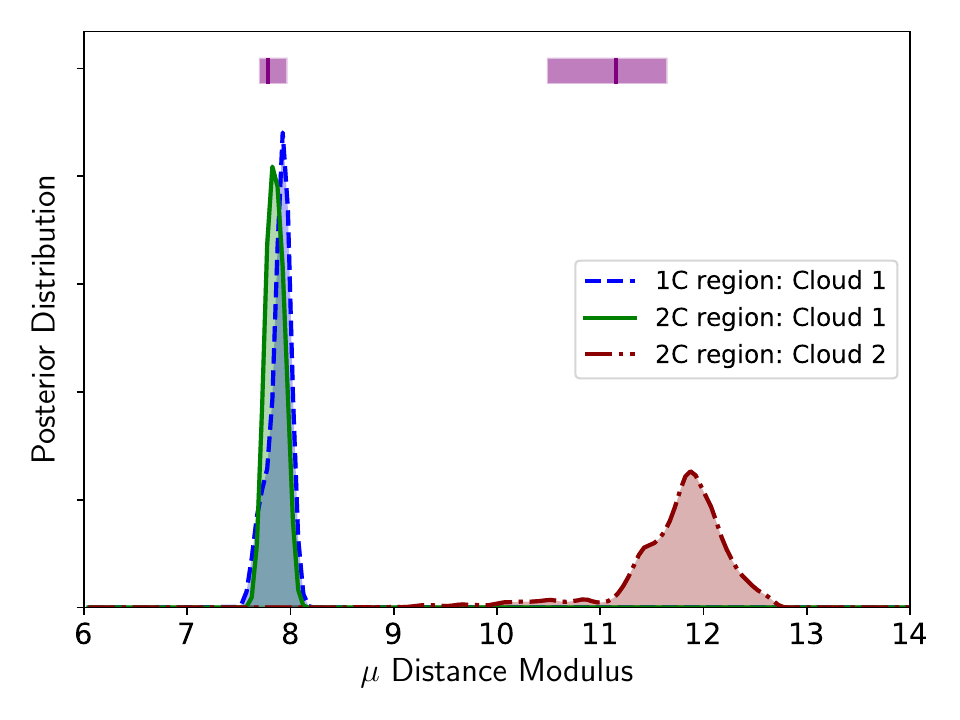}\\
    \includegraphics[trim={1.6cm 0.4cm 2.5cm 0.2cm},clip,width=.95\linewidth]{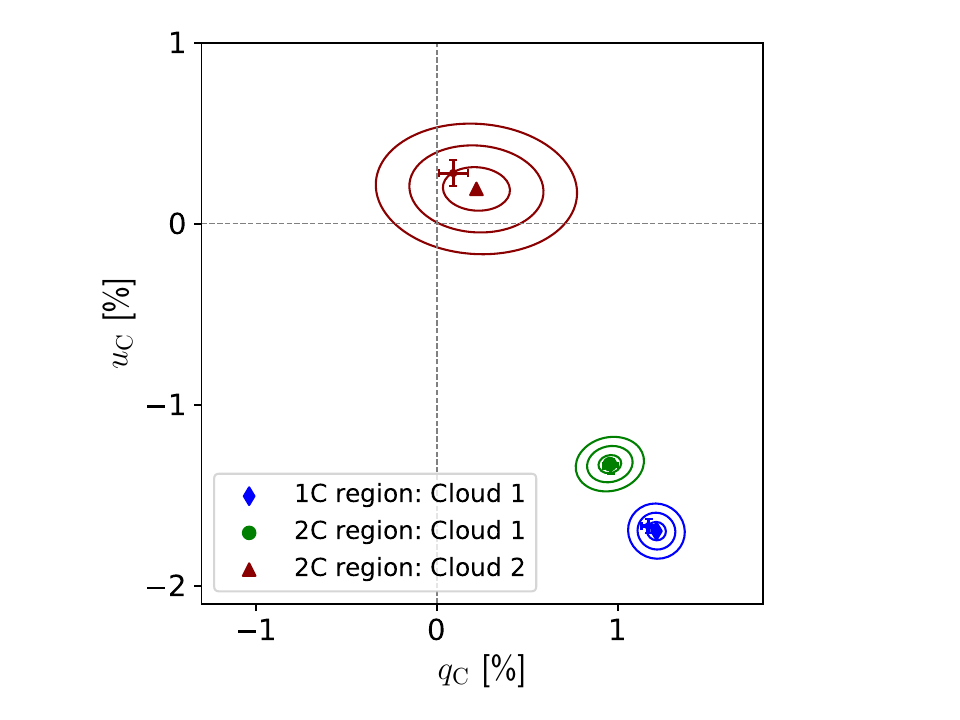}\\[-1.5ex]
    \caption{Posterior distributions for the cloud distance modulus (top) and cloud mean polarization (bottom) obtained for the 1-cloud region while fitted with the one-layer model (blue) and for the 2-cloud region while fitted with the two-layer model with posteriors in green and dark red for the nearby and faraway cloud, respectively. The contours indicate the 1, 2, and 3$\sigma$ confidence levels.
    The values obtained by \cite{Pan2019a} are reported, using purple horizontal bands for the distance modulus on the top panel and errorbars to report polarization in the bottom panel using the same colors as for the confidence contours we obtain. Values for the 2-cloud region correspond to their distance cut that maximizes the detection of the faraway cloud.
    }
    \label{fig:Panop_1C2C_posteriors}
\end{figure}

In Fig.~\ref{fig:Panop_1C2C_posteriors} we present and compare the posterior distributions for the cloud distance modulus and the cloud mean polarization that we obtain with the one-layer model applied to the 1-cloud region and the two-layer model applied to the 2-cloud region.
For comparison we also report the values obtained by \cite{Pan2019a}. It is seen that their results are compatible with our posterior distributions. The confidence intervals that we derive from our marginalized posterior distributions are larger than their uncertainties. This is true in particular for the polarization parameters of the faraway cloud. The reason is that our estimates include any possible correlation between parameters of the model while their estimates are conditional to the choice of both cloud distances.

We thus conclude that even applied to actual data, the method that we have developed makes it possible to perform a tomographic decomposition of the magnetized ISM in dusty regions.
We emphasize that at no time did we rely on the a priori assumption that there are one or two clouds along the given LOSs. This shows that our inversion method is a standalone method in the sense that it can be used blindly on stellar polarization and distance data, and independently of any other observable, and lead to a reliable decomposition of the ISM signal along the LOS.

\section{Conclusions and discussion}
\label{sec:conclusion}
Starlight polarization is a direct tracer of the orientation of the POS component of the magnetic field in the dusty ISM. When combined with measurements of stellar distances, starlight polarization has the potential to allow for a 3D reconstruction of the magnetized ISM in dusty regions. In this paper, and motivated by the forthcoming leap forward in available stellar polarization data from the \textsc{Pasiphae} survey, we have aimed to develop a robust method to perform such a 3D reconstruction.

We have developed a Bayesian method to reconstruct the POS magnetized ISM structure along the LOS through maximum-likelihood analysis of the stellar data alone. Our method, which relies on a generic model in which dust clouds have a thickness along the LOS smaller than the typical separation between stars, accounts for uncertainties in  stellar parallaxes and in the Stokes parameters. It further accounts, in a model-independent way, for the intrinsic scatter that is expected from turbulence within individual clouds. We obtained a likelihood that accounts for all sources of noise and scatter and implemented it in Python. Our code depends on the nested sampling code \texttt{dynesty} to maximize the log-likelihood function,  construct the posterior distributions of the model parameters, and estimate the evidence. The code, named \texttt{BISP-1}, for Bayesian Inference of Starlight Polarization in one dimension (along distance), is made public.

We tested our Bayesian inversion method on mock starlight polarization data obtained from a self-consistent toy model. We have demonstrated that our method is effective at recovering the cloud properties as soon as the polarization signal induced by a cloud to its background stars is higher than $\sim 0.1\%$. When the minimum (systematic) uncertainty on observed stellar polarization is assumed to be at the level of 0.1\% (in the degree of polarization) and the induced polarization is at a similar level, we found that $\approx 30$ stars in the background of a cloud are required to place useful constraints on the cloud properties. The larger the induced polarization signal is, the better the method's performance, and the lower the number of required stars. In addition, to accurately recover the distances and the mean polarization properties of clouds, we found that our method also makes it possible to constrain the parameters characterizing the turbulence-induced intrinsic scatter. This might open new avenues in the characterization of the ISM turbulence and estimation of the magnetic field strength. We will explore those avenues in future works.

We have further demonstrated that our Bayesian inversion method  efficiently recovers cloud properties (distance and polarization) when applied to the actual data sets that were first used to demonstrate that starlight polarization, coupled to distance measurements, can be used to decompose the magnetized ISM signal as a function of distance (\citealt{Pan2019a}).
We obtained results that are fully consistent with those from the original study but within a robust Bayesian framework which allowed us to build proper posterior distributions on our model parameters of our reconstruction and, therefore, to put our results on a more solid footing.
With this application we have shown that our method can work independently and blindly on star data to reconstruct the structure of the magnetized ISM. For example, we need not rely on external data to inform the method on the number of components along the LOS. This demonstrates the strength of our method, as well as the great potential of starlight polarization as a direct and fully independent probe to the 3D structure of the magnetized ISM.

Modifications of the sampling method implemented in this work are possible and could lead to speeding up the calculations. An example is a profiled-likelihood method that would estimate on the fly, and from the data, a certain number of the six free parameters per cloud. For example, the parameters related to the intrinsic scatter could be estimated from the scatter of the data points and their observational uncertainties once the cloud distances are fixed. The main advantage of such an implementation is the reduction of computing time owing to the reduction of the parameter-space dimensions. The main disadvantages are that the log-likelihood hyper-surface is no longer homogeneously sampled, preventing reliable estimates of the statistical evidence of a given model (number of clouds along the LOS), and that no proper posterior distributions of the parameters estimated on the fly can be safely reconstructed. We postpone the exploration of such alternative implementations 
to future work.

In its current implementation, our Bayesian inversion method relies on the dust-layer model that we have introduced.
We expect for our model assumptions to hold for small beam sizes, so that any variations in the magnetic field and dust density do not vary appreciably.
However, since a minimum number of stars is required to constrain the dust-cloud characteristics effectively, a minimum beam size is determined by both the true distribution of stars in space and the distance and multiplicity of clouds in the beam.
While the analysis presented here focused on two beam sizes, we note that the minimum beam size allowed likely varies with sky position.
One likely must evaluate, on a case-by-case basis, the trade-off between increasing the number of stars (widening the beam) and minimizing the variations in the ISM properties on the sky.

Furthermore, although the conceptual dust-layer model is supported by current observations of the high latitude sky, we must keep in mind that it might fail to account for all dust along a sightline and that careful analyses of the results and residuals will be mandatory, as it is generally the case in Bayesian modeling (\citealt{Romero-Shaw2022}). If, in the future, our simple layer-model prescription fails to fit the data, the underlying model will have to be changed, perhaps to be replaced by a smoother function of the LOS distance. In such a case, the generalization of the formalism that we have laid out will be straightforward and somewhat simpler than for the step model that we have assumed in this paper.

Finally, we must note that we have implicitly assumed in this paper that the polarization of stars is due to the magnetized ISM, only. However, stars of certain types may show intrinsic polarization,
likely related to the existence of a circum-stellar disk where planets form or other asymmetries in the object (e.g., \citealt{Cotton2016}; \citealt{Gontcharov2019}). These stars usually show a higher degree of polarization than neighboring stars with unrelated position angles.
In its current implementation, our method does not account for these intrinsically polarized stars. Therefore, to apply our decomposition method, the
input stellar sample must have first been cleaned from potentially intrinsically polarized stars.
Various techniques can be used for this purpose such as sigma clipping.
Alternatively, in a Bayesian framework, we could consider constructing a likelihood of a mixture model in which the polarization of a star would have a probability of being due to the ISM or of being intrinsic given some of the star's properties.
Such an approach to deal with outliers has been implemented in \cite{Zucker2019} to estimate the distance of molecular clouds based on the measured stellar extinction.
However, we must defer such efforts to future work, given that basic statistical information on these sources (e.g. number density of intrinsically polarized stars per sky location) is currently lacking, and will only be known with the data from future unbiased surveys.

\bigskip

\begin{acknowledgements}
We are grateful to our anonymous referee who provided us with a detailed and constructive report that helped us improve this paper.
The PASIPHAE program is supported by grants from the European Research Council (ERC) under grant agreements No. 771282 and No. 772253; by the National Science Foundation (NSF) award AST-2109127; by the National Research Foundation of South Africa under the National Equipment Programme; by the Stavros Niarchos Foundation under grant PASIPHAE; and by the Infosys Foundation.
GVP acknowledges support by NASA through the NASA Hubble Fellowship grant \#HST-HF2-51444.001-A awarded by the Space Telescope Science Institute, which is operated by the Association of Universities for Research in Astronomy, Incorporated, under NASA contract NAS5-26555.
VPa acknowledges support by the Hellenic Foundation for Research and Innovation (H.F.R.I.) under the “First Call for H.F.R.I. Research Projects to support Faculty members and Researchers and the procurement of high-cost research equipment grant” (Project 1552 CIRCE). VPa and AT acknowledge support from the Foundation of Research and Technology - Hellas Synergy Grants Program through project MagMASim, jointly implemented by the Institute of Astrophysics and the Institute of Applied and Computational Mathematics.
EN has received funding from the HFRI's 2nd Call for H.F.R.I. Research Projects to Support Post-Doctoral Researchers (Project number 224).
KT and AP acknowledge support from the Foundation of Research and Technology - Hellas Synergy Grants Program through project POLAR, jointly implemented by the Institute of Astrophysics and the Institute of Computer Science.
TG is grateful to the Inter-University Centre for Astronomy and Astrophysics (IUCAA), Pune, India for providing the Associateship programme under which a part of this work was carried out.
GVP would like to thank C. Zucker for helpful discussions in the early stages of this project.
\end{acknowledgements}

\bibliographystyle{aa}
\bibliography{myBiblio}

\begin{appendix}

\section{Mock starlight polarization data}
\label{sec:mock_data_appendix}

To test and validate the inversion method presented in this paper we developed a pipeline to generate mock samples of starlight polarization data. Our simulation pipeline has the main following characteristics: ($i$) it makes use of star samples extracted from the \textit{Gaia} data to guarantee realistic distributions of star distances, number density, brightness, etc., ($ii$) it relies on a self-consistent implementation of the thin-layer dust-cloud model discussed in Sect.~\ref{sec:model} and presented below which includes a prescription for the intrinsic scatter, and finally ($iii$) it computes and propagates to the simulated stellar polarization data realistic uncertainties as expected for the \textsc{Pasiphae} survey.
These three parts are described in the rest of this Appendix. We further explore our toy-model to simulate the polarization signal from the magnetized ISM in Appendix~\ref{sec:IntrSc}.

Our simulation pipeline works as follows. First, we choose a sample of stars in a small-aperture beam and extract parallaxes, parallax uncertainties and apparent magnitudes. Second, we simulate the stellar polarization measurements for any chosen setup of the magnetized and dusty ISM and estimate the polarization uncertainties. We `observe' the stars by randomly drawing the Stokes parameters and parallaxes within their respective uncertainties.

\subsection{Star samples from the \textit{Gaia} Universe Model Snapshot}
\label{sec:gums}
We seek to create realistic samples of stellar photometry and distance, based on the expected sky footprint of the \textsc{Pasiphae} survey. In this section we discuss how we generate and choose representative samples of stars from the GUMS database.

\begin{figure*}
    \centering
    \includegraphics[trim={0cm .1cm 0cm 0cm},clip,scale = 1]{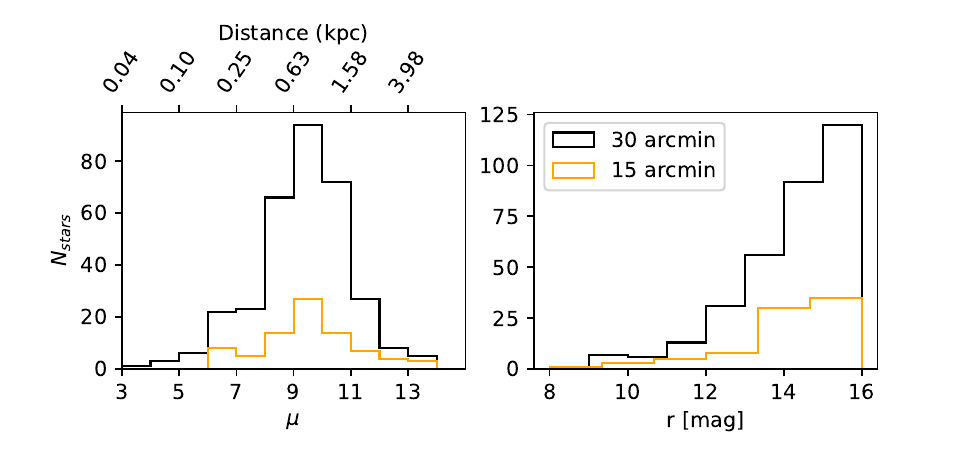}\\[-2.ex]
    \caption{Distribution of stellar distances (left) and apparent magnitudes (right) after the selection criteria have been applied. The distributions for the large ($0.5^\circ$ radius) and small ($0.25^\circ$ radius) beams are shown with black and orange lines, respectively.
    }
    \label{fig:R_dist_cuts}
\end{figure*}

\begin{figure*}
    \centering
    \includegraphics[scale = 1]{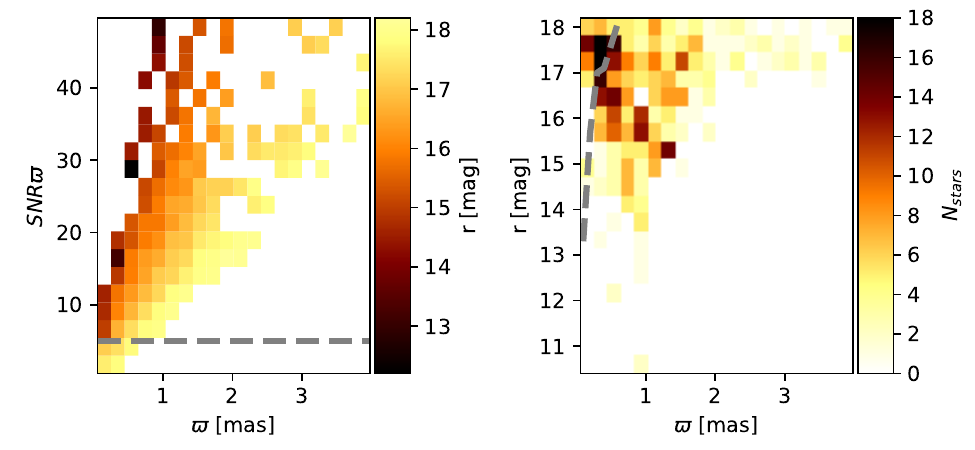}\\[-2.ex]
    \caption{Stellar properties and selection criteria. Left: Mean SDSS $r$-band magnitude in bins defined on the parallax S/N vs. parallax plane. Bins with no stars are shown in white.  Right: 2D distribution of the number of stars in the $r$ vs. $\varpi$ plane. The cut at ${\rm{SNR}}_{\varpi} = 5$ is shown with a dashed gray line in both panels. We note that the introduction of the parallax S/N cut imposes a biased selection in the $r - \varpi$ plane. The data shown are for the large (30-arcmin radius) beam. We do not show the full dynamic range of $\varpi$ and ${\rm{SNR}}_{\varpi}$ for better visualization of the key features of the dataset.}
    \label{fig:gums_before_cuts}
\end{figure*}

\subsubsection{Mock stellar catalog generation}

To obtain realistic samples of stellar parameters (photometry, distances) for the high Galactic latitude sky, we use the GUMS database (\citealt{Robin2012}) associated with Gaia Early Data Release 3  (EDR3, \citealt{Lindegren2021}), which provides some photometric information as well as parallaxes. We begin by selecting sightlines toward each of the HEALPix $N_{\rm{side}} = 4$ pixels at high Galactic latitude (with $|b| > 60^\circ$). We then query the \textit{Gaia} archive within a circular region centered on each of the aforementioned pixels, with two different radii of 0.5$^\circ$ and 0.25$^\circ$. For ease of computation, we exclude very faint stars with ${\rm{G}} > 18.5$ within the query. This brightness cut is later superseded by a stricter cut based on our derived $r$ band photometry, as explained next. We thus obtain 24 samples of stars for each selected beam size. 

Next, we wish to compute realistic uncertainties for the parallax and polarization of each star. The GUMS query returns the $V-I$ color, the $V$-band absolute magnitude $M_V$, the extinction $A_V$, the apparent magnitude in the $G$-band, the \textit{Gaia} broadband color $G_{BP}$ and the barycentric distance, $d$, in parsec. We determine the parallax errors as a function of $G$-band magnitude based on EDR3 performance (\citealt{Lindegren2021}). We use the quoted median parallax error per bin of $G$ of their table 4 (5-parameter solution sources) and create an interpolating function. Each star is thus assigned a true parallax by inverting the distance and converting to units of milli-arcseconds (mas) and an uncertainty based on its $G$ band magnitude.
The estimation of uncertainties in the polarization measurements from \textsc{Pasiphae} relies on (a) SDSS $r$-band photometry and (b) the expected performance of the survey.
We derive $r$-band magnitudes in the following section and briefly discuss the dependence of the uncertainty in the stellar Stokes parameters on $r$-band magnitude in Appendix~\ref{sec:mock_realnoise}.

The stellar properties of the samples cover a broad range of apparent magnitudes, parallaxes and S/N in parallax (Fig.~\ref{fig:gums_before_cuts}, left, shows a representative sample of size 1$^\circ$). 
We tailor the stellar samples to the expected limiting brightness of the \textsc{Pasiphae} survey, by applying a cut of $ r < 16 \,\rm mag$.
The total number of stars per sample (post-cut) varies in the range [225, 414] and [64, 118] for the 1$^\circ$ and 0.5$^\circ$ beams, respectively. For ease of computation, we choose one representative sample for each beam size on which to apply our analysis. The representative sample for each beam is that which has a total number of stars (post-brightness-cut) close to the average number of stars of the ensemble of sightlines. We further limit the samples to stars with a S/N in parallax higher than 5.

As shown in Fig.~\ref{fig:R_dist_cuts} (left), the distribution of stellar parallaxes (distances) peaks at $\sim 1 \,{\rm{mas}}$ (1000 pc), but this is brightness-dependent (see Fig.~\ref{fig:gums_before_cuts}, right). As one would expect, our brightness cut limits the number of stars at small parallaxes (large distances). We also note there is a less intuitive selection bias that results from our choice to use stars with $\varpi/\sigma_\varpi > 5$. Fig.~\ref{fig:gums_before_cuts} shows the line of $\varpi/\sigma_\varpi = 5$ in two different planes (SNR$_\varpi$ vs. $\varpi$ and $r$ vs. $\varpi$). As seen in the right panel, imposing the S/N cut results in a parallax-dependent brightness selection: stars to the left and top of the dashed line are excluded from our analysis. Our samples miss some faint and nearby stars. This is the result of the dependence of the parallax error on stellar brightness. In other words, the limiting magnitude of star samples analyzed in this work is parallax-dependent.

\subsection{Derivation of stellar $r$-band photometry given GUMS outputs}
\label{sec:rband}

We must determine the apparent magnitude in the SDSS-$r$ band. This requires some manipulation of different color transformations of the photometric data provided by GUMS. We begin by computing the apparent magnitude in the $V$ band:
\begin{equation}
m_V = M_V + 5 \, \log(d) - 5 + A_V    
.\end{equation}
We can connect the SDSS photometry with \textit{Gaia} $G$ band photometry using the relations from \cite{Jordi2010}, as initially referenced by the GUMS paper (\citealt{Robin2012}):
\begin{align}
    G - G_{BP} =& -0.1703 -1.0813 \, (r-i) \nonumber \\
        & \phantom{o} - 0.1424\,(r-i)^2 + 0.0271\,(r-i)^3   ,
\end{align}
where the uncertainty of the coefficients is $\sigma = 0.1$. We solve this equation for the color $r-i$. 
We can then use the color transformations by \cite{Jordi2006}:
\begin{align}
    &(i-I) = (0.247 \pm 0.003) \,(R - I) + (0.329 \pm 0.002) \\
    &(r-i) = (1.007 \pm 0.005) \,(R - I) - (0.236 \pm 0.003) \;.
\end{align}
Solving the second expression for $(R-I)$ and substituting for $(R-I)$ in the first expression, we obtain:
\begin{equation}
(i-I) = 0.0578 + 0.245 \,(r-i) + 0.329 \; . 
\end{equation}
Using the color $(r-I)$, we substitute $(i-I)$ from the previous expression and we solve for $m_r$:
\begin{align}
    &m_r - m_I = (r-I) = (r-i) + (i-I) \\
    &m_r = m_I + (r-i) + 0.0578 + 0.245\,(r-i) + 0.329.
\end{align}
We finally obtain an expression for SDSS $r$-band apparent magnitude:
\begin{equation}
m_r = m_I + 1.245\,(r-i) + 0.3868,
\end{equation}
where $m_I = m_V - (V-I)$ can be computed from the $V-I$ color provided in GUMS.

\subsection{ISM polarization signal: A toy model}
\label{sec:toymodel}
Both the 3D orientation of the magnetic field and the dust density distribution are expected to vary within a cloud, at least to some extent, due to compressible MHD turbulence. These two effects induce a spread in the polarization signal of stars in the background of a cloud (the intrinsic scatter).
This is implemented by considering the signal part to be made of two components: a regular component and a stochastic component.
To build intuition on the effect of intrinsic scatter on the Stokes parameters, we simplify the problem and only implement the scatter that arises from fluctuations in the 3D geometry of the magnetic field. We defer a treatment of fluctuations in the dust distribution to future work.
Hence, for a given cloud, a single value of the maximum degree of polarization ($P_{\rm{max}}$) is used to model the polarization signal of all background stars (Eq.~\ref{eq:stokes_reg}).
On the contrary, both the inclination angle and the position angle of the magnetic field are considered to vary from star to star because of fluctuations in magnetic field geometry.

We model the total 3D magnetic field as the sum of a regular component ($\mathbf{B}_{\rm{reg}}$) and a stochastic one ($\mathbf{B}_{\rm{sto}}$).
We consider  $\mathbf{B}_{\rm{reg}}$ to be uniform within a cloud and to have a norm of unity. We model $\mathbf{B}_{\rm{sto}}$ through 3D Gaussian realizations of white noise. That is, the stochastic component is built from realizations of 3D isotropic random vectors obtained from sampling an independent normal distribution for each of its three components. Assuming a large sample of random realizations for $\mathbf{B}_{\rm{sto}}$ we estimate the sample $rms$ and use it to normalized all the random draws. The distributions of the norms of $\mathbf{B}_{\rm{sto}}$ have a standard deviation of one. To reach a statistically stable normalization of $\mathbf{B}_{\rm{sto}}$, the $rms$ evaluation should be performed for a sufficiently large sample of realizations. We use at least 1000 realizations.
Finally, we model the 3D total magnetic field as the sum of a regular ($\mathbf{B}_{\rm{reg}}$) and a stochastic component:
\begin{equation}
    \mathbf{B}_{\rm{tot}} \propto \mathbf{B}_{\rm{reg}} + A_{\rm{turb}} \, \mathbf{B}_{\rm{sto}} \;,
    \label{eq:Btot}
\end{equation}
where $A_{\rm{turb}}$ quantifies the amplitude of the stochastic component with respect to the regular one.
A different realization of $\mathbf{B}_{\rm{sto}}$ is attributed to each star in the sample. As a result, to each star corresponds a different inclination and position angle. Therefore, by virtue of Eq.~(\ref{eq:stokes_reg}), different values of the Stokes parameters are obtained only from fluctuations in the 3D geometry of the magnetic field.
We emphasize that the parameter $A_{\rm{turb}}$ used above should not be confused with the turbulent-to-mean magnetic field ratio used in methods to estimate the strength of the magnetic field (e.g., \citealt{Skalidis2021}). $A_{\rm{turb}}$ is a metric of the statistical fluctuations of the magnetic field geometry in 3D and is model dependent. We find in Appendix~\ref{sec:IntrSc} that values in the range 0.1 to 0.3 may be representative of clouds at intermediate and high Galactic latitudes.

To summarize, our toy model has five free parameters per cloud: the cloud parallax ($\varpi_{\rm{C}} = 1/d_{\rm{C}})$, the maximum degree of polarization ($P_{\rm{max}}$), the inclination ($\gamma_{\mathbf{B}_{\rm{reg}}}$) and position ($\psi_{\mathbf{B}_{\rm{reg}}}$) angles of $\mathbf{B}_{\rm{reg}}$ and, finally, the relative amplitude of fluctuations in magnetic field orientation ($A_{\rm{turb}}$).
We notice that, 
apart from the cloud parallax, these parameters are not the same as the model parameters entering our data equation (Eq.~\ref{eq:data_equation}) and, therefore, are not the parameters being sampled in the maximization process.

Due to projection effects, and since the inclination angle is positively defined, the mean inclination angle of the total magnetic field may generally be larger than the inclination angle of the regular magnetic field given as input. In general this results in a depolarization of the cloud signal which may be stronger in one of the two polarization channels (Stokes $q_{\rm{V}}$ or $u_{\rm{V}}$) depending on the position angle $\psi_{\mathbf{B}}$. Therefore, in addition to producing a scatter in the polarization plane, the stochastic component in the 3D magnetic field geometry induces a bias in the mean Stokes parameters.
We further explore our toy model in Appendix~\ref{sec:IntrSc} with a particular emphasis on the bias and the covariance that magnetic-field fluctuations produce. Here, we note that the bias is not physical and simply originates from the construction of our model. What matters, and what should be modeled from real observations, are the mean values of the Stokes parameters of a cloud and the dispersion around these means.

\subsection{Realistic noise}
\label{sec:mock_realnoise}
\begin{figure}
    \centering
    \includegraphics[trim={0.6cm 1.2cm 0cm 0.2cm},clip,width=.95\linewidth]{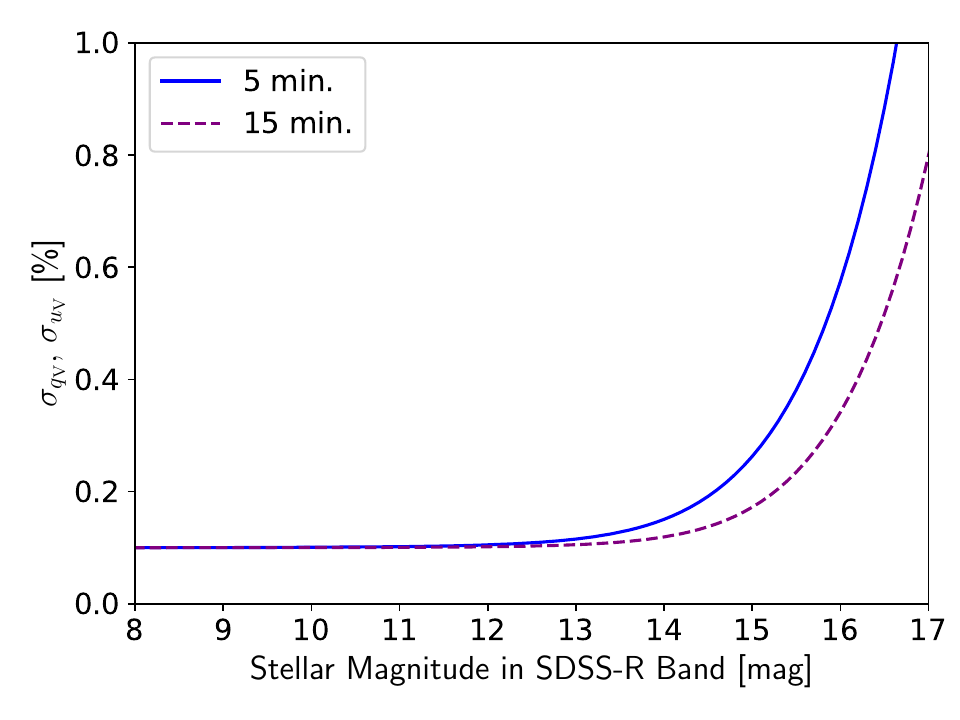}\\[-.2ex]
    {\hspace{.6cm} \small Stellar Magnitude in SDSS-$r$ Band [mag] \hspace{2.6cm}}\\[-1.ex]
    \caption{Uncertainties on individual measurements of the relative Stokes parameters in SDSS-r band as a function of star's magnitude as expected from \textsc{Pasiphae}'s northern instruments (WALOP-N) for 5 minutes and 15 minutes exposure times shown with solid blue and dashed purple lines, respectively.}
    \label{fig:Sigma_qu-vs-Rmag}
\end{figure}

Once the polarization signal from the input ISM setup is attributed to each star in the sample, we add noise in both polarization and parallax.
We take the values of parallax uncertainties from our samples of stars derived from the \textit{Gaia} data (Appendix~\ref{sec:gums}) and the star parallaxes are simply randomized within their uncertainty range following Gaussian distributions.
To give realistic uncertainties on the Stokes parameters of starlight polarization, we rely on current expectations of the performance of the WALOP-N instrument to be used for the \textsc{Pasiphae} survey in the northern hemisphere.
The observational uncertainties, which can be estimated from the optical modeling of the WALOP-N instrument (Maharana et al. {\textit{in prep.}}), depend primarily on the magnitude of the stars in the observation band (SDSS-$r$ band) and on the exposure time. To take into account  instrumental systematics, we add to our observational uncertainty budget a contribution of 0.1\% to the two relative Stokes values $q_{\rm{V}}$ and $u_{\rm{V}}$ (see \citealt{Maharana2020}; \citealt{Maharana2021}; \citealt{Maharana2022}; \citealt{Anche2022}).
Therefore, for each star, an estimate of the observational uncertainty is obtained and used to randomize the Stokes parameters obtained from the ISM configuration. The randomization of $q_{\rm{V}}$'s and $u_{\rm{V}}$'s is performed independently.
The uncertainty in the Stokes parameters as a function of star magnitude is shown in Fig.~\ref{fig:Sigma_qu-vs-Rmag} for two typical exposure times.
Examples of mock starlight polarization data obtained for a five minutes exposure time, for a beam of $0.5^\circ$ circular diameter, and a median number density are shown in Fig.~\ref{fig:1Layer_expl} for the cases of one and two clouds along the LOS.

\section{Polarization observables and intrinsic scatter}
\label{sec:IntrSc}
For the purpose of this work we have developed and implemented a toy model, described in Sect.~\ref{sec:toymodel}, that allows us to generate mock observations based on the multilayer magnetized ISM paradigm described in Sect.~\ref{sec:model}. In particular, our implementation makes it possible to account for a source of intrinsic scatter in the simulated data. We expect such scatter to exist because of small-scale variations in the ISM, that generically result from turbulence. Generally fluctuations are expected in both density distribution and magnetic field geometry but dust grain properties, dust temperature, etc. could also vary on small scales. In our modeling, and as described in Sect.~\ref{sec:toymodel}, we only consider fluctuations in 3D geometry of the magnetic field and thus ignore fluctuations in dust density distribution or other properties that may affect the value of the maximum degree of polarization $P_{\rm{max}}$ (see Eq.~\ref{eq:stokes_reg}).
In this appendix we explore our toy-model implementation to gain intuition on the effects of the intrinsic scatter on the Stokes parameters for different levels of intrinsic scatter (controlled through the value of $A_{\rm{turb}}$ -- see Eq.~\ref{eq:Btot}) and for several combinations of the inclination angle ($\gamma_{\mathbf{B}_{\rm{reg}}}$) and position angle ($\psi_{\mathbf{B}_{\rm{reg}}}$) of the regular component of the magnetic field.
In the adopted polarization convention (IAU), $q_{\rm{V}}$ is maximum for $\psi_{\mathbf{B}_{\rm{reg}}} = 0^\circ$ and zero at $\psi_{\mathbf{B}_{\rm{reg}}} = 45^\circ$ where $u_{\rm{V}}$ is maximum.

We begin by considering the relative Stokes parameters normalized by the value of $P_{\rm{max}}$. Therefore, in the absence of a stochastic component, the degree of polarization varies from 0 to 1 for the case of magnetic field pointing toward the observer ($\gamma_{\mathbf{B}_{\rm{reg}}} = 90^\circ$) to the case where the magnetic field lies in the POS ($\gamma_{\mathbf{B}_{\rm{reg}}} = 0^\circ$).
From geometrical considerations, we expect that the effect of the addition of a stochastic component on the values of the Stokes parameters $q_{\rm{V}}$, $u_{\rm{V}}$ will depend on the chosen value of $A_{\rm{turb}}$, on $\gamma_{\mathbf{B}_{\rm{reg}}}$, and possibly on $\psi_{\mathbf{B}_{\rm{reg}}}$.
To infer such dependence, we want to consider different orientations of the regular magnetic field as seen from the observer. 

\begin{figure}
    \centering
    \includegraphics[trim={0.4cm 0.4cm 0.4cm 1.cm},clip,width=.95\linewidth]{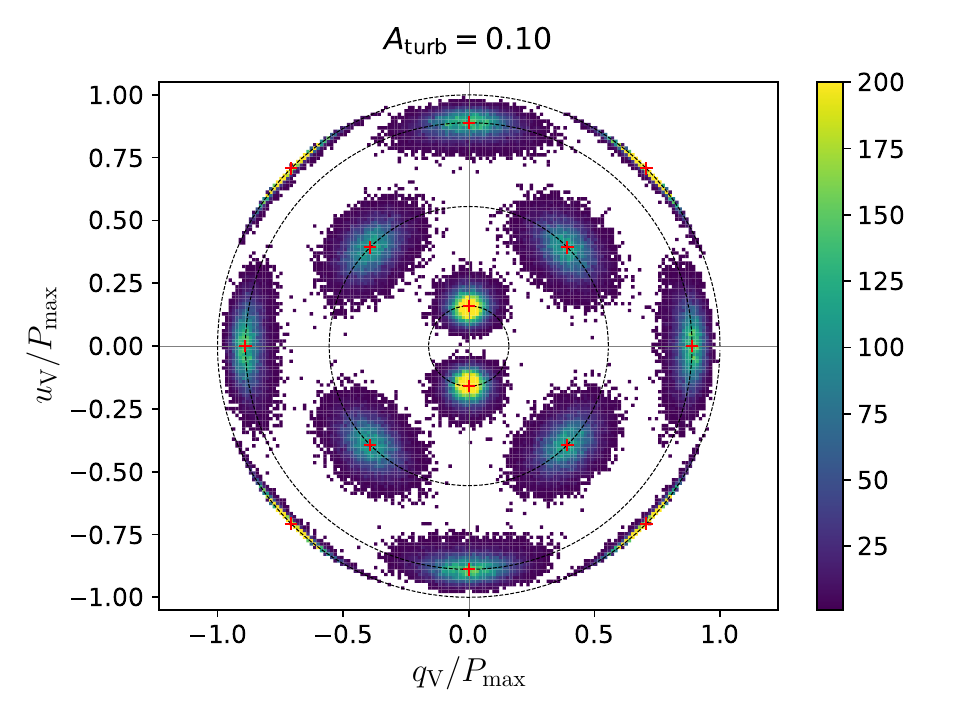}\\[-1.5ex]
    \caption{Examples of scatter produced in the $(q_{\rm{V}},\, u_{\rm{V}})$ plane by the addition of a stochastic component to a regular magnetic field for different inclination and position angle of the latter and for $A_{\rm{turb}} = 0.1$.
    Fourteen configurations are shown, as explained in the text. The red crosses indicate the $(q_{\rm{V}},\, u_{\rm{V}})$ obtained for $A_{\rm{turb}} = 0$ (regular field only). The scatter around each cross shows a 2D histogram of 10,000 realizations of $(q_{\rm{V}},\, u_{\rm{V}})$ values obtained with the addition of the stochastic component.
    We note that the $(q_{\rm{V}},\, u_{\rm{V}})$ are normalized by $P_{\rm{max}}$.
    }
    \label{fig:Covqu-B3D_Aturb01}
\end{figure}

We consider the following setup. 
We place an observer above the pole of the northern hemisphere ($b\geq0^\circ$) of a HEALPix map with $N_{\rm{side}} = 2$ (\citealt{Gorski2005}). This map contains 28 pixels to which correspond 3D orientations with colatitude and longitude coordinates. The former corresponds to the inclination of the magnetic field line with respect to the POS and latter to the position angle in the POS. By construction, these orientations are symmetric in longitude and therefore only half of the points do show different Stokes parameters given the definition of the polarization; $\psi_{\mathbf{B}}$ is a two-circular quantity defined in the range $[0^\circ,\, 180^\circ)$. Hence, we drop half of the northern hemisphere and thus consider 14 combinations of $(\gamma_{\mathbf{B}_{\rm{reg}}},\,\psi_{\mathbf{B}_{\rm{reg}}})$. To each, corresponds a pair of $(q_{\rm{V}}, \, u_{\rm{V}})$ at position $\cos^2 \gamma_\mathbf{B_{\rm{reg}}} \; (\cos[2\psi_\mathbf{B_{\rm{reg}}}],\,\sin[2\psi_\mathbf{B_{\rm{reg}}}])$ in the polarization plane (see Eq.~\ref{eq:stokes_reg}). Those positions are shown as red crosses in Fig.~\ref{fig:Covqu-B3D_Aturb01}. These crosses correspond to the Stokes parameters that would be observed for all background stars if the magnetic field was made of the regular component only. Then, we add the stochastic component in magnetic field geometry according to Eq.~\ref{eq:Btot}. This creates variations about those values. Example of variations obtained in the polarization plane for $A_{\rm{trub}} = 0.1$ is shown in Fig.~\ref{fig:Covqu-B3D_Aturb01}.

For most of the 14 configurations, the 2D scatter produced in the polarization plane cannot be considered as resulting from two independent normal distributions; the correlation coefficient is not zero in general. Instead, configurations with the regular magnetic field close to the POS are significantly asymmetric. Interestingly, depending on the position angle, the correlation coefficient can take large values. For example, in the case of $\gamma = 0^\circ$ and $\psi = 22.5^\circ$, the correlation coefficient is close to $-1$. The more the regular magnetic field points to the observer, the more the scatter becomes symmetric in $q_{\rm{V}}$ and $u_{\rm{V}}$.

At this stage, we notice that fluctuations in dust density distribution, or any other ISM parameters affecting the value of $P_{\rm{max}}$, would add radial scatter on this plot. These would then reduce the possible asymmetry and therefore would reduce the value of the correlation coefficient. From the shape of the 2D scatter seen in Fig.~\ref{fig:Covqu-B3D_Aturb01} we may conclude that modeling the effect of the intrinsic scatter through the characterization of a bivariate normal distribution with given covariance matrix is a fair approach which should be general enough to cover a large class of models of sources of fluctuations in the magnetized ISM.
This motivates our choice while writing the model equation in the main text (see Sect.~\ref{sec:model}). From this point, we therefore consider that the polarization signal in the $(q_{\rm{V}},\,u_{\rm{V}})$ plane, mean and intrinsic scatter, can be self-consistently described by a bivariate normal distribution centered on the mean polarization and with a covariance matrix potentially having a nonzero off-diagonal element.

\smallskip

A scatter in the $(q_{\rm{V}},\,u_{\rm{V}})$ plane corresponds to scatter in the degree of polarization and in the EVPA.
The scatter in EVPA is thought to inform on the degree of turbulence in clouds and to be related to the amplitude of the magnetic field (e.g., \citealt{Skalidis2021}), at least when the magnetic field lies mostly in the POS.
\begin{figure}
    \centering
    \includegraphics[trim={.4cm .4cm 0.2cm 0cm},clip,width=.95\linewidth]{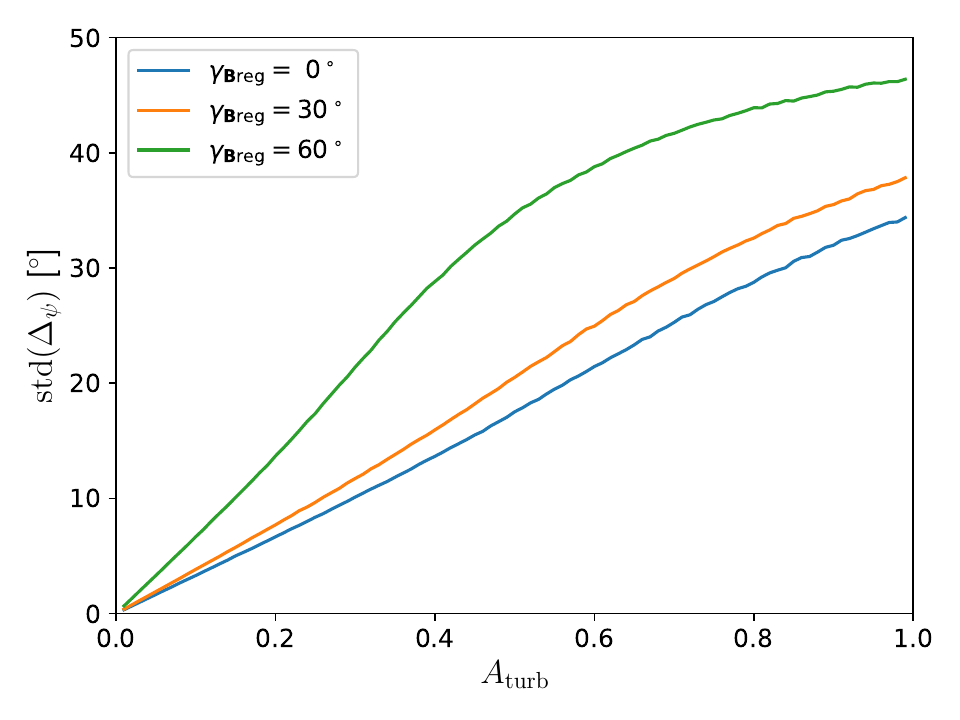}\\[-1.5ex]
    \caption{Scatter in EVPA as a function of $A_{\rm{turb}}$. For each value of $A_{\rm{turb}}$, $10^5$ realizations of $\mathbf{B}_{\rm{tot}}$ are computed. The Stokes parameters are computed and polarization position angles determined and compared to the polarization vector corresponding to the regular component only. The standard deviation of the polarization angle difference is then computed and plotted against $A_{\rm{turb}}$. We carry the analysis for $\gamma_{\mathbf{B}_{\rm{reg}}} = 0^\circ, 30^\circ$ and $60^\circ$.\\
    }
    \label{fig:stdDPPA-Aturb}
\end{figure}
We explore the dependence of the scatter in EVPA as a function of the parameter $A_{\rm{turb}}$. Because of projection effects, we carry this analysis for three values of $\gamma_{\mathbf{B}_{\rm{reg}}}$ ($0^\circ,\,30^\circ,\,60^\circ$). Our results are reported in Fig.~\ref{fig:stdDPPA-Aturb}. The EVPA scatter increases linearly at low values of $A_{\rm{turb}}$, then the increase slows down until the EVPA scatter reaches saturation.
The larger the angle of inclination, the greater the rate of increase and thus, breakup and saturation occur at lower values of $A_{\rm{turb}}$.
Given that there is currently a lack of starlight polarization data for diffuse sightlines, there is not much observational constraints on the scatter in EVPA and, hence, on the value that $A_{\rm{turb}}$ may take. In small angular regions toward denser clouds or denser sightlines, authors report values of the EVPA scatter to range from few degrees up to about fifteen degrees (\citealt{Soler2016}; \citealt{Planck2016XXXV}; \citealt{Panopoulou2016}; \citealt{Skalidis2022}), also consistent with numerical simulations of the ISM in sub- and trans-Alv{\'e}nic regimes (e.g., \citealt{Skalidis2021b}). We assume that those values are representative to our case. Therefore, we choose values of $A_{\rm{turb}}$ in the range 0.1 to 0.3 to be representative of degree of intrinsic scatter we may find at intermediate and high Galactic latitude, in general, from our future survey.

\smallskip

\begin{figure}
    \centering
    \includegraphics[trim={.4cm .4cm 0.2cm 0cm},clip,width=.95\linewidth]{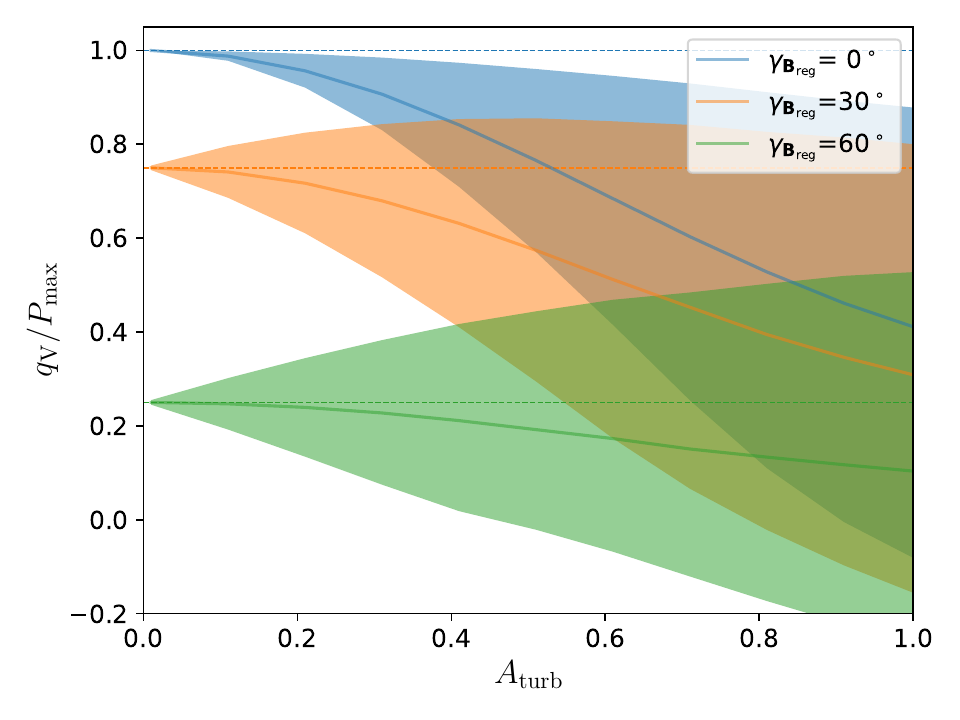}\\[-1.5ex]
    \caption{Distributions of $q_{\rm{V}}/P_{\rm{max}}$ corresponding to various inclination angle and amplitude of the intrinsic scatter.
    We carry the analysis out for $\gamma_{\mathbf{B}_{\rm{reg}}} = 0^\circ,\, 30^\circ$, and $60^\circ$ and $A_{\rm{turb}} \in [0,\,1]$.
    For each value pair, $10^5$ realizations of $\mathbf{B}_{\rm{tot}}$ are computed and $q_{\rm{V}}$ is evaluated.
    The continuous lines show the mean of the distribution for $\gamma_{\mathbf{B}_{\rm{reg}}} = 0^\circ,\,30^\circ,\,\text{and}\,60^\circ$ in blue, orange and green, respectively. The shaded areas indicate the range span between percentiles 16 and 84 of the distribution. The dashed lines indicate the $q_{\rm{V}}/P_{\rm{max}}$ values obtained when no stochastic term is added in the magnetic field. 
    }
    \label{fig:pV-Aturb}
\end{figure}

In Fig.~\ref{fig:pV-Aturb}, we show distributions of the Stokes parameter $q_{\rm{V}}$ (normalized to $P_{\rm{max}}$) as a function of $A_{\rm{turb}}$ and for the same three inclination angle values studied before. We show the mean (continuous lines) and the interval between percentiles 16 and 84 (shaded areas) of the distributions. We see that depending on the inclination angle there might be a nonzero difference between the mean of the distributions and the values corresponding to the regular field only (dashed lines). This is a bias that purely results from projection effects. This bias is larger for small inclination angles (i.e., magnetic field close to the POS) than for large inclination angles and vanishes when the field lines point toward the observer.
The same figure is obtained for Stokes $u_{\rm{V}}$ but with a position angle rotated by 45$^\circ$.
This result is consistent with what \cite{Hu2022} obtained from numerical simulations of incompressible MHD turbulence in molecular cloud.

\medskip

In the following we explore the effects of the intrinsic scatter on the element of the covariance matrix of the $(q_{\rm{V}},\,u_{\rm{V}})$ pairs that it generates. In particular, we are interested in exploring the dependence on the inclination and position angle of the regular magnetic field and on the dependence with the amplitude of the stochastic component relative to the regular one.

\begin{figure}
    \centering
    \includegraphics[trim={0.4cm 0.4cm 0.4cm 0cm},clip,width=.95\linewidth]{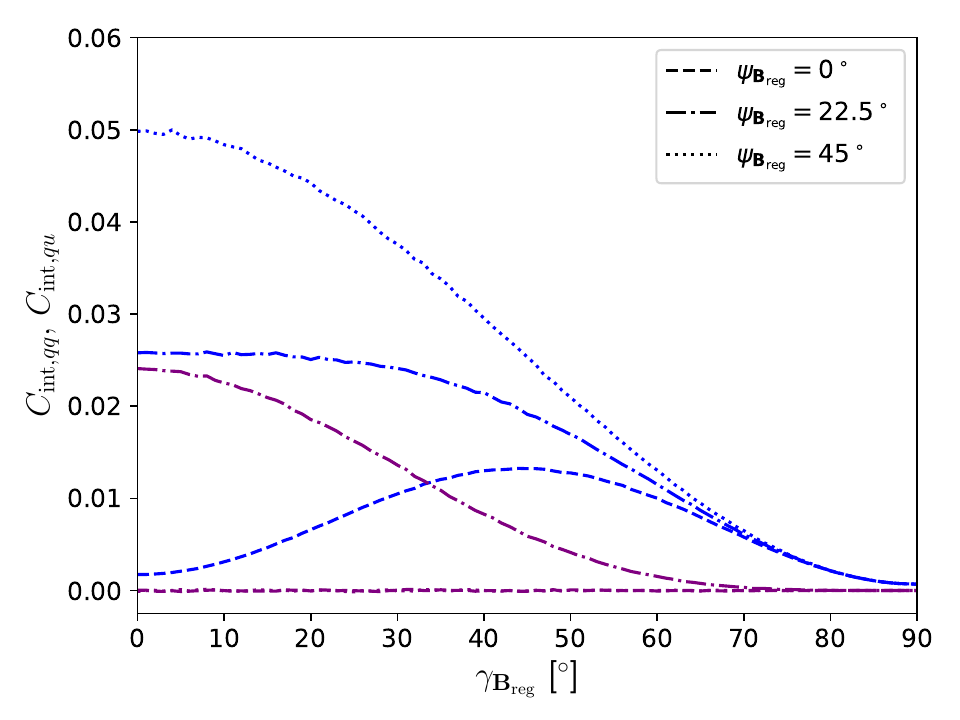}\\[-1.5ex]
    \caption{Dependence of $C_{{\rm{int}},qq}$ (blue) and $C_{{\rm{int}},qu}$ (purple) as a function of $\gamma_{\mathbf{B}_{\rm{reg}}}$ and for three values of the position angles: $\psi_{\mathbf{B}_{\rm{reg}}} = 0^\circ$ (dashed lines), $\psi_{\mathbf{B}_{\rm{reg}}} = 22.5^\circ$ (dashed-dotted lines), and $\psi_{\mathbf{B}_{\rm{reg}}} = 45^\circ$ (dotted lines).
    A value of $A_{\rm{turb}} = 0.2$ is chosen here.  
    The elements of the covariance matrix are given in units of $P_{\rm{max}}^2$. A value of 0.01 corresponds to a polarization uncertainty ($\sigma_{q}$ or $\sigma_{q}$) of 10 per cent of $P_{\rm{max}}$.
    }
    \label{fig:CqqCqu-gamma}
\end{figure}

In Fig.~\ref{fig:CqqCqu-gamma}, we present the evolution of $C_{\rm{int},qq}$ and $C_{\rm{int},qu}$ as a function of $\gamma_{\mathbf{B}_{\rm{reg}}}$, for the three position angles $\psi_{\mathbf{B}_{\rm{reg}}} = 0^\circ,\,22.5^\circ,\,\text{and}\, 45^\circ$ and for $A_{\rm{turb}} = 0.2$.
We do not show $C_{{\rm{int}},uu}$, as it is identical to $C_{{\rm{int}},qq}$ but for a position angle rotated by $45^\circ$.
We see that the covariance-matrix elements are nontrivial but certainly trigonometric functions of the inclination angle. The cross term vanishes when $\psi_{\mathbf{B}_{\rm{reg}}} = 0^\circ$ or any multiple of $45^\circ$ (i.e., when either $q_{\rm{V}}$ or $u_{\rm{V}}$ is zero).
If fluctuations in magnetic field geometry were the only sources of intrinsic scatter, a detailed analysis of the covariance-matrix elements from observational data points could potentially lead to constraints on inclination angle of the magnetic field with respect to the POS. This information is generally not accessible from dust-related polarization observables and providing a direct access to it would constitute a breakthrough in dust studies of the magnetized ISM. However, as we have already noticed, other sources of fluctuations are expected in the turbulent ISM and they directly affect the values of $P_{\rm{max}}$, adding a certainly non-negligible contribution to $C_{{\rm{int}},qq}$ and $C_{{\rm{int}},uu}$ and reducing the relative importance of $C_{{\rm{int}},qu}$, thus bringing some complexity to the situation depicted in Fig.~\ref{fig:CqqCqu-gamma}. While future work will be necessary to clarify whether those sources of scatter can be disentangled, we note that other sources of complication will come from both observational noise and from the a-priori unknown value of $A_{\rm{turb}}$.

\medskip

Finally, we address the question of the significance of the intrinsic scatter and how it compares to observational uncertainties for a large range of $A_{\rm{turb}}$ values and for different inclination and position angles of the regular component of the magnetic field.
For this investigation we consider a sample of 100 stars in the background of a cloud. We fix our observational uncertainty on the polarization as $\sigma_p = \sigma_q = \sigma_u = 0.2 \%$ and we fix a degree of polarization of the cloud to be $p_{\rm{C}} = 0.6\%$ in absence of scatter. We thus change the value of $P_{\rm{max}}$ to compensate for change of $\gamma_{\mathbf{B}_{\rm{reg}}}$.

\begin{figure*}
    \centering
    \includegraphics[trim={.9cm 0.2cm 1.cm 0cm},clip,width=.98\linewidth]{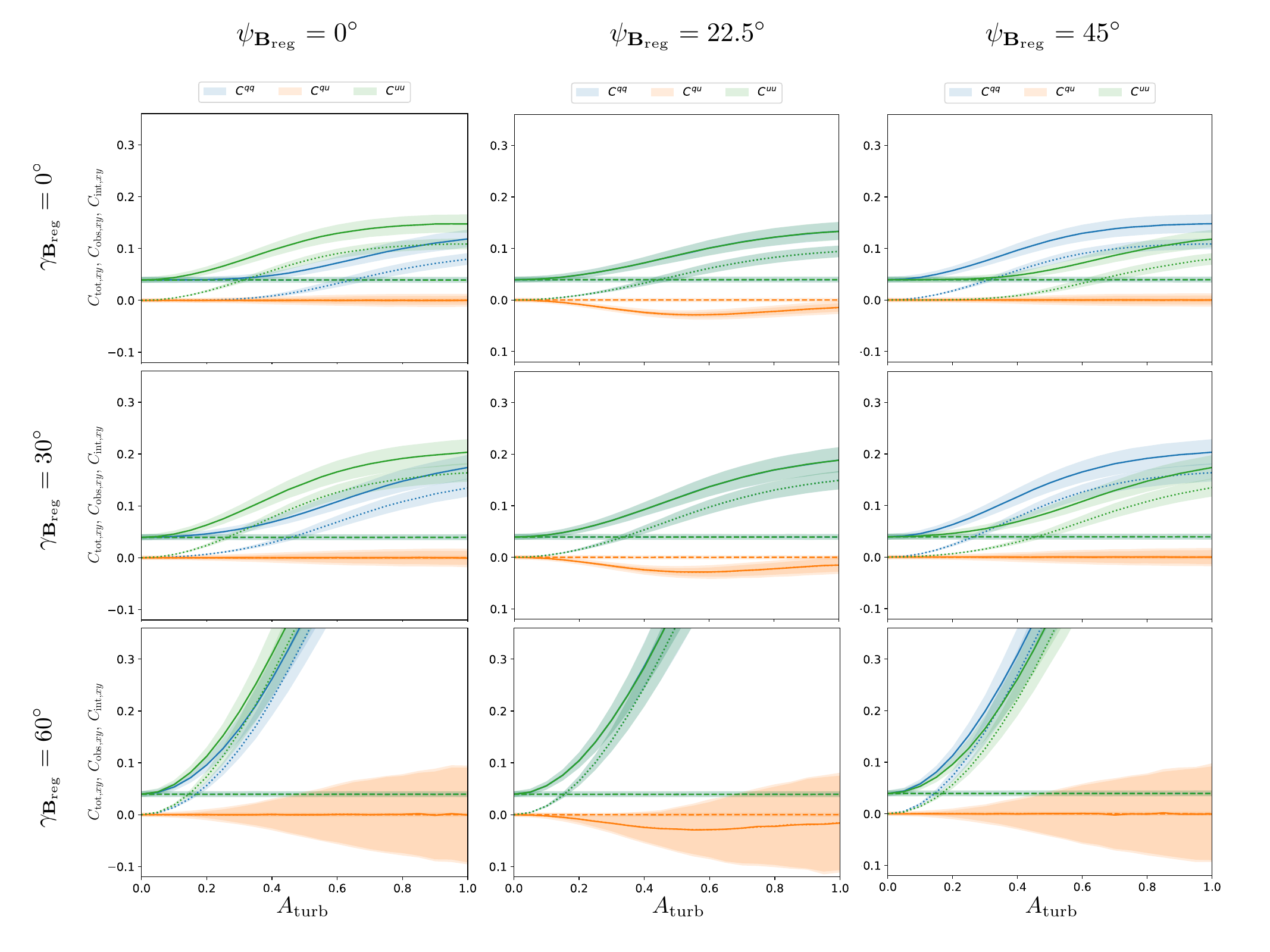}\\[-2.ex]
    \caption{Comparison of the contributions from the observational noise and from the intrinsic scatter to the different elements of the covariance matrix as a function of the amplitude of the stochastic component and for different configurations of $\mathbf{B}_{\rm{reg}}$. We show the total elements of the covariance matrix (solid lines) along with the contribution from the noise (dashed lines) and from the intrinsic scatter (dotted lines). Blue, green and orange correspond to $C_{{\rm{int}},qq}$, $C_{{\rm{int}},uu}$ and $C_{{\rm{int}},qu}$ respectively. The lines and shaded areas show the median and 16 and 84 percentile of the values obtained while repeating the analysis for 10,000 simulations in which both the noise and the intrinsic scatter vary. In this simulation we have set $p_{\rm{reg}} = 0.6\%$ and $\sigma_p = \sigma_q = \sigma_u = 0.2\%$ corresponding roughly to a S/N of 3 for the individual star polarization measurement.}
    \label{fig:COVqu_totnoiseturb}
\end{figure*}

For each set of ($A_{\rm{turb}}$,\,$\gamma_{\mathbf{B}_{\rm{reg}}}$,\,$\psi_{\mathbf{B}_{\rm{reg}}}$) values we can generate simulated Stokes parameters (i) computed only from $\mathbf{B}_{\rm{reg}}$, (ii) computed with $\mathbf{B}_{\rm{tot}}$ for specified $A_{\rm{turb}}$ and (iii) with the addition of the observational noise. For each simulated data set we estimate the total covariance matrix of the Stokes parameters and the contributions from the observational and from the intrinsic scatter taken separately. Denoting $C_{{\rm{tot}},xy} = C_{{\rm{obs}},xy} + C_{{\rm{int}},xy}$ with $x,\,y$ being either $q$ or $u$, we can measure the different terms in our simulations and thus compare the relative amplitude of the different contributions. We perform this exercise for 10,000 simulated data sets to build distributions of the elements of the covariance matrix.
The result is shown in Fig.~\ref{fig:COVqu_totnoiseturb} in which, for each matrix element taken separately and for each of their contributions (noise and intrinsic scatter), we show how the distributions change as a function of $A_{\rm{turb}}$ by reporting the median and the interval between 16 and 84 percentiles as shaded areas.

The relative importance of the contribution from the intrinsic scatter with respect to the contribution from the noise to the total scatter in the data points is well demonstrated by the different panels of Fig.~\ref{fig:COVqu_totnoiseturb}. Depending on the exact setup of the ISM and amplitude of the stochastic component in the magnetic field, the data scatter can be dominated either by the intrinsic scatter or by the noise. Keeping in mind that this also depends on the S/N level which we have fixed to $\approx 3$ for individual measurements, we see that the contribution from the intrinsic scatter exceeds the noise contribution in at least one of the polarization channels when $A_{\rm{turb}} \gtrsim 0.3$ and that this threshold reduces when $\gamma_{\mathbf{B}_{\rm{reg}}}$ increases. When either or both the amplitude of the cloud polarization gets lower or the overall uncertainty on individual star measurements gets larger, this threshold goes to lower $A_{\rm{turb}}$ values (not shown in the figure).

\end{appendix}

\label{lastpage}
\end{document}